\documentclass[a4paper,10pt]{article}
\pdfoutput=1 
\usepackage{jheppub} 
\usepackage{hyperref}
\hypersetup{
	bookmarks=true,         
	unicode=false,          
	pdftoolbar=true,        
	pdfmenubar=true,        
	pdffitwindow=false,     
	pdfstartview={FitH},    
	pdftitle={My title},    
	pdfauthor={Author},     
	pdfsubject={Subject},   
	pdfcreator={Creator},   
	pdfproducer={Producer}, 
	pdfkeywords={keyword1} {key2} {key3}, 
	pdfnewwindow=true,      
	colorlinks=true,       
	linkcolor=red,          
	citecolor=purple,        
	filecolor=magenta,      
	urlcolor=blue,           
	linktocpage=true
}
\usepackage{graphics}
\usepackage{rotating}
\usepackage{subfigure}
\usepackage{physics}
\usepackage{amssymb}
\usepackage{xcolor}
\usepackage{pstricks}
\usepackage{amsfonts}
\usepackage{mathrsfs,amsmath}
\usepackage{epsfig}
\usepackage{amsmath,amssymb,amsthm,graphicx,latexsym}
\usepackage{float}
\usepackage{textcomp}
\usepackage{array}
\usepackage{booktabs}
\usepackage{dsfont}

\usepackage{ulem}

\newcommand{\be}{\begin{equation}}
\newcommand{\ee}{\end{equation}}
\newcommand{\bc}{\begin{center}}
	\newcommand{\ec}{\end{center}}
\newcommand{\bea}{\begin{eqnarray}}
\newcommand{\eea}{\end{eqnarray}}
\newcommand{\bml}{\begin{subequations}}
	\newcommand{\eml}{\end{subequations}}
\newcommand{\bfig}{\begin{figure}}
	\newcommand{\efig}{\end{figure}}

\newcommand{\slashed}{\hspace{-1.1ex}/}

\newcommand{\bmat}{\begin{pmatrix}}
	\newcommand{\emat}{\end{pmatrix}}
\begin{document}
	\title{\textsc{\fontsize{25}{17}\selectfont \sffamily \bfseries \textcolor{purple}{Quantum entanglement in de Sitter space from Stringy Axion: An analysis using $\alpha$ vacua}}}

	\author[a,b]{Sayantan Choudhury,
		\footnote{\textcolor{purple}{\bf Alternative
				E-mail: sayanphysicsisi@gmail.com}. ${}^{}$}}
\author[c,d,e]{Sudhakar Panda
}
	\affiliation[a]{Quantum Gravity and Unified Theory and Theoretical Cosmology Group, Max Planck Institute for Gravitational Physics (Albert Einstein Institute),
	   Am M$\ddot{u}$hlenberg 1,
	   14476 Potsdam-Golm, Germany.}
	   \affiliation[b]{
	Inter-University Centre for Astronomy and Astrophysics, Post Bag 4,
	Ganeshkhind, Pune 411007, India.
	}
	\affiliation[c]{Institute of Physics, Sachivalaya Marg, Bhubaneswar, Odisha - 751005, India.
		}
		\affiliation[d]{	
		National Institute of Science Education and Research,
		Jatni, Bhubaneswar, Odisha - 752050, India.}
		\affiliation[e]{Homi Bhabha National Institute, Training School Complex,
		Anushakti Nagar, Mumbai-400085, India.
		}
	\emailAdd{sayantan@aei.mpg.de,sayantan.choudhury@aei.mpg.de,\\~~~~~~~panda@iopb.res.in}

	\abstract{In this work, we study the phenomena of quantum entanglement by computing de Sitter entanglement entropy from von Neumann measure. For this purpose we consider a bipartite quantum field theoretic set up for axion field, previously derived from ${\bf Type~ II~B}$ string theory compactified to four dimensions. We consider the initial vacuum to be CPT invariant non-adiabatic $\alpha$ vacua state under ${\bf SO(1,4)}$ isometry, which is characterized by a real one-parameter family. To implement this technique we use a ${\bf S^2}$ which divide the de Sitter into two exterior and interior sub-regions. First, we derive the wave function of axion in an open chart for $\alpha$ vacua by applying Bogoliubov transformation on the solution for Bunch-Davies vacuum state. Further, we quantify the density matrix by tracing over the contribution from the exterior region. Using this result we derive entanglement entropy, R$\acute{e}$nyi entropy and explain the long-range quantum effects in primordial cosmological correlations. Our results for $\alpha$ vacua provides the necessary condition for generating non zero entanglement entropy in primordial cosmology. 
	}

	\maketitle
	\flushbottom
	\section{\textcolor{blue}{Introduction}}
	
	It is well accepted fact that von Neumann entropy is a measure of quantum entanglement to quantify long range correlation in condensed matter physics \cite{Amico:2007ag,Horodecki:2009zz,Laflorencie:2015eck} and cosmology \cite{MartinMartinez:2012sg,Nambu:2008my,Campo:2005qn,Nambu:2011ae,VerSteeg:2007xs,Mazur:2008wa,Maldacena:2012xp,Maldacena:2015bha,Choudhury:2016cso,Choudhury:2016pfr,Choudhury:2017bou,Kanno:2014lma,Kanno:2017dci,Kanno:2016gas,Kanno:2014bma,Kanno:2014ifa,Fischler:2013fba,Fischler:2014tka}. In condensed matter physics entanglement entropy exactly mimics the role of an order parameter and the corresponding phase transition phenomena can be characterized by correlations at  quantum level. Also, it is expected that, from this understanding of long range effects in quantum correlations, we can understand the underlying physics of the theory of multiverse, bubble nucleation etc. in de Sitter space \cite{Coleman:1980aw}. As a consequence, we can observe a prompt response due to the local measurement in quantum physics, by violating causal structure of the space-time. In quantum theory such causality violation is known as Einstein-Podolsky-Rosen (EPR) paradox \cite{Bell:1964kc}. But in such type of local measurement  causality remains unaffected as the required quantum information is not propagating. In this context Schwinger effect in de Sitter space  \cite{Frob:2014zka,Fischler:2014ama} is one of the prominent examples of quantum entanglement. In Schwinger effect, particle pair creation takes place with a finite separation in de Sitter space-time in presence of a constant electric field \cite{Kanno:2014lma} and the quantum states exhibit long range correlation. 
	
	To quantify entanglement entropy in the context of quantum field theory, one requires to have a bipartite system. In strong coupling regime of a such a theory we can compute entanglement entropy by using the principles of gauge gravity duality in the bulk \cite{Ryu:2006bv,Ryu:2006ef,Nishioka:2009un,Rangamani:2016dms,Hubeny:2007xt,Dong:2013qoa,Camps:2013zua,Banerjee:2014oaa,Bhattacharyya:2014yga,Pal:2015mda}. Using this technique  many issues have been addressed in the context of holographic entanglement entropy. Further, in ref.~\cite{Maldacena:2012xp} the authors have constructed a completely different computational algorithm to quantify entanglement entropy using Bunch Davies initial state\footnote{We note that, Bunch Davies initial state is exact equivalent to the Euclidean or adiabatic vacuum state in quantum field theory.} in de Sitter space. Later, in ref.~\cite{Kanno:2014lma} this has been extended  to the computation of entanglement entropy using $\alpha$ vacua initial state in de Sitter space by following the techniques presented in ref.~\cite{Maldacena:2012xp}. Moreover, using Bunch Davies initial state in de Sitter space,  we have computed entanglement entropy in a field theory where the effective action contains a linear source term  \cite{Choudhury:2017bou}.  We have shown that this result complements the necessary condition for the violation of Bell's inequality in primordial cosmology.

	In quantum field theory in de Sitter background, one can  construct a one parameter family of initial vacuum states which are CPT invariant under ${\bf SO(1,4)}$ isometry group \cite{Mottola:1984ar,Allen:1985ux}. These class of states are characterized by  a real parameter $\alpha$, known as $\alpha$ vacua. For $\alpha=0$ these states reduce to the usual Bunch Davies vacuum state.  In a more technical terms, the $\alpha$ vacua can be treated as squeezed quantum states, which are created by an unitary operator acting on the Bunch Davies vacuum state. This leads to the generalisation of Wick's theorem in interacting quantum field theory, which allows us to describe any free quantum field theory Green's function computed  using $\alpha$ vacua in terms of the products of the Green's functions computed using Bunch Davies vacuum \cite{Goldstein:2003ut}. See refs.~\cite{Allen:1985ux,deBoer:2004nd,Brunetti:2005pr} for more details on the quantum field theory of $\alpha$ vacua. For a specific $\alpha$ vacuum state as a quantum initial condition, it is possible to describe the long range correlations within the framework of quantum field theory. As a consequence, the non-local quantum phenomena can be associated with the long range effects, which is described by quantum entanglement of vacuum state as an initial condition. We note that till date no such experiment is available using which one can able to test the local behavior of quantum field theory in cosmological scale (Hubble scale). However, it is expected that in future it may be possible to test such prescriptions.  Additionally, it is important to note that, propagators in free quantum field theory of de Sitter space computed in presence of adiabatic Bunch Davies vacuum state manifest Hadamard singularity which is consistent with the result obtained in the context of Minkowski flat space-time limit \cite{Kay:1988mu,Wald}.  
	
	However, for interacting quantum field theory in the de Sitter space-time background, such  singular propagators applicable for adiabatic Bunch Davies vacuum are dubious. Hence $\alpha$ vacuum state plays significant role, using which one can express the propagators in interaction picture. In the quantum field theory described by the $\alpha$ vacua state the real parameter $\alpha$ plays the role of super-selection number associated with a quantum state of a different bipartite Hilbert space. But it is still a debatable issue that whether the interaction picture of the quantum field theory with any arbitrary $\alpha$ vacua with any super-selection rule are consistent with the physical requirements of quantum mechanics or not \cite{deBoer:2004nd}. In general, one can treat the $\alpha$ vacua as a family of quantum initial state, where we have quantum fluctuation around an excited state. Here it might be possible that the Hilbert space corresponding to excited state (for $\alpha$ vacua) and the adiabatic Bunch Davies vacuum coincides with each other. In such a situation it is perfectly consistent to describe quantum field theory of excited state in de Sitter space in terms of the adiabatic Bunch Davies vacuum in the ultraviolet regime\footnote{In the infrared regime, due to the nonremoval of divergences appearing from various interaction in quantum field theory, explaining the physics of excited states with the adiabatic vacuum is not a good approximation.} As a result this specific identification allows us to write an effective field theory description in the ultraviolet regime . This implies that identifying the correct and more appropriate quantum $\alpha$ vacuum state is fine tuned. However, this fine tuning only allows us to describe the quantum field theory with any excited states compared to ground state described by the adiabatic Bunch Davies vacuum. Using this prescription apart from inflationary paradigm, one may be able to explain a lot of unexplored late time physical phenomena of nonstandard vacuum state. In this work, we further generalise the computational strategy of entanglement entropy for axion  field using $\alpha$ vacua initial state in de Sitter space. This result will establish the generation of quantum entanglement entropy in early universe in a more generalised fashion. In this setup, while the possibility of EPR pair creation increases, it appears that the quantum long range correlation will increase naturally. As a consequence, the amount of entanglement entropy increases as the parameter $\alpha$ increases. In this report, we have investigated this possibility with a specific model of axion field theory previously  derived from {\bf Type IIB} string theory \cite{McAllister:2008hb,Silverstein:2008sg,McAllister:2014mpa,Panda:2010uq} setup. Here, we will  demonstrate the Bell's inequality violation from nonzero entanglement entropy of axion field. This connection also will be helpful in future to provide a theoretical tool to compare various models of inflation \cite{Choudhury:2015hvr,Maharana:1997cz,Baumann:2009ds,Agarwal:2011wm}. We have also commented on R$\acute{e}$nyi entropy using the same setup which will finally give rise to a complete new interpretation to long range quantum correlation for the case of $\alpha$ vacua. 
	

	We note that for cosmology, it is crucial to know  the observational constraints on the $\alpha$ (non- Bunch Davies) vacua from CMB maps \cite{Kanno:2014ifa,Kanno:2015ewa,Dimitrakopoulos:2015yva}. It is expected that the (auto and cross) correlation functions of primordial fluctuations get modified significantly, which is  an important information to understand the underlying new physics of $\alpha$ vacua. Also this will help us to discriminate between the physical outcomes of $\alpha$ vacua and the adiabatic Bunch Davies vacuum state. 
		
 Similarly, for the case of gravity, it is also important to understand the physical implications of the new physics originated from $\alpha$ vacua described in a specific curved gravitational background. Note that Einstein General Theory of relativity is a classical field theoretic description, which describes the interactions in astrophysical scales and constrained by galaxy rotation curves, dynamics of clusters etc. \cite{deBoer:2004nd,Aguirre:2001xs}. However, in the infrared regime of the gravity sector, we do not have observational probes to test the infrared correction to the classical field theory of gravity.  From the theoretical perspective, if we describe the fluctuation in the metric in terms of spin 2 transverse, traceless degrees of freedom in de Sitter space-time, then using $\alpha$ vacua and the non local field redefinition in metric ,one can express the scalar degrees of freedom also. This scheme needs to be developed in future.
			
	This paper is organised as follows. In \underline{\textcolor{violet}{\bf section} \ref{2}}, we briefly review the basic set up using which we will compute the entanglement entropy and R$\acute{e}$nyi entropy using $\alpha$ vacua. In \underline{\textcolor{violet}{\bf section} \ref{3a1}}, we introduce the axion model from string theory. Then using this model we compute the expression for the wave function in a de Sitter hyperbolic open chart in presence of Bunch Davies vacuum in \underline{\textcolor{violet}{\bf section} \ref{3av1}}. Further using Bogoliubov transformation we express the solution in terms of  new basis, called $\alpha$ vacua in \underline{\textcolor{violet}{\bf section} \ref{3a2}}. After that in \underline{\textcolor{violet}{\bf section} \ref{3b}}, we construct the density matrix in presence of $\alpha$ vacua. Using this result further in \underline{\textcolor{violet}{\bf section} \ref{3c}}, we derive the expression for the von Neumann entropy which is the measure of entanglement entropy in presence of $\alpha$ vacua. Next in \underline{\textcolor{violet}{\bf section} \ref{3c}}, we compute R$\acute{e}$nyi entropy using the result of density matrix as derived in \underline{\textcolor{violet}{\bf section} \ref{3b}}. Finally we conclude in \underline{\textcolor{violet}{\bf section} \ref{4}} with some future prospects of the present work.

	\section{\textcolor{blue}{Basic setup: Brief review}}
	\label{2}	
	
		In this section we briefly review the computational method to derive entanglement entropy in de Sitter space following the work performed in ref.~\cite{Maldacena:2012xp} and ref.~\cite{Choudhury:2017bou}. We consider a time preserving space-like hypersurface ${\bf S^2}$ for this purpose. As a result ${\bf S^2}$ is divided into two sub regions, interior and exterior,  which are characterized by \textcolor{purple}{\bf RI} ($\equiv$\textcolor{purple}{\bf L}) and \textcolor{red}{\bf RII} ($\equiv$\textcolor{red}{\bf R}). In terms of the Lorentzian signature an open chart in de Sitter space is described by three different subregions Further\cite{Maldacena:2012xp,Choudhury:2017bou}:
		    \bea
		               \label{x2}
		      \displaystyle \textcolor{red}{\bf R(=RII)}/ \displaystyle \textcolor{purple}{\bf L(=RI)}&:&\footnotesize\displaystyle\left\{\begin{array}{ll}
		     \displaystyle \tau_{\rm E}=\pi/2-it_{\bf R}/\tau_{\rm E}=-\pi/2+it_{\bf L}~~~~~~~~~~~~ &
		                                                               \mbox{\small {\textcolor{red}{\bf for $t_{\bf R/L}\geq 0$}}}  
		                                                              \\ 
		              \displaystyle \rho_{\rm E}=-ir_{\bf R/L} & \mbox{\small { \textcolor{red}{\bf for $r_{\bf R/L}\geq 0$}}}.~~~~~~~~
		                                                                        \end{array}
		                                                              \right.\\
		\label{x3}
		      \displaystyle \textcolor{blue}{\bf C}&:&\footnotesize\displaystyle\left\{\begin{array}{ll}
		     \displaystyle \tau_{\rm E}=t_{\bf C}~~~~~~~~~~~~~~~~~~~~~~~~~~~~~~~~~~~~~~~~~ &
		                                                               \mbox{\small {\textcolor{blue}{\bf for $-\pi/2\leq t_{\bf C}\leq \pi/2$}}}  
		                                                              \\ 
		              \displaystyle \rho_{\rm E}=\pi/2-ir_{\bf C} & \mbox{\small { \textcolor{blue}{\bf for $-\infty<r_{\bf c}< \infty$}}}.~~~~~~~~
		                                                                        \end{array}
		                                                              \right. \eea
		 Also in open chart the metric with Lorentzian signature can be written as \cite{Maldacena:2012xp,Choudhury:2017bou}:                                                             
		  \bea
		                 \label{x2z}
		        \displaystyle \textcolor{red}{\bf R(=RII)}/ \displaystyle \textcolor{purple}{\bf L=(RI)}&:&\footnotesize\displaystyle\left\{\begin{array}{ll}
		       \displaystyle ds^2_{\bf R/L}=H^{-2}\left[-dt^2_{\bf R/L}+\sinh^2t_{\bf R/L}\left(dr^2_{\bf R/L}+\sinh^2r_{\bf R/L}~d\Omega^2_{\bf 2}\right)\right], 
		                                                                          \end{array}
		                                                                \right.\\
		  \label{xx3}
		        \displaystyle \textcolor{blue}{\bf C}&:&\displaystyle\left\{\begin{array}{ll}
		       \displaystyle  ds^2_{\bf C}=H^{-2}\left[dt^2_{\bf C}+\cos^2t_{\bf C}\left(-dr^2_{\bf C}+\cosh^2r_{\bf C}~d\Omega^2_{\bf 2}\right)\right], \end{array}
		                                                                \right. \eea
		where $H=\dot{a}/a$ is the Hubble parameter and $d\Omega^2_{\bf 2}$ represents angular part of the metric on ${\bf S}^2$.

		Now let us assume that the total Hilbert space of the local quantum mechanical system is described by ${\bf \cal H}$, which can be written using bipartite decomposition in a direct product space \cite{Callan:1994py} as,
		${\bf \cal H}={\bf \cal H}_{\bf INT}\otimes {\bf \cal H}_{\bf EXT}$.
		Here ${\bf \cal H}_{\bf INT}$ and ${\bf \cal H}_{\bf EXT}$ are the Hilbert space associated with interior and exterior region and describe the localised modes in \textcolor{purple}{\bf RI} and \textcolor{red}{\bf RII} respectively. Consequently, one can construct the reduced density matrix for the internal \textcolor{purple}{\bf RI} region by tracing over the external \textcolor{red}{\bf RII} region and the Von Neumann entropy measure, the entanglement entropy in de Sitter space can be expressed as:
		\bea \rho(\alpha)={\bf Tr}_{\bf R}|{\bf \alpha}\rangle \langle {\bf \alpha}| \Longrightarrow S(\alpha)= -{\bf Tr}\left[\rho(\alpha)\ln \rho(\alpha)\right].\eea
		Here the vacuum state $|{\bf \alpha}\rangle$ is the $\alpha$ vacuum. The reduced density matrix, which is a key ingredient  for computing entanglement entropy, is obtained by tracing over the exterior (\textcolor{red}{\bf R}) region. Also it is important to note that the total entanglement entropy can be expressed as a sum of UV divergent and finite contribution. In $3+1$ D, the UV-divergent part of the entropy can be written as \cite{Maldacena:2012xp,Kanno:2014lma,Choudhury:2017bou}:
	    \bea \label{ecvnt} S_{\bf UV-divergent}&=&{\bf c_1}\epsilon^{-2}_{\bf UV}{\cal A}_{\bf ENT}+\left[{\bf c_2}+\left({\bf c_3}m^2+{\bf c_4}H^2\right){\cal A}_{\bf ENT}\right]\ln\left(\epsilon_{\bf UV}H\right),\\
	     \label{aetw} S_{\bf UV-finite}&=&{\bf c_5}{\cal A}_{\bf ENT}H^2-{\bf c_6}\ln\left(\sqrt{{\cal A}_{\bf ENT}}H\right)+{\bf finite~terms}.\eea
	    where $\epsilon_{\bf UV}$ is the short distance lattice UV cut-off, ${\cal A}_{\bf ENT}$ is the proper area of the entangling region of ${\bf S^2}$ and ${\bf c_i}\forall i=1,2,3,4$ are the coefficients. Here we restrict ourself only within the UV-finite part which contains the information of long range effects of quantum state. Here 
	       ${\bf c_6}$ quantify the long range effect. In general, ${\bf c_6}$ can be expressed as \cite{Maldacena:2012xp,Kanno:2014lma,Choudhury:2017bou}, 
	      ${\bf c_6}\equiv S_{\bf intr},$ where $S_{\bf intr}$ is the UV-finite relevant part which we quantify in later sections. 
  
    \section{\textcolor{blue}{Quantum entanglement for axionic pair using $\alpha$ vacua}}
    \label{3}

    \subsection{Wave function of axion in open chart}
    \label{3a}
    
       \subsubsection{Model for axion effective potential}
       \label{3a1}
        In this section our prime objective is to compute de Sitter entanglement entropy 
            for axion field. Such axion field  appears from RR sector of {\bf Type IIB} string theory compactified on ${\bf CY^3}$ in presence of ${\bf NS~5}$ brane. For details, see refs.~\cite{McAllister:2008hb,Silverstein:2008sg,McAllister:2014mpa,Panda:2010uq,Svrcek:2006yi}. Let us start with the following effective action for axion field:
        \bea\label{axi}  S_{axion}&=& \int d^{4}x \sqrt{-g}\left[-\frac{1}{2}(\partial \phi)^2 +\mu^3\phi+\Lambda^4_{G}\cos\left(\frac{\phi}{f_{a}}\right)\right]= \int d^{4}x \sqrt{-g}\left[-\frac{1}{2}(\partial \phi)^2 +\mu^3\left[\phi+bf_{a}\cos\left(\frac{\phi}{f_{a}}\right)\right]\right],~~~~~\eea
 where $\mu^3$ is the mass scale, $f_a$ is the decay constant of axion and we introduce a parameter $b$, is defined as,  
        $b= \frac{\Lambda^4_{G}}{\mu^3 f_{a}}.$ Here $\Lambda_{G}$ can be expressed  as,  $ \Lambda_{G}=\sqrt{\frac{m_{SUSY} L^3}{ \sqrt{\alpha^{'}}g_{s}}}~e^{-cS_{inst}},$
                   where $S_{inst}$ is the instanton action, factor $c\sim{\cal O}(1)$, $m_{SUSY}$ is SUSY breaking scale, $\alpha^{'}$ represents Regge slope parameter, $g_s$ characterises the string coupling constant and $L^6$ is the world volume factor. Here we restrict up to the linear term of the effective potential as given by $V(\phi)\approx\mu^3\phi,$ which can be interpreted as a massless source in the equation of motion. In the limit $\phi<<f_a$, the total effective potential for axion can be approximated as, $V(\phi)\approx\mu^3\left(bf_{a}+\phi\right)-\frac{m^2_{axion}}{2}\phi^2,$
where we introduce the effective mass of the axion as,
                            $ m^2_{axion}=\frac{\mu^3 b}{f_{a}}=\frac{\Lambda^4_{G}}{f^2_{a}}.$ 

     \subsubsection{Wave function for Axion using $\alpha$ vacua}
     \label{3a2}                               
    Here our prime objective is to derive results for $\alpha$-vacua, which can be interpreted as a quantum state filled with particles defined by some hypothetical observer who  initially belongs to the Bunch Davies vacuum state ($\alpha=0$). Here the $\alpha$ vacua are invariant under ${\bf SO(1,4)}$ isometry group of de Sitter space.  Consequently we use the equivalent prescription followed in case of Bunch Davies vacuum by defining two subspaces in de Sitter space,  \textcolor{red}{\bf RI} and \textcolor{red}{\bf RII} respectively. In general $\alpha$-vacua is CPT invariant, which is parametrised by a single real positive parameter $\alpha$ which plays the role of super-selection quantum number. We use the results obtained for the solution of the EOM where we expand the field in terms of creation and annihilation operators in Bunch Davies vacuum, and further using Bogoliubov transformation the mode functions for the $\alpha$-vacua can be written as:
\bea \label{ass5} \Phi(r,t,\theta,\phi)&=&\int^{\infty}_{0} dp \sum_{\sigma=\pm 1}\sum^{\infty}_{l=0}\sum^{+l}_{m=-l}\left[d_{\sigma plm}{\cal E}^{(\alpha)}_{\sigma plm}(r,t,\theta,\phi)+d^{\dagger}_{\sigma plm}({\cal E}^{(\alpha)}_{\sigma plm})^{*}(r,t,\theta,\phi)\right],~~~~~~ \eea
where the $\alpha$-vacua state are defined as,
$d_{\sigma p l m}|\alpha\rangle=0 \forall \sigma=(+1,-1);0<p<\infty;l=0,\cdots,\infty,m=-l,\cdots,+l$.
In this context, the $\alpha$-vacua mode function ${\cal E}^{(\alpha)}_{\sigma plm}$ can be expressed in terms of Bunch Davies mode function ${\cal U}_{\sigma plm}(r,t,\theta,\phi)$ using Bogoliubov transformation as:
\bea \label{ass3} {\cal E}^{(\alpha)}_{\sigma plm}&=&\left[\cosh\alpha~{\cal U}_{\sigma plm}(r,t,\theta,\phi)+\sinh\alpha~{\cal U}^{*}_{\sigma plm}(r,t,\theta,\phi)\right].\eea  
After substituting Eq~(\ref{ass3})  in Eq~(\ref{ass5}) we get the following expression for the wave function:
\bea \Phi(r,t,\theta,\phi)&=&\frac{H}{\sinh t}\int^{\infty}_{0} dp \sum^{\infty}_{l=0}\sum^{+l}_{m=-l}\sum_{\sigma=\pm 1}\left[d_{\sigma plm}\cosh\alpha~\chi_{p,\sigma}(t)+d^{\dagger}_{\sigma plm}\sinh\alpha~\chi^{*}_{p,\sigma}(t)\right]{\cal Y}_{plm}(r,\theta,\phi),~~~~~~ \eea
Finally, the solution of the time dependent part of the wave function can be recast as:
     \bea
                                             \label{cvvcx}
                                    \displaystyle \chi_{p,\sigma}(t)=\sum_{q={\bf R},{\bf L}}\left\{\frac{1}{{\cal N}_{p}}\left[\alpha^{\sigma}_{q}~{\cal P}^{q}+\beta^{\sigma}_{q}~{\cal P}^{q*}\right]+\sum^{\infty}_{n=0}\frac{1}{{\cal N}_{p_n}\left(p^2_n-p^2\right)}\left[\bar{\alpha}^{\sigma}_{q,n}~\bar{\cal P}^{q,n}+\bar{\beta}^{\sigma}_{q,n}~\bar{\cal P}^{q*,n}\right]\right\},~~~~~~~~~\eea 
                                    where we use the following shorthand notation, 
                                    $\bar{\cal P}^{q,n}= \mu^3\sinh^2t~ \int dt^{'}~\chi^{(c)}_{p_n,\sigma,q}(t^{'})~ {\cal P}^{q,n}$.
    Additionally, here we use the shorthand notations ${\cal P}^{q}$, ${\cal P}^{*q}$, ${\cal P}^{q,n}$, ${\cal P}^{*q,n}$ for the Legendre polynomial,  which is defined in ref.~\cite{Choudhury:2017bou}. Also the coefficient functions $(\alpha^{\sigma}_{q}, \beta^{\sigma}_{q})$ and $(\alpha^{\sigma}_{q,n}, \beta^{\sigma}_{q,n})$, normalisation constants ${\cal N}_{p}$, ${\cal N}_{p_n}$ are explicitly mentioned in ref.~\cite{Choudhury:2017bou} .
    
For further computation $\alpha$-vacua are defined in terms of Bunch Davies vacuum state as:
 \bea \label{sdss} |\alpha\rangle &=& \exp\left(\displaystyle\frac{1}{2} \tanh\alpha~\sum_{\sigma=\pm 1}a^{\dagger}_{\sigma}a_{\sigma}\right)\exp\left(\frac{1}{2}\sum_{i,j={\bf R},{\bf L}}m_{ij}~b^{\dagger}_{i}~b^{\dagger}_{j}+\frac{1}{2}\sum_{i,j={\bf R},{\bf L}}\sum^{\infty}_{n=0}\bar{m}_{ij,n}~\bar{b}^{\dagger}_{i,n}~\bar{b}^{\dagger}_{j,n},\right)(|{\bf R}\rangle \otimes |{\bf L}\rangle),~~~\eea
Further one can also write the the ${\bf R}$ and ${\bf L}$ vacua as \cite{Choudhury:2017bou}:
 \bea   |{\bf R}\rangle&=& |{\bf R}\rangle_{(c)}+\sum^{\infty}_{n=0}|{\bf R}\rangle_{(p),n},~~~  |{\bf L}\rangle= |{\bf L}\rangle_{(c)}+\sum^{\infty}_{n=0}|{\bf L}\rangle_{(p),n},\eea 
  with $(c)$ and $(p)$ representing the complementary and particular part respectively. Here the matrices $m_{ij}$ and $\bar{m}_{ij,n}$ corresponding to complementary and particular part of the solution are explicitly computed in ref.~\cite{Choudhury:2017bou} for Bunch Davies vacuum.
 Also the creation and annihilation operators $a^{\dagger}_{\sigma}$ and $a_{\sigma}$ for the ${\bf R}$ and ${\bf L}$ vacuum are defined as \cite{Choudhury:2017bou}:
         \bea\label{xxxq1} a_{\sigma}&=&\sum_{q={\bf R},{\bf L}}\left\{\left[\gamma_{q\sigma}b_{q}+\delta^{*}_{q\sigma}b^{\dagger}_{q}\right]+\sum^{\infty}_{n=0}\left[\bar{\gamma}_{q\sigma,n}\bar{b}_{q,n}+\bar{\delta}^{*}_{q\sigma,n}\bar{b}^{\dagger}_{q,n}\right]\right\},\\ \label{xxxq2} a^{\dagger}_{\sigma}&=&\sum_{q={\bf R},{\bf L}}\left\{\left[\gamma^{*}_{q\sigma}b^{\dagger}_{q}+\delta_{q\sigma}b_{q}\right]+\sum^{\infty}_{n=0}\left[\bar{\gamma}^{*}_{q\sigma,n}\bar{b}^{\dagger}_{q,n}+\bar{\delta}_{q\sigma,n}\bar{b}_{q,n}\right]\right\},\eea   
         with $\sigma=\pm 1$. Here it is important to note that, the coefficient matrices for the Bogoliubov transformation $\gamma_{q\sigma}$, $\delta_{q\sigma}$, $\bar{\gamma}_{q\sigma,n}$ and $\bar{\delta}_{q\sigma,n}$ helps us to write the $a$ type of oscillators in terms of a new $b$ type of oscillators. For more details see ref.~\cite{Choudhury:2017bou} where all the symbols are explicitly defined. Here it is important to note that, the newly introduced $b$ type of oscillators exactly satisfy the harmonic oscillator algebra, provided the oscillators corresponding to the solution of complementary and particular part of the time dependent solution of the wave function are not interacting with each other. This surely helps us to set up the rules for the operation of creation and annihilation operators of these oscillators in this context \cite{Choudhury:2017bou}.

  Below, we use the definition of $\alpha$-vacuum state as given in Eq~(\ref{sdss}), which is very useful to compute entanglement entropy in de Sitter space. However, note that the technical steps for the computation of the entanglement entropy in de Sitter space from $\alpha$-vacua are exactly similar to the steps followed for Bunch Davies vacuum. The difference will only appear when we use the creation and annihilation operators in the context of $\alpha$-vacua, which can be written in terms of the creation and annihilation operators defined for ${\bf R}$ or ${\bf L}$ vacuum state as:
 \bea d_{\sigma}&=&\sum_{q={\bf R},{\bf L}}\left\{\left[\left(\cosh\alpha~\gamma_{q\sigma}-\sinh\alpha~\delta_{q\sigma}\right)b_{q}+\left(\cosh\alpha~\delta^{*}_{q\sigma}-\sinh\alpha~\gamma^{*}_{q\sigma}\right)b^{\dagger}_{q}\right]\right.\nonumber\\ && \left.~~~~+\left[\left(\cosh\alpha~\sum^{\infty}_{n=0}\bar{\gamma}_{q\sigma,n}\bar{b}_{q,n}-\sinh\alpha~\sum^{\infty}_{n=0}\bar{\delta}_{q\sigma,n}\bar{b}_{q,n}\right)+\left(\cosh\alpha~\sum^{\infty}_{n=0}\bar{\delta}^{*}_{q\sigma,n}\bar{b}^{\dagger}_{q,n}-\sinh\alpha~\sum^{\infty}_{n=0}\bar{\gamma}^{*}_{q\sigma,n}\bar{b}^{\dagger}_{q,n}\right)\right]\right\},\\
 d^{\dagger}_{\sigma}&=&\sum_{q={\bf R},{\bf L}}\left\{\left[\left(\cosh\alpha~\gamma^{*}_{q\sigma}-\sinh\alpha~\delta^{*}_{q\sigma}\right)b^{\dagger}_{q}+\left(\cosh\alpha~\delta_{q\sigma}-\sinh\alpha~\gamma_{q\sigma}\right)b_{q}\right]\right.\nonumber\\ && \left.~~~~ +\left[\left(\cosh\alpha~\sum^{\infty}_{n=0}\bar{\gamma}^{*}_{q\sigma,n}\bar{b}^{\dagger}_{q,n}-\sinh\alpha~\sum^{\infty}_{n=0}\bar{\delta}^{*}_{q\sigma,n}\bar{b}^{\dagger}_{q,n}\right)+\left(\cosh\alpha~\sum^{\infty}_{n=0}\bar{\delta}_{q\sigma,n}\bar{b}_{q,n}-\sinh\alpha~\sum^{\infty}_{n=0}\bar{\gamma}_{q\sigma,n}\bar{b}_{q,n}\right)\right]\right\},~~~~~~~~~~\eea
 where we use the definition of creation and annihilation operators in Bunch Davies vacuum as mentioned in Eq~(\ref{xxxq2}) and Eq~(\ref{xxxq1}). In this computation it is important to note that, under Bogoliubov transformation the original matrix $\gamma_{q\sigma}$, $\delta_{q\sigma}$, $\bar{\gamma}_{q\sigma,n}$ and $\bar{\delta}_{q\sigma,n}$ used for Bunch Davies vacuum are transformed in the context of $\alpha$-vacua as:
 \bea \gamma_{q\sigma}&\longrightarrow& \left(\cosh\alpha~\gamma_{q\sigma}-\sinh\alpha~\delta_{q\sigma}\right),~~
 \delta_{q\sigma}\longrightarrow\left(\cosh\alpha~\delta_{q\sigma}-\sinh\alpha~\gamma_{q\sigma}\right),\\
 \bar{\gamma}_{q\sigma,n}&\longrightarrow&\left(\cosh\alpha~\bar{\gamma}_{q\sigma,n}-\sinh\alpha~\bar{\delta}_{q\sigma,n}\right),~~
 \bar{\delta}_{q\sigma,n}\longrightarrow\left(\cosh\alpha~\bar{\delta}_{q\sigma,n}-\sinh\alpha~\bar{\gamma}_{q\sigma,n}\right).\nonumber\eea
 Considering this fact, after Bogoliubov transformation $\alpha$-vacua state can be written in terms of ${\bf R}$ and ${\bf L}$ vacua as:
 \bea |\alpha\rangle &=&\exp\left(\frac{1}{2}\sum_{i,j={\bf R},{\bf L}}\tilde{m}_{ij}~b^{\dagger}_{i}~b^{\dagger}_{j}+\frac{1}{2}\sum_{i,j={\bf R},{\bf L}}\sum^{\infty}_{n=0}\bar{\tilde{m}}_{ij,n}~\bar{b}^{\dagger}_{i,n}~\bar{b}^{\dagger}_{j,n}\right)(|{\bf R}\rangle \otimes |{\bf L}\rangle),\eea
   Here $\tilde{m}_{ij}$ and $\bar{\tilde{m}}_{ij,n}$ represents the entries of the  matrices corresponding to the complementary and particular solution in presence of $\alpha$ vacuum which we will compute in this paper. 
   
   Further one can write the annihilation of $\alpha$ vacuum in terms of the annihilations of the direct product state of ${\bf R}$ and ${\bf L}$ vacuum as:
             \bea d_{\sigma}|\alpha\rangle=\sum_{q={\bf R},{\bf L}}\sum^{4}_{s=1}{\cal J}^{(q)}_{s}=0,\eea
             where neglecting contribution from the powers of creation operators, ${\cal J}^{(q)}_{s}\forall s=1,2,3,4, q={\bf R},{\bf L}$ are defined as:
             \bea \sum_{q={\bf R},{\bf L}}{\cal J}^{(q)}_{1}&=&\sum_{q={\bf R},{\bf L}}\left(\cosh\alpha~\gamma_{q\sigma}-\sinh\alpha~\delta_{q\sigma}\right)b_q~e^{{\hat{\tilde{\cal O}}}}\left(|{\bf R}\rangle\otimes |{\bf L}\rangle\right)\approx\sum_{i,j={\bf R},{\bf L}}\tilde{m}_{ij}\left(\cosh\alpha~\gamma_{j\sigma}-\sinh\alpha~\delta_{j\sigma}\right)b^{\dagger}_{i}\left(|{\bf R}\rangle\otimes |{\bf L}\rangle\right),~~~~~~~\\
             \sum_{q={\bf R},{\bf L}}{\cal J}^{(q)}_{2}&=&\sum_{q={\bf R},{\bf L}}\left(\cosh\alpha~\delta^{*}_{q\sigma}-\sinh\alpha~\gamma^{*}_{q\sigma}\right)b^{\dagger}_q~e^{{\hat{\tilde{\cal O}}}}\left(|{\bf R}\rangle\otimes |{\bf L}\rangle\right)\approx\sum_{q={\bf R},{\bf L}}\left(\cosh\alpha~\delta^{*}_{q\sigma}-\sinh\alpha~\gamma^{*}_{q\sigma}\right)b^{\dagger}_q\left(|{\bf R}\rangle\otimes |{\bf L}\rangle\right),~~~\\
             \sum_{q={\bf R},{\bf L}}{\cal J}^{(q)}_{3}&=&\sum_{q={\bf R},{\bf L}}\left(\cosh\alpha~\sum^{\infty}_{n=0}\bar{\gamma}_{q\sigma,n}\bar{b}_{q,n}-\sinh\alpha~\sum^{\infty}_{n=0}\bar{\delta}_{q\sigma,n}\bar{b}_{q,n}\right)~e^{{\hat{\tilde{\cal O}}}}\left(|{\bf R}\rangle\otimes |{\bf L}\rangle\right)\nonumber\\
                         &\approx&\sum_{i,j={\bf R},{\bf L}}\left(\cosh\alpha~\sum^{\infty}_{n=0}\bar{\tilde{m}}_{ij,n}\bar{\gamma}_{j\sigma,n}\bar{b}^{\dagger}_{i,n}-\sinh\alpha~\sum^{\infty}_{n=0}\bar{\tilde{m}}_{ij,n}\bar{\delta}_{j\sigma,n}\bar{b}^{\dagger}_{i,n}\right)\left(|{\bf R}\rangle\otimes |{\bf L}\rangle\right),~~~~~~~~~~~~~~~\\
                             \sum_{q={\bf R},{\bf L}}{\cal J}^{(q)}_{4}&=&\sum_{q={\bf R},{\bf L}}\left(\cosh\alpha~\sum^{\infty}_{n=0}\bar{\delta}^{*}_{q\sigma,n}\bar{b}^{\dagger}_{q,n}-\sinh\alpha~\sum^{\infty}_{n=0}\bar{\gamma}^{*}_{q\sigma,n}\bar{b}^{\dagger}_{q,n}\right)~e^{{\hat{\tilde{\cal O}}}}\left(|{\bf R}\rangle\otimes |{\bf L}\rangle\right)
                                     \nonumber\\
                                                 &\approx&\sum_{q={\bf R},{\bf L}}\sum^{\infty}_{n=0}\left(\cosh\alpha~\sum^{\infty}_{n=0}\bar{\delta}^{*}_{q\sigma,n}\bar{b}^{\dagger}_{q,n}-\sinh\alpha~\sum^{\infty}_{n=0}\bar{\gamma}^{*}_{q\sigma,n}\bar{b}^{\dagger}_{q,n}\right)\left(|{\bf R}\rangle\otimes |{\bf L}\rangle\right).~~~~~~~~~~~~\eea
    This directly implies that:
    \bea &&\left[\tilde{m}_{ij}\left(\cosh\alpha~\gamma_{j\sigma}-\sinh\alpha~\delta_{j\sigma}\right)+\left(\cosh\alpha~\delta^{*}_{i\sigma}-\sinh\alpha~\gamma^{*}_{i\sigma}\right)\right]b^{\dagger}_{i}\nonumber~~~~~~~~~~~~~~~~~\\
    &&+\left[\left(\cosh\alpha~\sum^{\infty}_{n=0}\bar{\tilde{m}}_{ij,n}\bar{\gamma}_{j\sigma,n}\bar{b}^{\dagger}_{i,n}-\sinh\alpha~\sum^{\infty}_{n=0}\bar{m}_{ij,n}\bar{\delta}_{j\sigma,n}\bar{b}^{\dagger}_{i,n}\right)+\left(\cosh\alpha~\sum^{\infty}_{n=0}\bar{\delta}^{*}_{i\sigma,n}\bar{b}^{\dagger}_{i,n}-\sinh\alpha~\sum^{\infty}_{n=0}\bar{\gamma}^{*}_{i\sigma,n}\bar{b}^{\dagger}_{i,n}\right)\right]= 0.~~~~~~~~~~\eea
    As we have already mentioned that the complementary and  particular part of the solutions are completely independent of each other and hence vanish individually. Consequently, we get the following constraints in case of $\alpha$ vacuum:
          \bea \label{qz1} \left[\tilde{m}_{ij}\left(\cosh\alpha~\gamma_{j\sigma}-\sinh\alpha~\delta_{j\sigma}\right)+\left(\cosh\alpha~\delta^{*}_{i\sigma}-\sinh\alpha~\gamma^{*}_{i\sigma}\right)\right]&=& 0,\\
          \label{qz2} \left[\left(\cosh\alpha~\bar{\tilde{m}}_{ij,n}\bar{\gamma}_{j\sigma,n}-\sinh\alpha~\bar{m}_{ij,n}\bar{\delta}_{j\sigma,n}\right) 
          +\left(\cosh\alpha~\bar{\delta}^{*}_{i\sigma,n}-\sinh\alpha~\bar{\gamma}^{*}_{i\sigma,n}\right)\right]&=& 0\forall ~n.\eea
            Further using Eq~(\ref{qz1}) and Eq~(\ref{qz2}), the matrices corresponding to the complementary and particular part of the solution can be expressed as: 
            \bea \label{qz1a} \tilde{m}_{ij}&=& -\left(\cosh\alpha~\delta^{*}_{i\sigma}-\sinh\alpha~\gamma^{*}_{i\sigma}\right)\left(\cosh\alpha~\gamma-\sinh\alpha~\delta\right)^{-1}_{\sigma j}~~~~~\nonumber\\
   &=& -\frac{\Gamma\left(\nu+\frac{1}{2}-ip\right)}{\Gamma\left(\nu+\frac{1}{2}+ip\right)}\frac{2~{\cal D}^{(\nu)}_{ij}}{e^{2\pi p}\left(\cosh\alpha-\sinh\alpha~e^{-2\pi p}\right)^2+e^{2i\pi\nu}\left(\cosh\alpha+\sinh\alpha~e^{-2i\pi\nu}\right)^2},\\        
             \label{qz22a} \bar{\tilde{m}}_{ij,n}&=&-\left(\cosh\alpha~\bar{\delta}^{*}_{i\sigma,n}-\sinh\alpha~\bar{\gamma}^{*}_{i\sigma,n}\right)\left(\cosh\alpha~\bar{\gamma}-\sinh\alpha~\bar{\delta}\right)^{-1}_{\sigma j,n}~~~~~~~~~~\nonumber\\
             &=& -\frac{\Gamma\left(\nu+\frac{1}{2}-ip_n\right)}{\Gamma\left(\nu+\frac{1}{2}+ip_n\right)}\frac{2~{\cal D}^{(\nu,n)}_{ij}}{e^{2\pi p_n}\left(\cosh\alpha-\sinh\alpha~e^{-2\pi p_n}\right)^2+e^{2i\pi\nu}\left(\cosh\alpha+\sinh\alpha~e^{-2i\pi\nu}\right)^2}~~~\forall~(i,j)={\bf R},{\bf L}.~~~~~~~~~~~~\eea 
Here we define the ${\cal D}$ matrices as:
\bea ~~~~~~{\cal D}^{(\nu)}_{ij}&=&\left(\begin{array}{ccc} {\cal D}^{(\nu)}_{\bf RR} &~~~ {\cal D}^{(\nu)}_{\bf RL} \\ {\cal D}^{(\nu)}_{\bf LR} &~~~ {\cal D}^{(\nu)}_{\bf LL}  \end{array}\right),~~~~~~~~~~{\cal D}^{(\nu,n)}_{ij}=\left(\begin{array}{ccc} {\cal D}^{(\nu,n)}_{\bf RR} &~~~ {\cal D}^{(\nu,n)}_{\bf RL} \\ {\cal D}^{(\nu,n)}_{\bf LR} &~~~ {\cal D}^{(\nu,n)}_{\bf LL}  \end{array}\right).~~~~~\eea
and the corresponding entries of the ${\cal D}$ matrices are given by: 
\bea 
{\cal D}^{(\nu)}_{\bf RR}&=&{\cal D}^{(\nu)}_{\bf LL}=\left(\cosh^2\alpha~e^{i\pi\nu}+\sinh^2\alpha~e^{-i\pi\nu}\right)\cos\pi\nu-\sinh 2\alpha~ \sinh^2 \pi p,\\
{\cal D}^{(\nu)}_{\bf RL}&=&{\cal D}^{(\nu)}_{\bf LR}=i\left(\cosh^2\alpha~e^{i\pi \nu}+\sinh^2\alpha~e^{-i\pi\nu}+\sinh 2\alpha~\cos\pi\nu\right)\sinh\pi p,\\
{\cal D}^{(\nu,n)}_{\bf RR}&=&{\cal D}^{(\nu,n)}_{\bf LL}=\left(\cosh^2\alpha~e^{i\pi\nu}+\sinh^2\alpha~e^{-i\pi\nu}\right)\cos\pi\nu-\sinh 2\alpha~ \sinh^2 \pi p_n,\\
{\cal D}^{(\nu,n)}_{\bf RL}&=&{\cal D}^{(\nu,n)}_{\bf LR}=i\left(\cosh^2\alpha~e^{i\pi \nu}+\sinh^2\alpha~e^{-i\pi\nu}+\sinh 2\alpha~\cos\pi\nu\right)\sinh\pi p_n.~~~~~~\eea                                                        
 Before further discussion here we point out few important features from the obtained results:
 \begin{itemize}
 \item We see that for the complementary and particular part of the solution  \bea \label{ss1a}\tilde{m}_{\bf RR}&=& \tilde{m}_{\bf LL}= -\frac{\Gamma\left(\nu+\frac{1}{2}-ip\right)}{\Gamma\left(\nu+\frac{1}{2}+ip\right)}\frac{2~\left[\left(\cosh^2\alpha~e^{i\pi\nu}+\sinh^2\alpha~e^{-i\pi\nu}\right)\cos\pi\nu-\sinh 2\alpha~ \sinh^2 \pi p\right]}{e^{2\pi p}\left(\cosh\alpha-\sinh\alpha~e^{-2\pi p}\right)^2+e^{2i\pi\nu}\left(\cosh\alpha+\sinh\alpha~e^{-2i\pi\nu}\right)^2},~~~~~~~~~~~\\
 \label{ss2a}\bar{\tilde{m}}_{{\bf RR},n}&=& \bar{\tilde{m}}_{{\bf LL},n}= -\frac{\Gamma\left(\nu+\frac{1}{2}-ip_n\right)}{\Gamma\left(\nu+\frac{1}{2}+ip_n\right)}\frac{2~\left[\left(\cosh^2\alpha~e^{i\pi\nu}+\sinh^2\alpha~e^{-i\pi\nu}\right)\cos\pi\nu-\sinh 2\alpha~ \sinh^2 \pi p_n\right]}{e^{2\pi p_n}\left(\cosh\alpha-\sinh\alpha~e^{-2\pi p_n}\right)^2+e^{2i\pi\nu}\left(\cosh\alpha+\sinh\alpha~e^{-2i\pi\nu}\right)^2}.~~~~~\eea
 which is non vanishing for $0<\nu\leq 3/2$ and $\nu>3/2$. For $\nu=3/2$ we get the non vanishing result using $\alpha$-vacuum and this result is significantly different from the result obtained for Bunch Davies vacuum state. 
 
 Finally to implement numerical analysis we use the following approximated expressions for the entries of the coefficient matrices as given by~\footnote{For rest of the analysis we absorb this overall phase factor $e^{i\theta}$.}:
   \bea \label{ssccd1} \tilde{m}_{\bf RR}&=& e^{i\theta}~\frac{\sqrt{2}~e^{-p\pi}\cos\pi\nu}{\sqrt{\cosh 2\pi p+\cos 2\pi \nu}}\frac{\left[\left(\cosh^2\alpha+\sinh^2\alpha~e^{-2i\pi\nu}\right)-\sinh 2\alpha~ \sinh^2 \pi p~e^{-i\pi\nu}\sec\pi\nu\right]}{\left(\cosh^2\alpha+\sinh^2\alpha~e^{-2\pi( p+i\nu)}\right)},~~~~~~~~~~~\\
   \label{ssa2a}\bar{\tilde{m}}_{{\bf RR},n}&=& e^{i\theta}~\frac{\sqrt{2}~e^{-p_n\pi}\cos\pi\nu}{\sqrt{\cosh 2\pi p_n+\cos 2\pi \nu}}\frac{\left[\left(\cosh^2\alpha+\sinh^2\alpha~e^{-2i\pi\nu}\right)-\sinh 2\alpha~ \sinh^2 \pi p_n~e^{-i\pi\nu}\sec\pi\nu\right]}{\left(\cosh^2\alpha+\sinh^2\alpha~e^{-2\pi( p_n+i\nu)}\right)}.\eea
   
 \item We see that for the complementary and particular part of the solution: 
 \bea \label{gg1a}\tilde{m}_{\bf RL}&=&\tilde{m}_{\bf LR}=-\frac{\Gamma\left(\nu+\frac{1}{2}-ip\right)}{\Gamma\left(\nu+\frac{1}{2}+ip\right)}\frac{2~i\left[\left(\cosh^2\alpha~e^{i\pi \nu}+\sinh^2\alpha~e^{-i\pi\nu}+\sinh 2\alpha~\cos\pi\nu\right)\sinh\pi p\right]}{e^{2\pi p}\left(\cosh\alpha-\sinh\alpha~e^{-2\pi p}\right)^2+e^{2i\pi\nu}\left(\cosh\alpha+\sinh\alpha~e^{-2i\pi\nu}\right)^2},~~~~~~~~~~\\
  \label{gg2a}\bar{\tilde{m}}_{{\bf RL},n}&=& \bar{\tilde{m}}_{{\bf LR},n}=-\frac{\Gamma\left(\nu+\frac{1}{2}-ip\right)}{\Gamma\left(\nu+\frac{1}{2}+ip\right)}\frac{2~i\left[\left(\cosh^2\alpha~e^{i\pi \nu}+\sinh^2\alpha~e^{-i\pi\nu}+\sinh 2\alpha~\cos\pi\nu\right)\sinh\pi p_n\right]}{e^{2\pi p_n}\left(\cosh\alpha-\sinh\alpha~e^{-2\pi p_n}\right)^2+e^{2i\pi\nu}\left(\cosh\alpha+\sinh\alpha~e^{-2i\pi\nu}\right)^2}.~~~~~~~~~~\eea
  Additionally, the non vanishing entries of the off diagonal components of the coefficient matrix for both of the cases in presence of $\alpha$-vacuum indicates the existence of quantum entanglement in the present computation, which we will explicitly show that finally give rise to a non vanishing entanglement entropy. 
 
Finally to interpret the result numerically we use the following approximated expressions for the entries of the coefficient matrices as given by:
   \bea \label{assccd1}\tilde{m}_{\bf RL}&=& e^{i\left(\theta+\frac{\pi}{2}\right)}~\frac{\sqrt{2}~e^{-p\pi}\sinh\pi p}{\sqrt{\cosh 2\pi p+\cos 2\pi \nu}}\frac{\left[\cosh^2\alpha+\sinh^2\alpha~e^{-2i\pi\nu}+\sinh 2\alpha~\cos\pi\nu~ e^{-i\pi\nu}\right]}{\left(\cosh^2\alpha+\sinh^2\alpha~e^{-2\pi( p+i\nu)}\right)},~~~~~~~~~~~\\
   \label{assa2a}\bar{\tilde{m}}_{{\bf RL},n}&=& e^{i\left(\theta+\frac{\pi}{2}\right)}~\frac{\sqrt{2}~e^{-p_n\pi}\sinh\pi p_n}{\sqrt{\cosh 2\pi p_n+\cos 2\pi \nu}}\frac{\left[\cosh^2\alpha+\sinh^2\alpha~e^{-2i\pi\nu}+\sinh 2\alpha~\cos\pi\nu~ e^{-i\pi\nu}\right]}{\left(\cosh^2\alpha+\sinh^2\alpha~e^{-2\pi( p_n+i\nu)}\right)}.\eea

 \end{itemize}

 To find a suitable basis first of all we trace over all possible contributions from ${\bf R}$ and ${\bf L}$ region. To implement this we need to perform another Bogoliubov transformation introducing new sets of operators as given by:
    \bea 
    \label{kc1} \tilde{c}_{\bf R}&=& \tilde{u}~b_{\bf R}+\tilde{v}~b^{\dagger}_{\bf R},~~ \tilde{c}_{\bf L}= \bar{\tilde{u}}~b_{\bf L}+\bar{\tilde{v}}~b^{\dagger}_{\bf L},~ \tilde{C}_{{\bf R},n}= \tilde{U}_n~b_{{\bf R},n}+\tilde{V}_n~b^{\dagger}_{{\bf R},n},~~ \tilde{C}_{{\bf L},n}= \bar{\tilde{U}}_n~b_{{\bf L},n}+\bar{\tilde{V}}_n~b^{\dagger}_{{\bf L},n},\eea   
       where following conditions are satisfied:
         \bea  
            |\tilde{u}|^2-|\tilde{v}|^2&=& 1,~|\bar{\tilde{u}}|^2-|\bar{\tilde{v}}|^2= 1,~~  
                       |\tilde{U}_n|^2-|\tilde{V}_n|^2= 1,~~|\bar{\tilde{U}}_n|^2-|\bar{\tilde{V}}_n|^2= 1.\eea 
  Using these new sets of operators one can write the $\alpha$-vacuum state in terms of new basis represented by the direct product of ${\bf R}^{'}$ and ${\bf L}^{'}$ vacuum state as:
  \bea |\alpha\rangle &=&e^{\hat{\cal \tilde{O}}}\left(|{\bf R}\rangle\otimes |{\bf L}\rangle\right)=\left({\cal \widetilde{N}}^{(\alpha)}_{p}\right)^{-1}e^{\hat{\cal \tilde{Q}}}\left(|{\bf R}^{'}\rangle\otimes |{\bf L}^{'}\rangle\right)^{(\alpha)},\eea  
  where we introduce a new composite operator $\hat{\cal \tilde{Q}}$ which is defined in the new transformed basis as:
    \bea \label{xqaq2} \hat{\cal \tilde{Q}}&=&\gamma^{(\alpha)}_{p}~\tilde{c}^{\dagger}_{\bf R}~\tilde{c}^{\dagger}_{\bf L}+\sum^{\infty}_{n=0}\Gamma^{(\alpha)}_{p,n}~\tilde{C}^{\dagger}_{{\bf R},n}~\tilde{C}^{\dagger}_{{\bf L},n},\eea
      where $\gamma^{(\alpha)}_{p}$ and $\Gamma^{(\alpha)}_{p,n}$ are defined corresponding to the complementary and particular solution, which we will explicitly compute further for $\alpha$ vacuum. Additionally, it is important to note that the overall normalization factor $\widetilde{\cal N}^{(\alpha)}_{p}$ is defined as:
      \bea \widetilde{\cal N}^{(\alpha)}_{p}&=& \left|e^{\hat{\cal \tilde{Q}}}\left(|{\bf R}^{'}\rangle\otimes |{\bf L}^{'}\rangle\right)^{(\alpha)}\right|\approx\left[1-\left(|\gamma^{(\alpha)}_p|^2+\sum^{\infty}_{n=0}|\Gamma^{(\alpha)}_{p,n}|^2\right)\right]^{-1/2},\eea
      which reduces to the result obtained for Bunch Davies vacuum in ref.~\cite{Choudhury:2017bou} for $\alpha=0$. In this calculation due to the second Bogoliubov transformation the direct product of the ${\bf R}$ and ${\bf L}$ vacuum state is connected to the direct product of the new ${\bf R}^{'}$ and ${\bf L}^{'}$ vacuum state as:
      \bea \left(|{\bf R}\rangle\otimes |{\bf L}\rangle\right)\rightarrow
      \left(|{\bf R}^{'}\rangle\otimes |{\bf L}^{'}\rangle\right)^{(\alpha)}=\widetilde{\cal N}^{(\alpha)}_p~e^{-\hat{\cal \tilde{Q}}}~e^{\hat{\cal \tilde{O}}}\left(|{\bf R}\rangle\otimes |{\bf L}\rangle\right).\eea
      Let us now mention the commutation relations of the creation and annihilation operators corresponding to the new sets of oscillators describing the ${\bf R}^{'}$ and ${\bf L}^{'}$ vacuum state as:
            \bea 
            \left[ \tilde{c}_i, \tilde{c}^{\dagger}_j\right]&=&\delta_{ij},~~~~ \left[  \tilde{c}_i, \tilde{c}_j\right]=0=
            \left[  \tilde{c}^{\dagger}_i, \tilde{c}^{\dagger}_j\right],  \left[ \tilde{C}_{i,n},\tilde{C}^{\dagger}_{j,m}\right]=\delta_{ij}{\delta}_{nm},~~~~\left[ \tilde{C}_{i,n},\tilde{C}_{j,m}\right]= 0=
                  \left[ \tilde{C}^{\dagger}_{i,m}\tilde{C}^{\dagger}_{j,m}\right].~~~~~~~~~~~~~~~
                  \eea
                  Here, for $\alpha$ vacuum, the oscillator algebra is exactly same as that obtained for Bunch Davies vacuum. However for $\alpha$ vacuum the structure of these operators are completely different and also they are acting in a different Hilbert space ${\cal H}_{\alpha}$, which is characterised by one parameter $\alpha$. Here it is important to note that for $\alpha=0$ these oscillators will act on Bunch Davies vacuum state where the corresponding Hilbert space, ${\cal H}_{\bf BD}$ is the subclass of ${\cal H}_{\alpha}$.
                  
       The action of creation and annihilation operators defined  on the $\alpha$ vacuum state are appended bellow:
       \bea 
       \label{gv1} \tilde{c}_{\bf R}|\alpha\rangle &=&\gamma^{(\alpha)}_{p}~\tilde{c}^{\dagger}_{\bf L}|\alpha\rangle,~~
    \tilde{c}_{\bf L}|\alpha\rangle =\gamma^{(\alpha)}_{p}~\tilde{c}^{\dagger}_{\bf R}|\alpha\rangle,~ \tilde{C}_{{\bf R},n}|\alpha\rangle =\Gamma^{(\alpha)}_{p,n}~\tilde{C}^{\dagger}_{{\bf L},n}|\alpha\rangle,~~ \tilde{C}_{{\bf L},n}|\alpha\rangle =\Gamma^{(\alpha)}_{p,n}~\tilde{C}^{\dagger}_{{\bf R},n}|\alpha\rangle.\eea Further, one can express the new $c$ type annihilation operators in terms of the old $b$ type annihilation operators as: 
             \bea \tilde{c}_{J}&=& b_{I}\tilde{\cal G}^{I}_{J}=b_{I}\left(\begin{array}{ccc} \tilde{U}_q &~~~ \tilde{V}^{*}_q \\ \tilde{V}_q &~~~ \tilde{U}^{*}_q  \end{array}\right)
                  ,~~~ 
             \tilde{C}_{J(n)}=\bar{b}_{J(n)}\left(\tilde{\cal G}_{(n)}\right)^{I}_{J}=\bar{b}_{J(n)}\left(\begin{array}{ccc} \bar{\tilde{U}}_{ q,n} &~~~ \bar{\tilde{V}}^{*}_{\sigma q,n} \\ \bar{\tilde{V}}_{ q,n} &~~~ \bar{\tilde{U}}^{*}_{ q,n}  \end{array}\right).\eea
                     Here the entries of the matrices for $\alpha$ vacuum are given by, $\tilde{U}_q \equiv {\rm \bf diag}\left(\tilde{u},\bar{\tilde{u}}\right)$,$\tilde{V}_q \equiv {\rm \bf diag}\left(\tilde{v},\bar{\tilde{v}}\right)$,$ \bar{\tilde{U}}_{q,n} \equiv {\rm \bf diag}\left(\tilde{U}_n,\bar{\tilde{U}}_n\right)$,$ \bar{\tilde{V}}_{q,n} \equiv {\rm \bf diag}\left(\tilde{V}_n,\bar{\tilde{V}}_n\right).$
             Further using Eq~(\ref{kc1}), in Eq~(\ref{gv1}), we get the following sets of homogeneous equations: 
  \bea 
   \underline{\textcolor{red}{\bf For~complementary~solution:}}~~~~~~~~~~~~~~~~~~~~~~~\nonumber\\ 
  \label{a1} \tilde{m}_{\bf RR}\tilde{u}+\tilde{v}-\gamma^{(\alpha)}_{p} \tilde{m}_{\bf RL}\bar{\tilde{v}}^{*}&=& 0,~~~
  \tilde{m}_{\bf RR}\bar{\tilde{u}}+\bar{\tilde{v}}-\gamma^{(\alpha)}_{p} \tilde{m}_{\bf RL}\tilde{v}^{*}= 0,\\ 
  \label{a3}  \tilde{m}_{\bf RL}\tilde{u}-\gamma^{(\alpha)}_{p} \bar{\tilde{u}}^{*}-\gamma^{(\alpha)}_{p}\tilde{m}_{\bf RR}\bar{\tilde{v}}^{*}&=& 0,~~~
  \tilde{m}_{\bf RL}\bar{\tilde{u}}-\gamma^{(\alpha)}_{p} \tilde{u}^{*}-\gamma^{(\alpha)}_{p}\tilde{m}_{\bf RR}\tilde{v}^{*}= 0,\\
     \underline{\textcolor{red}{\bf For~~particular~solution:}}~~~~~~~~~~~~~~~~~~~~~~~\nonumber\\ 
     \tilde{m}_{{\bf RR},n}\tilde{U}_n+\tilde{V}_n-\Gamma^{(\alpha)}_{p,n} \tilde{m}_{{\bf RL},n}\bar{\tilde{V}}^{*}_n&=& 0,~~~
        \tilde{m}_{{\bf RR},n}\bar{\tilde{U}}_n+\bar{\tilde{V}}_n-\Gamma^{(\alpha)}_{p,n} \tilde{m}_{{\bf RL},n}\tilde{V}^{*}_n= 0,\\ 
        \tilde{m}_{{\bf RL},n}\tilde{U}_n-\Gamma^{(\alpha)}_{p,n} \bar{\tilde{U}}^{*}_n-\Gamma^{(\alpha)}_{p,n} \tilde{m}_{{\bf RR},n}\bar{\tilde{V}}^{*}_n&=& 0,~~~
        \tilde{m}_{{\bf RL},n}\bar{\tilde{U}}_n-\Gamma^{(\alpha)}_{p,n}  \tilde{U}^{*}_n-\Gamma^{(\alpha)}_{p,n} \tilde{m}_{{\bf RR},n}\tilde{V}^{*}_n= 0,\eea
          Now with $\alpha$ vacuum, it is not sufficient to use $\tilde{v}^{*}=\bar{\tilde{v}}, 
    \tilde{u}^{*}=\bar{\tilde{u}}$ for particular part and also $\tilde{V}^{*}_n=\bar{\tilde{V}}_n, 
       \tilde{U}^{*}_n=\bar{\tilde{U}}_n$ for the complementary part. In this case, the system of four equations, each for complementary and particular part will not be reduced to two sets of simplified equations. This is an outcome of the fact that in case of $\alpha$ vacuum, the entries of the coefficient matrices $\tilde{m}_{ij}$ and $\tilde{\bar{m}}_{ij,n}$ are complex in nature. On the other hand, they are either real or imaginary for Bunch Davies vacuum state. To solve these equations for $\gamma^{(\alpha)}_p$ and $\Gamma^{(\alpha)}_{p,n}$, we also need to use the normalisation conditions, $|\tilde{u}|^2-|\tilde{v}|^2=1$ and $|\tilde{U}_n|^2-|\tilde{V}_n|^2=1$.

        Finally, the non trivial solutions obtained from these system of equations can be expressed as: 
         \bea
          \label{as1}\gamma^{(\alpha)}_{p}&=&\frac{1}{\sqrt{2}|\tilde{m}_{\bf RL}|}\left[\left(1+|\tilde{m}_{\bf RL}|^4+|\tilde{m}_{\bf RR}|^4-2|\tilde{m}_{\bf RR}|^2-\tilde{m}^2_{\bf RR}(\tilde{m}^{*}_{\bf RL})^2-\tilde{m}^2_{\bf RL}(\tilde{m}^{*}_{\bf RR})^2\right) 
          \pm \left\{\left(-1-|\tilde{m}_{\bf RL}|^4-|\tilde{m}_{\bf RR}|^4\right.\right.\right.\nonumber\\ && \left.\left.\left.+2|\tilde{m}_{\bf RR}|^2+\tilde{m}^2_{\bf RR}(\tilde{m}^{*}_{\bf RL})^2+\tilde{m}^2_{\bf RL}(\tilde{m}^{*}_{\bf RR})^2\right)^2 
          -4|\tilde{m}_{\bf RL}|^4\right\}^{\frac{1}{2}}\right]^{\frac{1}{2}},~~~~~~~~~~\\
           \label{as2}\Gamma^{(\alpha)}_{p,n}&=&\frac{1}{\sqrt{2}|\tilde{m}_{{\bf RL},n}|}\left[\left(1+|\tilde{m}_{{\bf RL},n}|^4+|\tilde{m}_{{\bf RR},n}|^4-2|\tilde{m}_{{\bf RR},n}|^2-\tilde{m}^2_{{\bf RR},n}(\tilde{m}^{*}_{{\bf RL},n})^2-\tilde{m}^2_{{\bf RL},n}(\tilde{m}^{*}_{{\bf RR},n})^2\right) 
\right.\nonumber\\ && \left. \pm \left\{\left(-1-|\tilde{m}_{{\bf RL},n}|^4-|\tilde{m}_{{\bf RR},n}|^4+2|\tilde{m}_{{\bf RR},n}|^2+\tilde{m}^2_{{\bf RR},n}(\tilde{m}^{*}_{{\bf RL},n})^2+\tilde{m}^2_{{\bf RL},n}(\tilde{m}^{*}_{{\bf RR},n})^2\right)^2 
           -4|\tilde{m}_{{\bf RL},n}|^4\right\}^{\frac{1}{2}}\right]^{\frac{1}{2}},~~~~~~~~~~\eea 
           where the components $\tilde{m}_{\bf RR}=\tilde{m}_{\bf LL}$, $\tilde{m}_{\bf RL}=\tilde{m}_{\bf LR}$ and  $\tilde{m}_{{\bf RR},n}=\tilde{m}_{{\bf LL},n}$, $\tilde{m}_{{\bf RL},n}=\tilde{m}_{{\bf LR},n}$ are defined  in  Eqn~(\ref{ss1a}), Eqn~(\ref{ss2a}) and  Eqn~(\ref{gg1a}), Eqn~(\ref{gg2a}) respectively. Note that, in both the solutions for $\gamma^{(\alpha)}_{p}$ and $\Gamma^{(\alpha)}_{p,n}$ we absorb the overall phase factor.
           
           After further simplification we get the following expressions for the non trivial solutions for arbitrary $\nu$ can be written as:
           \bea
            \label{as3}\gamma^{(\alpha)}_{p}&\approx& i\frac{\sqrt{2}}{\sqrt{\cosh 2\pi p +\cos 2\pi \nu}\pm\sqrt{\cosh 2\pi p +\cos 2\pi \nu+2}}\frac{ \left[\cosh^2\alpha+\sinh^2\alpha~e^{2i\pi\nu}+\sinh 2\alpha\cos \pi \nu~e^{i\pi\nu}\right]}{\left(\cosh^2\alpha+\sinh^2\alpha~e^{-2\pi( p-i\nu)}\right)},~~~~\\
            \label{as4}\Gamma^{(\alpha)}_{p,n}&\approx& i\frac{\sqrt{2}}{\sqrt{\cosh 2\pi p_n +\cos 2\pi \nu}\pm\sqrt{\cosh 2\pi p_n +\cos 2\pi \nu+2}}\frac{ \left[\cosh^2\alpha+\sinh^2\alpha~e^{2i\pi\nu}+\sinh 2\alpha\cos \pi \nu~e^{i\pi\nu}\right]}{\left(\cosh^2\alpha+\sinh^2\alpha~e^{-2\pi( p_n-i\nu)}\right)}.~~~~
             \eea                                                                                          
                                                                                                                                                                                      
    \subsection{Construction of density matrix using $\alpha$ vacua}
  \label{3b}
         In this subsection our prime objective is construct the density matrix using the $\alpha$ vacuum state which is expressed in terms of another set of annihilation and creation operators in the Bogoliubov transformed frame. Here the Bunch Davies vacuum state can be expressed as a product of the quantum state for each oscillator in the new frame after Bogoliubov transformation. Each oscillator is labeled by the quantum numbers $p, l$ and $m$ in this calculation. After tracing over the right part of the Hilbert space we get the following expression for the density matrix and the left part of the Hilbert space can be written as, 
       $(\rho_{\bf L}(\alpha))_{p,l,m}={\bf \rm Tr}_{\bf R}|\alpha\rangle \langle \alpha|$, where the $\alpha$ vacuum state can be written in terms of $\tilde{c}$ type of oscillators as:
         \bea \label{ddqz1} |\alpha\rangle &\approx&\left[1-\left(|\gamma^{(\alpha)}_p|^2+\sum^{\infty}_{n=0}|\Gamma^{(\alpha)}_{p,n}|^2\right)\right]^{1/2}\exp\left[\gamma^{(\alpha)}_{p}~\tilde{c}^{\dagger}_{\bf R}~\tilde{c}^{\dagger}_{\bf L}+\sum^{\infty}_{n=0}\Gamma^{(\alpha)}_{p,n}~\tilde{C}^{\dagger}_{{\bf R},n}~\tilde{C}^{\dagger}_{{\bf L},n}\right]\left(|{\bf R}^{'}\rangle\otimes |{\bf L}^{'}\rangle\right)^{(\alpha)},~~~~~~~~~~\eea
         which is already derived in the earlier section. Further using Eq~(\ref{ddqz1}), we find the following simplified expression for the density matrix for the left part of the Hilbert space for $\alpha$ vacuum as:
      \be \label{ff1x}(\rho_{\bf L}(\alpha))_{p,l,m}=\left(1-|\gamma^{(\alpha)}_{p}|^2\right)\sum^{\infty}_{k=0}|\gamma^{(\alpha)}_{p}|^{2k}\widetilde{|k;p,l,m\rangle}\widetilde{\langle k;p,l,m|}
      +(f^{(\alpha)}_{p})^2\sum^{\infty}_{n=0}\sum^{\infty}_{r=0}|\Gamma^{(\alpha)}_{p,n}|^{2r}\widetilde{|n,r;p,l,m\rangle}\widetilde{\langle n,r;p,l,m|},\ee
      where $\gamma^{(\alpha)}_{p}$ and $\Gamma^{(\alpha)}_{p,n}$ are derived in the earlier section. Also we define the $\alpha$ parameter dependent source normalisation factor ${\it f}^{(\alpha)}_{p}$ as, ${\it f}^{(\alpha)}_{p}=\left(\sum^{\infty}_{n=0}\left(1-|\Gamma^{(\alpha)}_{p,n}|^2\right)^{-1}\right)^{-1}$.
      In this computation the states $\widetilde{|k;p,l,m\rangle}$ and $\widetilde{|n,r;p,l,m\rangle}$ are defined in terms of the quantum state $|{\bf L}^{'}\rangle$ as:
      \bea \widetilde{|k;p,l,m\rangle}&=& \frac{1}{\sqrt{n!}}(\tilde{c}^{\dagger}_{\bf L})^{k}|{\bf L}^{'}\rangle,~~~~
             \widetilde{|n,r;p,l,m\rangle}= \frac{1}{\sqrt{r!}}(\tilde{C}^{\dagger}_{{\bf L},n})^{r}|{\bf L}^{'}\rangle.\eea
             Here we note that:
             \begin{enumerate}
             \item For $\alpha$ vacuum density matrix is diagonal for a given set of the ${\bf SO(1,3)}$ quantum numbers $p,l,m$ and additionally depends on the parameter $\alpha$ explicitly. This leads to the total density matrix to take the following simplified form as:
             \bea \label{ff1xs}\rho_{\bf L}(\alpha)=\left(1-|\gamma^{(\alpha)}_{p}|^2\right){\bf diag}\left(1,|\gamma^{(\alpha)}_{p}|^2,|\gamma^{(\alpha)}_{p}|^{4},|\gamma^{(\alpha)}_{p}|^{6}\cdots\right)+(f^{(\alpha)}_{p})^2\sum^{\infty}_{n=0}{\bf diag}\left(1,|\Gamma^{(\alpha)}_{p,n}|^2,|\Gamma^{(\alpha)}_{p,n}|^{4},|\Gamma^{(\alpha)}_{p,n}|^{6}\cdots\right),~~~~~~\eea
            
             \item 
             To find out an acceptable normalization of the total density matrix in presence of $\alpha$ vacuum state, we use the following limiting results:
                    \bea \sum^{\infty}_{k=0}|\gamma^{(\alpha)}_{p}|^{2k}&=&\lim_{k\rightarrow \infty}\frac{1-|\gamma^{(\alpha)}_{p}|^{2k}}{1-|\gamma^{(\alpha)}_{p}|^2}~~~~~\underrightarrow{|\gamma^{(\alpha)}_{p}|<1\forall \alpha}~~~~~ \frac{1}{1-|\gamma^{(\alpha)}_{p}|^2},\\
                    \sum^{\infty}_{n=0}\sum^{\infty}_{r=0}|\Gamma^{(\alpha)}_{p,n}|^{2r}&=& \sum^{\infty}_{n=0}\lim_{r\rightarrow \infty}\frac{1-|\Gamma^{(\alpha)}_{p,n}|^{2r}}{1-|\Gamma^{(\alpha)}_{p,n}|^2}~~~~~\underrightarrow{|\Gamma^{(\alpha)}_{p,n}|<1 \forall n,\alpha}~~~~~ \sum^{\infty}_{n=0}\frac{1}{1-|\Gamma^{(\alpha)}_{p,n}|^2}=\left(f^{(\alpha)}_{p}\right)^{-1}.~~~~~~~~~~~~\eea
                    Consequently using these results for $\alpha$ vacuum we get: 
                    \bea {\bf Tr}\left[\left(1-|\gamma^{(\alpha)}_{p}|^2\right){\bf diag}\left(1,|\gamma^{(\alpha)}_{p}|^2,|\gamma^{(\alpha)}_{p}|^{4},|\gamma^{(\alpha)}_{p}|^{6}\cdots\right)\right]&=&\left(1-|\gamma^{(\alpha)}_{p}|^2\right)\sum^{\infty}_{k=0}|\gamma^{(\alpha)}_{p}|^{2k}=1,~~~~~~~~~~~~\\  {\bf Tr}\left[(f^{(\alpha)}_{p})^2\sum^{\infty}_{n=0}{\bf diag}\left(1,|\Gamma^{(\alpha)}_{p,n}|^2,|\Gamma^{(\alpha)}_{p,n}|^{4},|\Gamma^{(\alpha)}_{p,n}|^{6}\cdots\right)\right]&=&(f^{(\alpha)}_{p})^2\sum^{\infty}_{n=0}\sum^{\infty}_{r=0}|\Gamma^{(\alpha)}_{p,n}|^{2r}=f^{(\alpha)}_p,~~~~~\eea
                    Consequently the normalisation condition of this total density matrix in presence of $\alpha$ vacuum state is given by, 
                           ${\bf Tr}\rho_{\bf L}(\alpha)
                           =1+f^{(\alpha)}_p$. 
                          This result is consistent with the ref.~\cite{Maldacena:2012xp} where
                           $f^{(\alpha)}_p=0\forall \alpha$ and also ref.~\cite{Choudhury:2017bou} where $\alpha=0$ and $f^{(0)}_p=f_p$. But for simplicity it is better to maintain always ${\bf Tr}\rho_{\bf L}(\alpha)= 1$ and to get this result for $\alpha$ vacuum the total density matrix can be redefined by changing the normalisation constant as:
                           \bea  (\rho_{\bf L}(\alpha))_{p,l,m}&=&\frac{\left(1-|\gamma^{(\alpha)}_{p}|^2\right)}{1+f^{(\alpha)}_{p}}\sum^{\infty}_{k=0}|\gamma^{(\alpha)}_{p}|^{2k}\widetilde{|k;p,l,m\rangle}\widetilde{\langle k;p,l,m|}+\frac{(f^{(\alpha)}_{p})^2}{1+f^{(\alpha)}_p}\sum^{\infty}_{n,r=0}|\Gamma^{(\alpha)}_{p,n}|^{2r}\widetilde{|n,r;p,l,m\rangle}\widetilde{\langle n,r;p,l,m|},~~~~~~~~~\eea
                         In this context equivalent convention for normalisation factors can also be chosen such that it always satisfies ${\bf Tr}\rho_{\bf L}(\alpha)= 1$, even in the presence of source contribution~\footnote{ Here one can choose the following equivalent ansatz for total density matrix in presence of $\alpha$ vacuum as:
                                                                   \bea \label{cffcc1sx}(\rho_{\bf L}(\alpha))_{p,l,m}&=&\left[\frac{1}{1-|\gamma^{(\alpha)}_{p}|^2}+f^{(\alpha)}_{p}\right]^{-1}\left[\sum^{\infty}_{k=0}|\gamma^{(\alpha)}_{p}|^{2k}\widetilde{|k;p,l,m\rangle}\widetilde{\langle k;p,l,m|}+(f^{(\alpha)}_{p})^2\sum^{\infty}_{n=0}\sum^{\infty}_{r=0}|\Gamma^{(\alpha)}_{p,n}|^{2r}\widetilde{|n,r;p,l,m\rangle}\widetilde{\langle n,r;p,l,m|}\right].~~~~~~~~~~~\eea
      }.
                         
                           \item For each set of values of the ${\bf SO(1,3)}$ quantum numbers $p,l,m$, the density matrix yields $(\rho_{\bf L})_{p,l,m}$ and so that the total density matrix can be expressed as a product of all such possible contributions:
                           \bea \rho_{\bf L}(\alpha)&=& \prod^{\infty}_{p=0} \prod^{p-1}_{l=0} \prod^{+l}_{m=-l}(\rho_{\bf L}(\alpha))_{p,l,m}.~~~~~~~~~  \eea 
                           This also indicates that in such a situation entanglement is absent among all states which carries non identical ${\bf SO(1,3)}$ quantum numbers $p,l,m$.
                           
                           \item Finally, the total density matrix can be written in terms of entanglement modular Hamiltonian of the axionic pair as,  
                           $\rho_{\bf L}(\alpha)=e^{-\beta {\cal H}_{\bf ENT}},$
                           where at finite temperature $T_{\bf dS}$ of de Sitter space $\beta=2\pi/T_{\bf dS}$. If we further assume that the dynamical Hamiltonian in de Sitter space is represented by entangled Hamiltonian then for a given principal quantum number $p$ the Hamiltonian for axion can be expressed as: 
                           \bea {\cal H}_{p}(\alpha)&=& \left[E^{(\alpha)}_{p}\tilde{c}^{\dagger}_p \tilde{c}_{p}+\sum^{\infty}_{n=0}{\cal E}^{(\alpha)}_{p,n}\tilde{C}^{\dagger}_{p,n}\tilde{C}_{p,n}\right].\eea
                           Acting this Hamiltonian on the $\alpha$ vacuum state we find:
                           \bea {\cal H}_{p}(\alpha)|\alpha\rangle &\approx&\left[1-\left(|\gamma^{(\alpha)}_p|^2+\sum^{\infty}_{n=0}|\Gamma^{(\alpha)}_{p,n}|^2\right)\right]^{1/2}\left[E^{(\alpha)}_{p}\tilde{c}^{\dagger}_p \tilde{c}_{p}+\sum^{\infty}_{n=0}{\cal E}^{(\alpha)}_{p,n}\tilde{C}^{\dagger}_{p,n}\tilde{C}_{p,n}\right]\nonumber\times\\
                           &&\exp\left[\gamma^{(\alpha)}_{p}~\tilde{c}^{\dagger}_{\bf R}~\tilde{c}^{\dagger}_{\bf L}+\sum^{\infty}_{m=0}\Gamma^{(\alpha)}_{p,m}~\tilde{C}^{\dagger}_{{\bf R},m}~\tilde{C}^{\dagger}_{{\bf L},m}\right]\left(|{\bf R}^{'}\rangle\otimes |{\bf L}^{'}\rangle\right)^{(\alpha)}=E^{(\alpha)}_{{\bf T},p}|\alpha\rangle, \eea
                           where 
                            the total energy spectrum of this system for $\alpha$ vacuum can be written as:
                                                  \bea E^{(\alpha)}_{{\bf T},p}= E^{(\alpha)}_p+\sum^{\infty}_{n=0}{\cal E}^{(\alpha)}_{p,n}~~~\forall \alpha~~,{\rm with} ~~E^{(\alpha)}_p = -\frac{1}{2\pi}\ln (|\gamma^{(\alpha)}_p|^2),~~
                           {\cal E}^{(\alpha)}_{p,n} = -\frac{1}{2\pi}\ln \left(\frac{1}{1-|\Gamma^{(\alpha)}_{p,n}|^2}.\right) \eea
                           where spectrum for the complementary and particular part for $\alpha$ vacuum state is defined by $E^{(\alpha)}_p$ and ${\cal E}^{(\alpha)}_{p,n} $.

Now if we consider any arbitrary mass parameter $\nu$ and any arbitrary value of the parameter $\alpha$,  the ${\bf SO(1,3)}$ principal quantum number $p$ dependent spectrum can be expressed as:
       \bea\label{xs1} E^{(\alpha)}_p &=&-\frac{1}{2\pi}\ln \left\{\frac{1}{2|\tilde{m}_{\bf RL}|^2}\left[\left(1+|\tilde{m}_{\bf RL}|^4+|\tilde{m}_{\bf RR}|^4\right.\right.\right.\nonumber\\ && \left.\left.\left.-2|\tilde{m}_{\bf RR}|^2-\tilde{m}^2_{\bf RR}(\tilde{m}^{*}_{\bf RL})^2-\tilde{m}^2_{\bf RL}(\tilde{m}^{*}_{\bf RR})^2\right) 
                 \pm \left\{\left(-1-|\tilde{m}_{\bf RL}|^4-|\tilde{m}_{\bf RR}|^4\right.\right.\right.\right.\nonumber\\ && \left.\left.\left.\left.+2|\tilde{m}_{\bf RR}|^2+\tilde{m}^2_{\bf RR}(\tilde{m}^{*}_{\bf RL})^2+\tilde{m}^2_{\bf RL}(\tilde{m}^{*}_{\bf RR})^2\right)^2 
                 -4|\tilde{m}_{\bf RL}|^4\right\}^{\frac{1}{2}}\right]\right\}, \\
                                                        \label{xs2}    {\cal E}^{(\alpha)}_{p,n} &=& \frac{1}{2\pi}\ln\left\{1-
  \frac{1}{2|\tilde{m}_{{\bf RL},n}|^2}\left[\left(1+|\tilde{m}_{{\bf RL},n}|^4+|\tilde{m}_{{\bf RR},n}|^4\right.\right.\right.\nonumber\\ && \left.\left.\left.-2|\tilde{m}_{{\bf RR},n}|^2-\tilde{m}^2_{{\bf RR},n}(\tilde{m}^{*}_{{\bf RL},n})^2-\tilde{m}^2_{{\bf RL},n}(\tilde{m}^{*}_{{\bf RR},n})^2\right) 
             \pm \left\{\left(-1-|\tilde{m}_{{\bf RL},n}|^4-|\tilde{m}_{{\bf RR},n}|^4\right.\right.\right.\right.\nonumber\\ && \left.\left.\left.\left.+2|\tilde{m}_{{\bf RR},n}|^2+\tilde{m}^2_{{\bf RR},n}(\tilde{m}^{*}_{{\bf RL},n})^2+\tilde{m}^2_{{\bf RL},n}(\tilde{m}^{*}_{{\bf RR},n})^2\right)^2 
             -4|\tilde{m}_{{\bf RL},n}|^4\right\}^{\frac{1}{2}}\right]\right\}.\eea
                               
 Here the components $\tilde{m}_{\bf RR}=\tilde{m}_{\bf LL}$, $\tilde{m}_{\bf RL}=\tilde{m}_{\bf LR}$ and  $\tilde{m}_{{\bf RR},n}=\tilde{m}_{{\bf LL},n}$, $\tilde{m}_{{\bf RL},n}=\tilde{m}_{{\bf LR},n}$ are defined in Eqn~(\ref{ss1a}), Eqn~(\ref{ss2a}) and Eqn~(\ref{gg1a}), Eqn~(\ref{gg2a}) respectively.   

Further using Eq~(\ref{as1}) and Eq~(\ref{as2}), we get the following simplified expressions:
    \be \begin{array}{lll}
                            \label{xaxsa3}\footnotesize E^{(\alpha)}_p \approx -\frac{1}{2\pi}\ln\left[\frac{2}{\left(\sqrt{\cosh 2\pi p +\cos 2\pi \nu}\pm\sqrt{\cosh 2\pi p +\cos 2\pi \nu+2}\right)^2}\left|\frac{ \left[\cosh^2\alpha+\sinh^2\alpha~e^{2i\pi\nu}+\sinh 2\alpha\cos \pi \nu~e^{i\pi\nu}\right]}{\left(\cosh^2\alpha+\sinh^2\alpha~e^{-2\pi( p-i\nu)}\right)}\right|^2\right],\\
                                                  {\cal E}^{(\alpha)}_{p,n}\approx \frac{1}{2\pi}\ln\left[1-\frac{2}{\left(\sqrt{\cosh 2\pi p_n +\cos 2\pi \nu}\pm\sqrt{\cosh 2\pi p_n +\cos 2\pi \nu+2}\right)^2}\left|\frac{ \left[\cosh^2\alpha+\sinh^2\alpha~e^{2i\pi\nu}+\sinh 2\alpha\cos \pi \nu~e^{i\pi\nu}\right]}{\left(\cosh^2\alpha+\sinh^2\alpha~e^{-2\pi( p_n-i\nu)}\right)}\right|^2\right].~~~~~
       \end{array}        \ee 
                                                         
                These results imply that for arbitrary parameter $\nu$ and $\alpha$ the entangled Hamiltonian $({\cal H}_{\bf ENT})$ and the Hamiltonian for axion ${({\cal H}_p})_{{\bf R \times H^3}}$ are significantly different,  compared to the result obtained in absence of linear source term in case of Bunch Davies vacuum.

       \end{enumerate}
       
    \subsection{Computation of entanglement entropy using $\alpha$ vacua}
    \label{3c}
          In this subsection our prime objective is to derive the expression for entanglement entropy in de Sitter space in presence of $\alpha$ vacuum state. In general the entanglement entropy with arbitrary $\alpha$ can be expressed as:
          \bea S(p,\nu,\alpha)&=& -{\rm \bf Tr }\left[\rho_{\bf L}(p,\alpha)\ln \rho_{\bf L}(p,\alpha)\right],\eea
         where the parameter $\nu$ and the corresponding $\alpha$ vacuum state are defined in the earlier section. In this context the expression for  entanglement entropy for a given ${\bf SO(1,3)}$ principal quantum number $p$ can be expressed as~\footnote{If we follow the equivalent ansatz of density matrix as mentioned in Eq~(\ref{cffcc1sx}), the expression for entanglement entropy for a given ${\bf SO(1,3)}$ principal quantum number $p$ can be expressed as:\bea S(p,\nu,\alpha)&=& -\ln a^{(\alpha)}_p-a^{(\alpha)}_p\frac{|\gamma^{(\alpha)}_{p}|^2}{\left(1-|\gamma^{(\alpha)}_{p}|^2\right)^2}\ln\left(|\gamma^{(\alpha)}_{p}|^2\right)\left(1+f^{(\alpha)}_p\left(1-|\gamma^{(\alpha)}_{p}|^2\right)\right)\nonumber\\
                   &&-a^{(\alpha)}_p f^{(\alpha)}_p\ln\left(1+f^{(\alpha)}_p\left(1-|\gamma^{(\alpha)}_{p}|^2\right)\right)-a^{(\alpha)}_p(f^{(\alpha)}_p)^2\left(1-|\gamma^{(\alpha)}_{p}|^2\right)\ln\left(1+f^{(\alpha)}_p\right).~~~~~~~~~~~~~\eea
                    For our computation we will not use this ansatz any further.}:
\bea\label{oqw1} S(p,\nu,\alpha)&=&- \left(1+\frac{f^{(\alpha)}_p}{1+f^{(\alpha)}_p}\right)\left[\ln\left(1-|\gamma^{(\alpha)}_{p}|^2\right)+\frac{|\gamma^{(\alpha)}_{p}|^2}{\left(1-|\gamma^{(\alpha)}_{p}|^2\right)}\ln\left(|\gamma^{(\alpha)}_{p}|^2\right)\right]-\left(1-f^{(\alpha)}_p\right)\ln\left(1+f^{(\alpha)}_p\right).~~~~~~~~~~~~~\eea             Then the final expression for the entanglement entropy in de Sitter space can be expressed as a sum over all possible quantum states which carries ${\bf SO(1,3)}$ principal quantum number $p$. Consequently, the final expression for the entanglement entropy in de Sitter space is given by the following expression:
       \bea S(\nu,\alpha)= \sum_{\bf States}\sum^{\infty}_{p=0}S(p,\nu,\alpha)\rightarrow V_{\bf H^3} \int^{\infty}_{p=0}~dp~{\cal D}_{3}(p)~S(p,\nu,\alpha)={\bf c_6}(\alpha,\nu)V_{\bf H^3}/V^{\bf REG}_{\bf H^3},~~~~~~~~\eea 
       where ${\cal D}_3(p)=p^2/2\pi^2$ characterize the density of states for radial functions on the Hyperboloid ${\bf H^3}$. Additionally, it is important to note that the volume of the hyperboloid ${\bf H^3}$ is denoted by the overall factor $V_{\bf H^3}$. Here the regularized volume of the hyperboloid ${\bf H^3}$ for $r\leq  L_c$ can be written as:
              \bea V_{\bf H^3}= V_{\bf S^2}\int^{L_c}_{r=0}dr~\sinh^2r~~\underrightarrow{\rm {\bf large}~L_c}~~\frac{\pi}{2}\left[e^{2L_c} -4L_c\right]&=&\left[{\bf A_{ENT}} -\pi \ln {\bf A_{ENT}}+\pi\ln\left(\frac{\pi}{2}\right)\right]=V^{\bf REG}_{\bf H^3}\left[\frac{1}{4\eta^2} +\ln\eta\right].~~~~~~~~~\eea  
              where ${\bf A_{ENT}}$ is the entangling area and we use $V_{\bf S^2}=4\pi$. 
              Here the cutoff $L_c$ can be written as,
              $L_c\sim -\ln \eta$. 
               In this context we define regularized volume of the hyperboloid ${\bf H^3}$ as, 
               $V^{\bf REG}_{\bf H^3}=V_{\bf S^3}/2=2\pi$. 
               
               In $3+1$ D , for the case of $\alpha$ vacuum, long range quantum correlation is measured by ${\bf c_6}(\alpha,\nu)$, which is defined as:
                      \bea \label{po1} {\bf c_6}(\alpha,\nu)\equiv S_{\bf intr}(\alpha,\nu)
                      = \left[\left(1+\frac{f^{(\alpha)}_p}{1+f^{(\alpha)}_p}\right){\cal I}^{(\alpha)}+\left(1-f^{(\alpha)}_p\right)\ln\left(1+f^{(\alpha)}_p\right){\cal V}\right],\eea 
                    where the integrals ${\cal I}^{(\alpha)}$ and ${\cal V}$ can be written in $3+1$ dimensional space-time  as:
                                     \bea {\cal I}^{(\alpha)}&=&-\frac{1}{\pi}\int^{\infty}_{p=0}~dp~p^2~\left[\ln\left(1-|\gamma^{(\alpha)}_{p}|^2\right)+\frac{|\gamma^{(\alpha)}_{p}|^2}{\left(1-|\gamma^{(\alpha)}_{p}|^2\right)}\ln\left(|\gamma^{(\alpha)}_{p}|^2\right)\right],~~~~
                                     {\cal V}=-\frac{1}{\pi}\int^{\infty}_{p=0}~dp~p^2.\eea
        \begin{figure*}[htb]
       \centering
         \subfigure[Normalized entanglement entropy vs $\nu^2$ in $3+1$ D de Sitter space in presence of $\alpha$ vacuum and in absence axionic source ($f^{(\alpha)}_p=0$).]{
    \includegraphics[width=7.2cm,height=4cm] {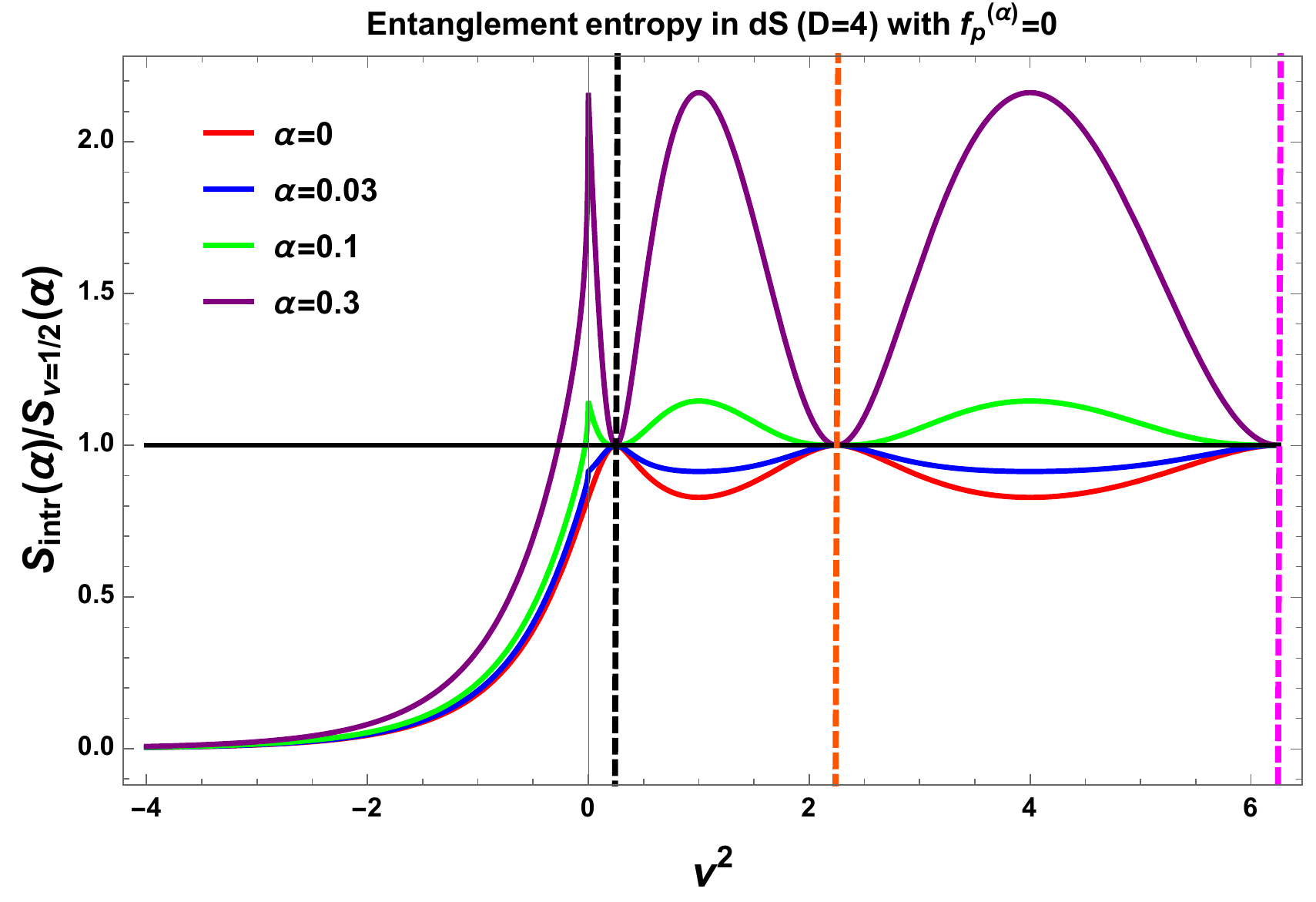}\label{azgax5}
        }
    \subfigure[Normalized entanglement entropy vs $\nu^2$ in $3+1$ D de Sitter space in presence of $\alpha$ vacuum and axionic source ($f^{(\alpha)}_p=10^{-7}$).]{
   \includegraphics[width=7.2cm,height=4cm] {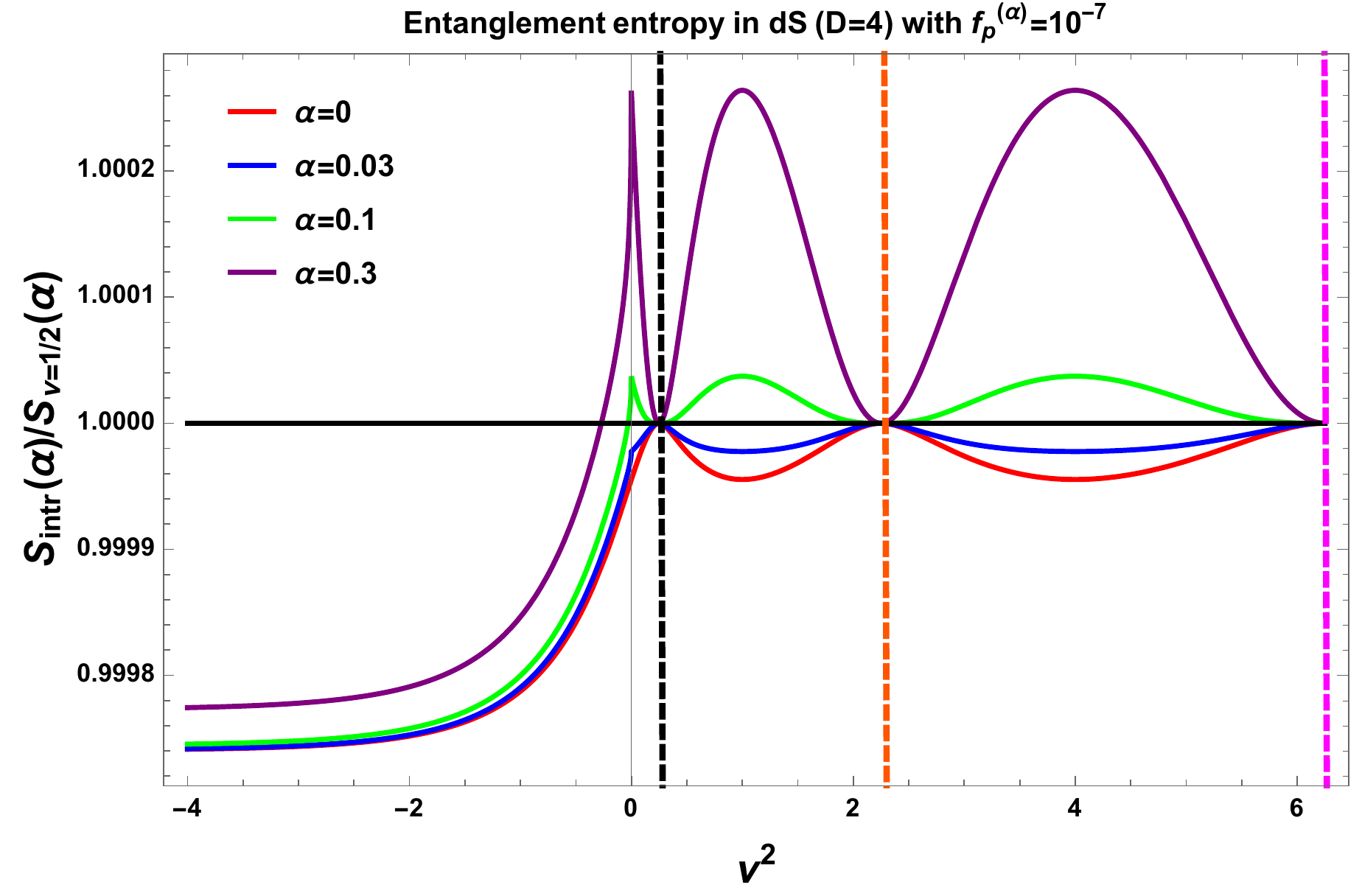} \label{azgax6}
   }
     \caption[Optional caption for list of figures]{Normalized entanglement entropy $S_{intr}/S_{\nu=1/2}$ vs mass parameter $\nu^2$ in $3+1$ D de Sitter space in absence of axionic source ($f^{(\alpha)}_p=0$) and in presence of axionic source ($f^{(\alpha)}_p=10^{-7}$) for $`+'$ branch of solution of $\alpha$ vacuum i.e $|\gamma^{(\alpha)}_p|$ and $|\Gamma^{(\alpha)}_{p,n}|$. In both the situations we have normalized with conformal $\nu=1/2$ result in presence of $\alpha$ vacuum.} 
                                                                    \label{szpxxgg2}
                                                                    \end{figure*}                                                                                                                         
       \begin{figure*}[htb]
      \centering
        \subfigure[Entanglement entropy vs $\alpha$ in $3+1$ D de Sitter space without axionic source ($f^{(\alpha)}_p=0$).]{
     \includegraphics[width=7.2cm,height=3.8cm] {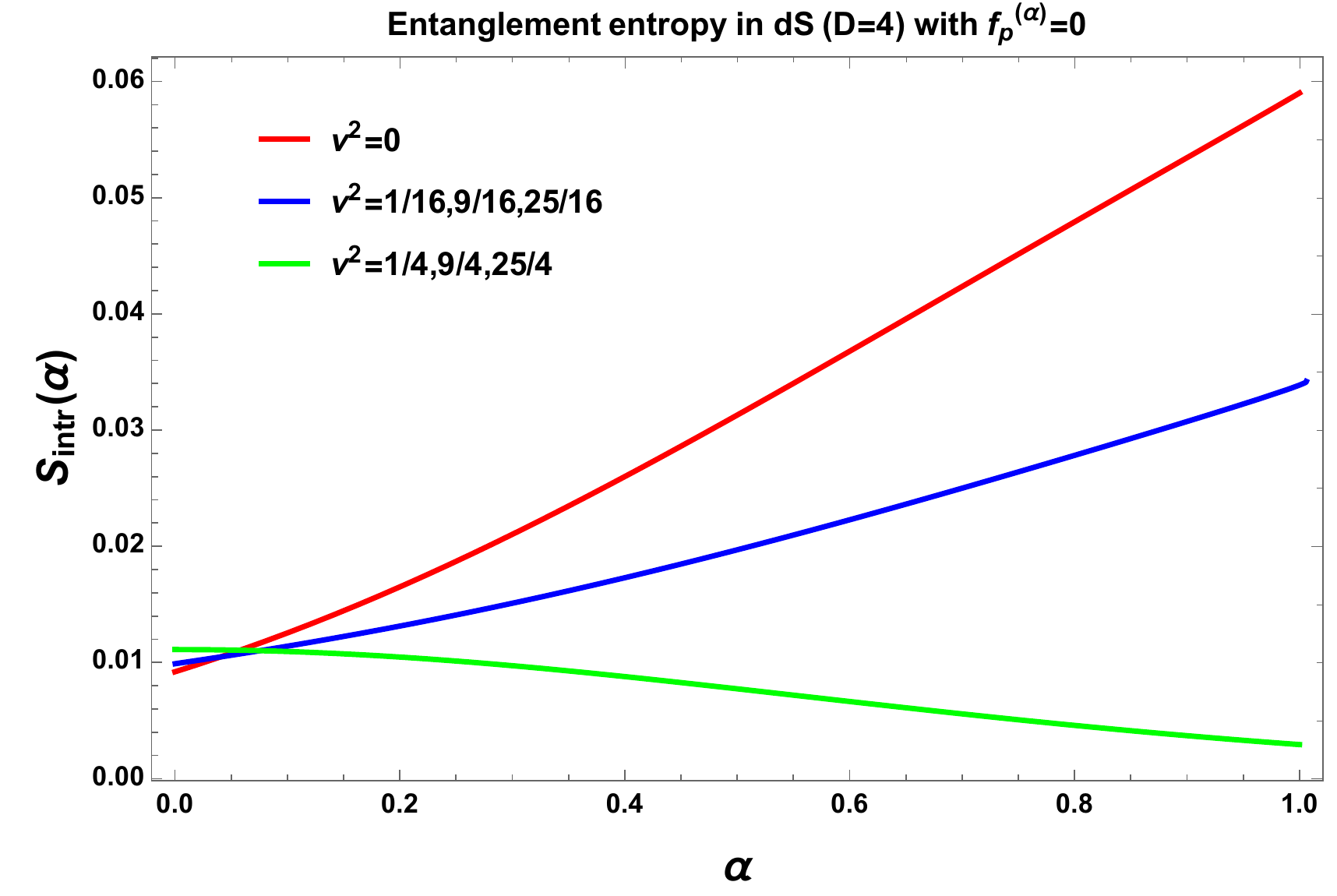}
    \label{zgax5}
        }
   \subfigure[Entanglement entropy vs  $\alpha$ in $3+1$ D de Sitter space with axionic source ($f^{(\alpha)}_p=10^{-7}$).]{
        \includegraphics[width=7.2cm,height=3.8cm] {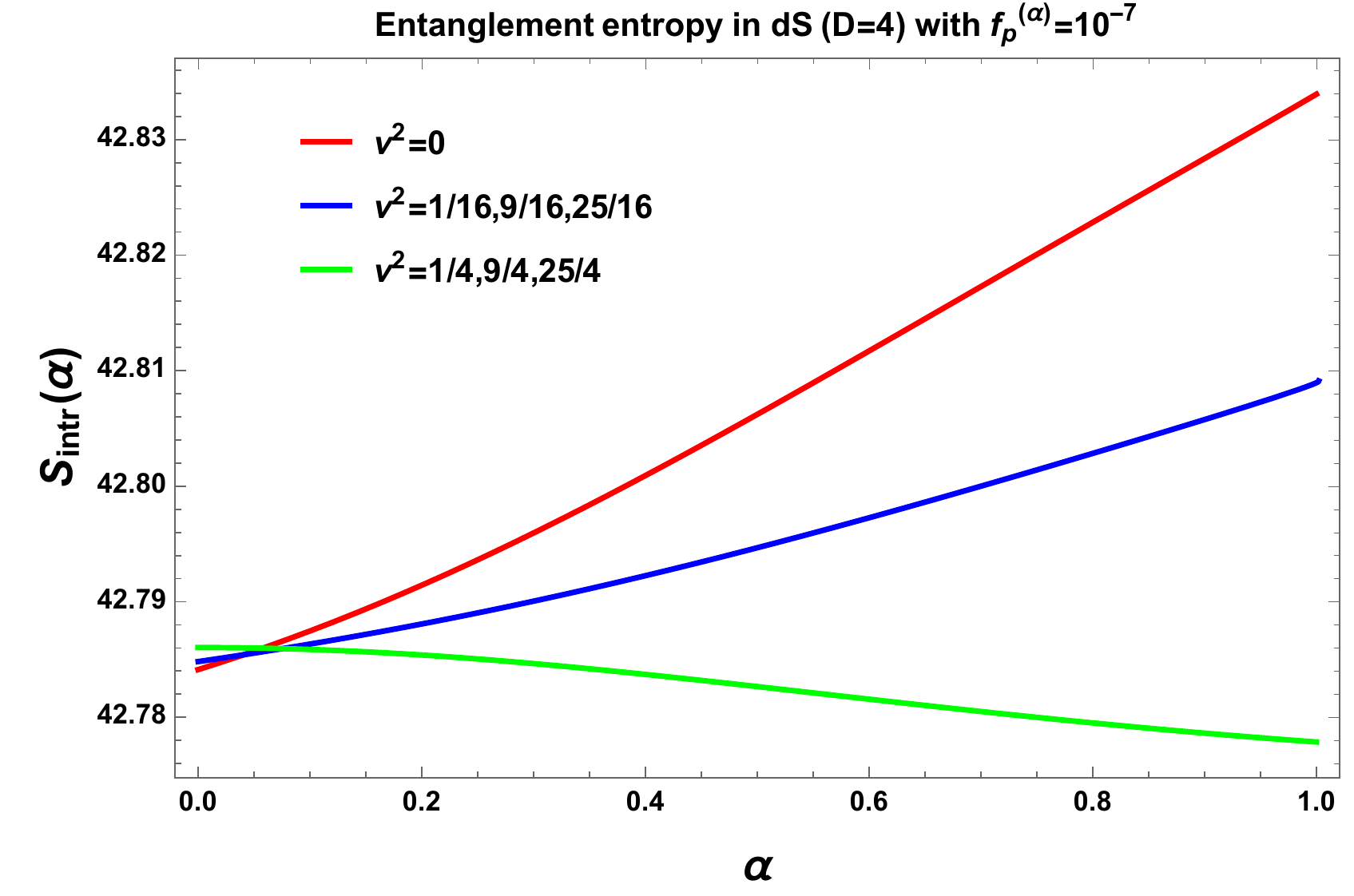}
    \label{zgax6}
            } 
    \subfigure[Entanglement entropy vs $\alpha$ in $3+1$ D de Sitter space with axionic source ($f^{(\alpha)}_p=10^{-7}$).]{
          \includegraphics[width=7.2cm,height=3.8cm] {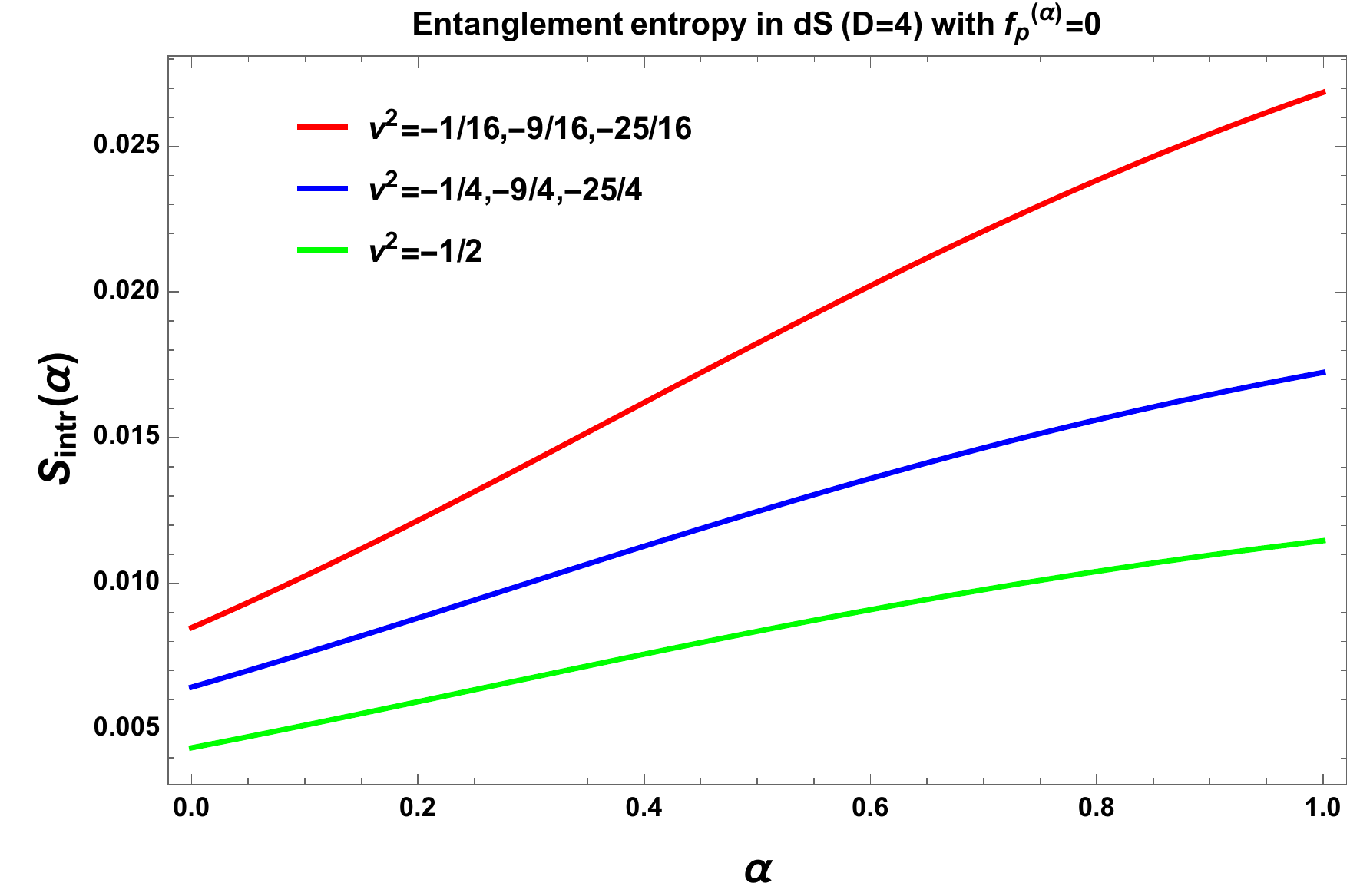}
      \label{zgax7}
              }   
   \subfigure[Entanglement entropy vs  $\alpha$ in $3+1$ D de Sitter space with axionic source ($f^{(\alpha)}_p=10^{-7}$).]{
           \includegraphics[width=7.2cm,height=3.8cm] {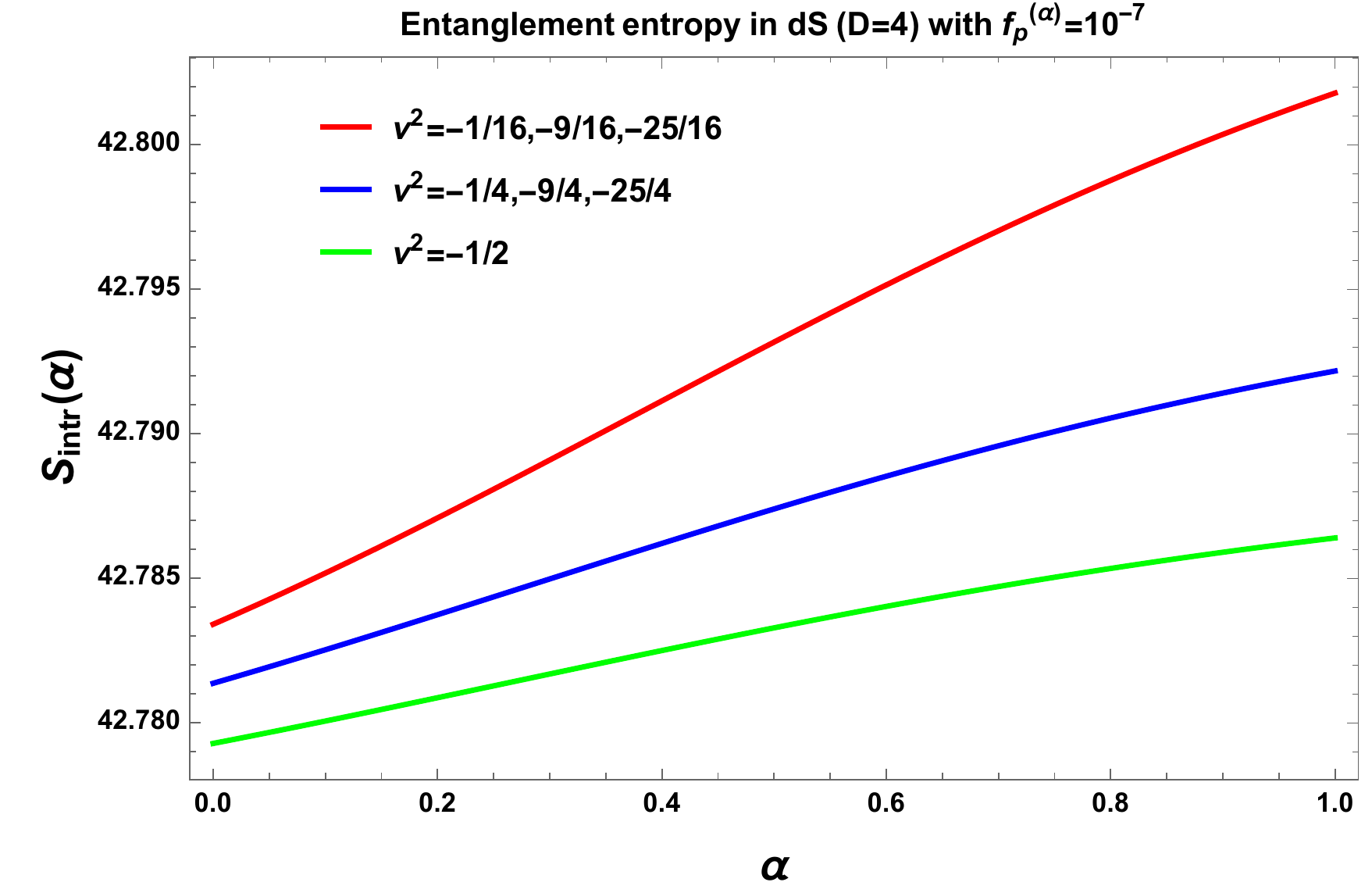}
       \label{zgax8}
               }                                                                         \caption[Optional caption for list of figures]{Entanglement entropy $S_{intr}(\alpha)$ vs  parameter $\alpha$ in $3+1$ D de Sitter space in absence of axionic source ($f^{(\alpha)}_p=0$) and in presence of axionic source ($f^{(\alpha)}_p=10^{-7}$) for $`+'$ branch of solution of $|\gamma^{(\alpha)}_p|$ and $|\Gamma^{(\alpha)}_{p,n}|$. Here we fix the value of the parameter $\nu^2$ at different positive and negative values including zero.} 
                                                                   \label{zpxxgg2}
                                                                   \end{figure*}                                                                                                                                               

               Here it is important to mention that:
               \begin{itemize}
               \item The integral ${\cal V}$ is divergent. To make it finite, we need to regularise it by introducing a change in variable by using $x=2\pi p$ and by introducing a UV cut off $\Lambda_{\bf C}$  leading to:
               \bea  {\cal V}&=&-\frac{1}{8\pi^4}\int^{\Lambda_{\bf C}}_{x=0}~dx~x^2=-\frac{\Lambda^3_{\bf C}}{24\pi^4}=-\frac{L^3}{3\pi}=-\frac{1}{4\pi^2}V_{\bf S^2}~~~~{\rm where}~~~V_{\bf S^2}=\frac{4}{3}\pi L^3=\frac{1}{6\pi^2}\Lambda^3_{\bf C}.\eea
               The magnitude of this integral represents the finite volume of the configuration space in which we are computing the entanglement entropy from Von Neumann measure.  In principle, this volume can be infinite, but after fixing the cut-off, this integral actually proportional to a finite volume of a sphere of radius $\Lambda_{\bf C}$.  Actually, the cut-off of the parameter $x$ fix the highest accessible value of the characteristic length scale $p$ , which mimics the role of some sort of principal quantum number as appearing in hydrogen atom problem. Only the difference is, here  $p$ is restricted within the window, $0<p<\Lambda_{\bf C}/2\pi$  and it takes all possible values within this window (i.e. continuous). After fixing the cut-off, the length of the system under consideration can be expressed as, $L=\int^{\Lambda_{\bf C}/2\pi}_{0}dp=\Lambda_{\bf C}/2\pi$. Now it is obvious that if  the cut-off $\Lambda_{\bf C}\rightarrow \infty$, the entanglement entropy is diverging, which physically implies that the effect of long range quantum correlation is very large at very large length scale, $L=\int^{\infty}_{0}dp\rightarrow \infty$ i.e. at  late time scales in de Sitter space.

               \item Further we analyse the integral ${\cal I}^{(\alpha)}$ using both of the solutions obtained for arbitrary $\nu$ and $\nu=3/2$. Following the previous argument, we also put a cut-off $\Lambda_{\bf C}$ to perform the integral on the rescaled ${\bf SO(1,3)}$ quantum number $x=2\pi p$ and after performing the integral we can study the behaviour of both of the results.
               First of all we start with the following integral with $``\pm"$ signature, as given by:
               \be{\cal I}^{(\alpha)}=-\frac{1}{8\pi^4}\int^{\Lambda_{\bf C}}_{x=0}~dx~x^2 
             \left[\ln\left(1-2G_{\pm}(x,\nu,\alpha)\right)+\frac{2G_{\pm}(x,\nu,\alpha)}{\left(1-2G_{\pm}(x,\nu,\alpha)\right)}\ln\left(2G_{\pm}(x,\nu,\alpha)\right)\right],~~~~~~~~~\ee
               where $G_{\pm}(x,\nu,\alpha)$ for any arbitrary value of the parameter $\alpha$ is defined as:
               \bea \label{ccx} G_{\pm}(x,\nu,\alpha)&=&\frac{1}{4|\tilde{m}_{\bf RL}|^2}\left[\left(1+|\tilde{m}_{\bf RL}|^4+|\tilde{m}_{\bf RR}|^4\right.\right.\nonumber\\ && \left.\left.-2|\tilde{m}_{\bf RR}|^2-\tilde{m}^2_{\bf RR}(\tilde{m}^{*}_{\bf RL})^2-\tilde{m}^2_{\bf RL}(\tilde{m}^{*}_{\bf RR})^2\right) 
                                \pm \left\{\left(-1-|\tilde{m}_{\bf RL}|^4-|\tilde{m}_{\bf RR}|^4\right.\right.\right.\nonumber\\ && \left.\left.\left.+2|\tilde{m}_{\bf RR}|^2+\tilde{m}^2_{\bf RR}(\tilde{m}^{*}_{\bf RL})^2+\tilde{m}^2_{\bf RL}(\tilde{m}^{*}_{\bf RR})^2\right)^2 
                                -4|\tilde{m}_{\bf RL}|^4\right\}^{\frac{1}{2}}\right],\eea
where the components $\tilde{m}_{\bf RR}=\tilde{m}_{\bf LL}$ and $\tilde{m}_{\bf RL}=\tilde{m}_{\bf LR}$ are redefined in terms of the new variable $x=2\pi p$.

  Here small axion mass ($\nu^2>0$) limiting situations are considered in $\nu=1/2$  conformal mass as well in $\nu=3/2$ case in presence of an additional arbitrary parameter $\alpha$. Additionally, we consider large axion mass ($\nu^2<0$ where $\nu\rightarrow -i|\nu|$) limiting situation. In this large axion mass limiting situation we consider the window of ${\bf SO(1,3)}$ principal quantum number is $0<p<|\nu|$. Consequently, the entries of the coefficient matrix $\tilde{m}$ can be approximated as:
 \bea 
                 \label{ar2}
        \displaystyle \tilde{m}_{\bf RR} &=& -\sqrt{\frac{\cosh(|\nu|-p)}{\cosh(|\nu|+p)}} 
                                        \frac{2~\left[\cosh2\alpha\cosh^2\pi |\nu|-\sinh2\alpha\sinh^2\pi p+\frac{1}{2}\sinh 2\pi|\nu|\right]}{(e^{2\pi p}+e^{2\pi |\nu|})\cosh^2\alpha+(e^{2\pi p}+e^{2\pi |\nu|})\sinh^2\alpha},~~~~~~~~~~ \\
                   \label{arx2}
          \displaystyle \tilde{m}_{\bf RL} &=&-\sqrt{\frac{\cosh(|\nu|-p)}{\cosh(|\nu|+p)}} 
                                                  \frac{2~i~\left[\left(\cosh2\alpha+\sinh2\alpha\right)\cosh\pi |\nu|+\sinh\pi|\nu|\right]}{(e^{2\pi p}+e^{2\pi |\nu|})\cosh^2\alpha+(e^{2\pi p}+e^{2\pi |\nu|})\sinh^2\alpha}.~~~~~~~~~\eea      
This implies that for $\alpha$ vacuum if we consider the large axion mass ($\nu^2<0$ where $\nu\rightarrow -i|\nu|$) limiting situation we get always real value for $\tilde{m}_{\bf RR}$ and  imaginary value for $\tilde{m}_{\bf RL}$. Consequently  one can esaily reduce the four sets of Eqn.~(\ref{a1}) and Eqn.~(\ref{a3}) into two sets of equations as exactly we have done in ref.~\cite{Choudhury:2017bou} for Bunch Davies vacuum. In this large axion mass ($\nu^2<0$ where $\nu\rightarrow -i|\nu|$) limiting situation the two solutions for the $\gamma^{(\alpha)}_p$ for $\alpha$ vacuum are given by: 
\bea \gamma^{(\alpha)}_p&\approx&\frac{1}{2|\tilde{m}_{\bf RL}|}\left[\left(1+|\tilde{m}_{\bf RL}|^2-\tilde{m}^2_{\bf RR}\right) 
         \pm \sqrt{\left(1+|\tilde{m}_{\bf RL}|^2-\tilde{m}^2_{\bf RR}\right)^2-4|\tilde{m}_{\bf RL}|^2}\right].~~~~~~~~~~~\eea                                                                                                      
 Small mass limiting situations are explicitly appearing in $\nu=1/2$ and  $\nu=3/2$ case. For our study here we consider large mass limiting situation which is important to study the physical outcomes. In this situation we divide the total window of $p$ into two regions, as given by $0<p<|\nu|$ and $|\nu|<p<\Lambda_{\bf C}$. Here in these region of interests the two solutions for $\gamma^{(\alpha)}_p$ in presence of $\alpha$ vacuum can be approximately written as:
                     \bea
                                    \label{r2zzzzz}
                           \displaystyle |\gamma^{(\alpha)}_p| &\approx&\displaystyle\left\{\begin{array}{ll}
                          \displaystyle e^{-\pi |\nu|}\left(1+\tan\alpha\right)~~~~~~~~~~~~ &
                                                                                    \mbox{\small {\textcolor{red}{\bf for $0<p<|\nu|$}}}  
                                                                                   \\ 
                                   \displaystyle \frac{e^{-\pi p}\left(1+\tan\alpha\right)\left(1+\tan\alpha~e^{2\pi|\nu|}\right)}{\left(1+\tan^2\alpha~e^{-2\pi p}\right)} & \mbox{\small { \textcolor{red}{\bf for $|\nu|<p<\Lambda_{\bf C}/2\pi$}}}.~~~~~~~~
                                                                                             \end{array}
                                                                                   \right.\eea
                        and
                        \bea
                                       \label{r2zzzzaz}
                              \displaystyle |\gamma^{(\alpha)}_p| &=&\displaystyle\left\{\begin{array}{ll}
                             \displaystyle e^{\pi |\nu|}\left(1+\tan\alpha\right)~~~~~~~~~~~~ &
                                                                                       \mbox{\small {\textcolor{red}{\bf for $0<p<|\nu|$}}}  
                                                                                      \\ 
                                      \displaystyle \frac{e^{\pi p}\left(1+\tan\alpha\right)\left(1+\tan\alpha~e^{2\pi|\nu|}\right)}{\left(1+\tan^2\alpha~e^{-2\pi p}\right)} & \mbox{\small { \textcolor{red}{\bf for $|\nu|<p<\Lambda_{\bf C}/2\pi$}}}.~~~~~~~~
                                                                                                \end{array}
                                                                                      \right.\eea
                      As a result, for large mass limiting range the $\alpha$ parameter dependent regularised integral ${\cal I}^{(\alpha)}_1$ for the first solution for $|\gamma^{(\alpha)}_p|$ can be written as:
                                          \bea \label{Ar2}
                                 \displaystyle {\cal I}^{(\alpha)}_1 &=&\displaystyle\footnotesize\left\{\begin{array}{ll}
                                \displaystyle -\frac{{\cal A}(\nu)}{8\pi^4}\left[\ln\left(1-e^{-2\pi\nu}\left(1+\tan\alpha\right)^2\right)+\frac{\left(2\ln\left(1+\tan\alpha\right)-2\pi\nu\right)~e^{-2\pi\nu}\left(1+\tan\alpha\right)^2}{\left(1-e^{-2\pi\nu}\left(1+\tan\alpha\right)^2\right)}\right]~~~ &
                                                                                          \mbox{\small {\textcolor{red}{\bf for $0<x<2\pi|\nu|$}}}  
                                                                                         \\ 
                                         \displaystyle -\frac{{\cal B}(\nu,\alpha,\Lambda_{\bf C})}{8\pi^4} & \mbox{\small { \textcolor{red}{\bf for $2\pi|\nu|<x<\Lambda_{\bf C}$}}}.~~~~~~~~
                                                                                                   \end{array}
                                                                                         \right.\eea
                              and for the second solution for $|\gamma^{(\alpha)}_p|$ we get:
                             \bea \label{Ar3}
                                   \displaystyle {\cal I}^{(\alpha)}_1 &=&\displaystyle\footnotesize\left\{\begin{array}{ll}
                                  \displaystyle -\frac{{\cal A}(\nu)}{8\pi^4}\left[\ln\left(1-e^{2\pi\nu}\left(1+\tan\alpha\right)^2\right)+\frac{\left(2\ln\left(1+\tan\alpha\right)+2\pi\nu\right)~e^{2\pi\nu}\left(1+\tan\alpha\right)^2}{\left(1-e^{2\pi\nu}\left(1+\tan\alpha\right)^2\right)}\right]~ &
                                                                                            \mbox{\small {\textcolor{red}{\bf for $0<x<2\pi|\nu|$}}}  
                                                                                           \\ 
                                           \displaystyle -\frac{{\cal C}(\nu,\alpha,\Lambda_{\bf C})}{8\pi^4} & \mbox{\small { \textcolor{red}{\bf for $2\pi|\nu|<x<\Lambda_{\bf C}$}}}.~~~~~~
                                                                                                     \end{array}
                                                                                           \right.\eea
                                                                                           In Eq.~(\ref{Ar2}) and Eq~(\ref{Ar3}) coefficients ${\cal A}(\nu)$, ${\cal B}(\nu,\alpha,\Lambda_{\bf C})$ and ${\cal C}(\nu,\alpha,\Lambda_{\bf C})$ are defined by the following expressions:
                                \bea \label{p1} {\cal A}(\nu)&=&\int^{2\pi\nu}_{x=0}~dx~x^2=\frac{8\pi^3}{3}\nu^3,\\
                               \label{p2} {\cal B}(\nu,\alpha,\Lambda_{\bf C})&\approx&\int^{\Lambda_{\bf C}}_{x=2\pi\nu}~dx~x^2~\left[\ln\left(1-\frac{e^{-x}\left(1+\tan\alpha\right)^2\left(1+\tan\alpha~e^{2\pi|\nu|}\right)^2}{\left(1-\tan^2\alpha~e^{-x}\right)^2} \right)\right.\nonumber\\&& \left.+\frac{\frac{e^{-x}\left(1+\tan\alpha\right)^2\left(1+\tan\alpha~e^{2\pi|\nu|}\right)^2}{\left(1-\tan^2\alpha~e^{-x}\right)^2}}{\left(1-\frac{e^{-x}\left(1+\tan\alpha\right)^2\left(1+\tan\alpha~e^{2\pi|\nu|}\right)^2}{\left(1-\tan^2\alpha~e^{-x}\right)^2} \right)}\left\{2\ln(\left(1+\tan\alpha\right)\left(1+\tan\alpha~e^{2\pi|\nu|}\right))\right.\right.\nonumber\\&&\left.\left.~~~~~~~~~~~~~~~~~~~~~~~~~~~~~~~~~~~~~-2\ln\left(1-\tan^2\alpha~e^{-x}\right)-x\right\}\right],~~~~~~~~~~~~\eea\bea
                          \label{p3} {\cal C}(\nu,\alpha,\Lambda_{\bf C})&=&\int^{\Lambda_{\bf C}}_{x=2\pi\nu}~dx~x^2~\left[\ln\left(1-\frac{e^{x}\left(1+\tan\alpha\right)^2\left(1+\tan\alpha~e^{2\pi|\nu|}\right)^2}{\left(1-\tan^2\alpha~e^{-x}\right)^2} \right)\right.\nonumber\\&& \left.+\frac{\frac{e^{x}\left(1+\tan\alpha\right)^2\left(1+\tan\alpha~e^{2\pi|\nu|}\right)^2}{\left(1-\tan^2\alpha~e^{-x}\right)^2}}{\left(1-\frac{e^{x}\left(1+\tan\alpha\right)^2\left(1+\tan\alpha~e^{2\pi|\nu|}\right)^2}{\left(1-\tan^2\alpha~e^{-x}\right)^2} \right)}\left\{2\ln(\left(1+\tan\alpha\right)\left(1+\tan\alpha~e^{2\pi|\nu|}\right))\right.\right.\nonumber\\&&\left.\left.~~~~~~~~~~~~~~~~~~~~~~~~~~~~~~~~~~~~~-2\ln\left(1-\tan^2\alpha~e^{-x}\right)+x\right\}\right].                         \eea 
                           
     Further within the window $0<x<2\pi |\nu|$ we take the large mass limit $|\nu|>>1$ in the first solution for $|\gamma^{(\alpha)}_p|$ in presence of $\alpha$ vacuum:
      \be \label{vb}
           \displaystyle \lim_{|\nu|>>1}{\cal I}^{(\alpha)}_1 \approx\displaystyle
          \displaystyle  \frac{2\nu^4}{3}e^{-2\pi\nu}\left(1+\tan\alpha\right)^2\left\{1-\frac{1}{\pi\nu}\ln\left(1+\tan\alpha\right)\right\}\left[1+\left(1+\tan\alpha\right)^2{\cal O}\left(\nu^{-1}\right)\right].\ee
        
                    Similarly the integral ${\cal V}$ can be written as:
 \bea \label{zz}
                                   \displaystyle {\cal V} &=&\displaystyle\left\{\begin{array}{ll}
                                  \displaystyle -\frac{1}{8\pi^4}\int^{2\pi|\nu|}_{x=0}dx~x^2=-\frac{|\nu|^3}{3\pi}~~~~~~ &
                                                                                            \mbox{\small {\textcolor{red}{\bf for $0<x<2\pi|\nu|$}}}  
                                                                                           \\ 
                                           \displaystyle -\frac{1}{8\pi^4}\int^{\Lambda_{\bf C}}_{x=2\pi|\nu|}dx~x^2=-\frac{1}{24\pi^4}\left( \Lambda^3_{\bf C}-8\pi^3|\nu|^3\right). & \mbox{\small { \textcolor{red}{\bf for $2\pi|\nu|<x<\Lambda_{\bf C}$}}}.~~~~~~
                                                                                                     \end{array}
                                                                                           \right.\eea  
                                                                                           Consequently in the large mass limiting situation ($0<x<2\pi|\nu|$) we get the following expression for the entanglement entropy:
 \bea &&\lim_{|\nu|>>1}{\bf c_6}(\alpha,\nu)\approx \displaystyle
                       \displaystyle  \left(1+\frac{f^{(\alpha)}_p}{1+f^{(\alpha)}_p}\right)\frac{2\nu^4}{3}e^{-2\pi\nu}\left(1+\tan\alpha\right)^2\nonumber\\
                       &&~~~~~~~~~~~~~~~~~~~~\times\left\{1-\frac{1}{\pi\nu}\ln\left(1+\tan\alpha\right)\right\}\left[1+\left(1+\tan\alpha\right)^2{\cal O}\left(\nu^{-1}\right)\right]-\left(1-f^{(\alpha)}_p\right)\ln\left(1+f^{(\alpha)}_p\right)\frac{\nu^3}{3\pi},~~~~~~~~~~\eea
                                                Further in absence of the source contribution in the large mass limit the long range quantum correlation can be expressed as: 
            \bea \lim_{|\nu|>>1, f_p\rightarrow 0}{\bf c_6}(\alpha,\nu)&\approx&\displaystyle
                      \displaystyle  \frac{2\nu^4}{3}e^{-2\pi\nu}\left(1+\tan\alpha\right)^2\left\{1-\frac{1}{\pi\nu}\ln\left(1+\tan\alpha\right)\right\}\left[1+\left(1+\tan\alpha\right)^2{\cal O}\left(\nu^{-1}\right)\right].~~~~~~\eea
                                     
           For the second solution of $|\gamma^{(\alpha)}_p|$ in presence of $\alpha$ vacuum, we get:
           \bea \label{we}
                        \displaystyle  \lim_{|\nu|>>1}{\cal I}^{(\alpha)}_1 &=&\displaystyle
                       \displaystyle -\frac{\nu^3}{3\pi}\left[\ln\left(1-e^{2\pi\nu}\left(1+\tan\alpha\right)^2\right)+\frac{\left(2\ln\left(1+\tan\alpha\right)+2\pi\nu\right)~e^{2\pi\nu}\left(1+\tan\alpha\right)^2}{\left(1-e^{2\pi\nu}\left(1+\tan\alpha\right)^2\right)}\right].~~~~~~ \eea
 
 Consequently in the large mass limiting situation ($0<x<2\pi|\nu|$) we get the following expression for the entanglement entropy:
  \bea \lim_{|\nu|>>1}{\bf c_6}(\alpha,\nu)&\approx&\displaystyle
                        \displaystyle  -\left(1+\frac{f^{(\alpha)}_p}{1+f^{(\alpha)}_p}\right)\frac{\nu^3}{3\pi}\left[\ln\left(1-e^{2\pi\nu}\left(1+\tan\alpha\right)^2\right)\right.\nonumber\\ &&\left.~\displaystyle ~~+\frac{\left(2\ln\left(1+\tan\alpha\right)+2\pi\nu\right)~e^{2\pi\nu}\left(1+\tan\alpha\right)^2}{\left(1-e^{2\pi\nu}\left(1+\tan\alpha\right)^2\right)}\right]-\left(1-f^{(\alpha)}_p\right)\ln\left(1+f^{(\alpha)}_p\right)\frac{\nu^3}{3\pi}.~~~~~~\eea
                                                 Further in absence of the source contribution in the large mass limit the long range quantum correlation can be expressed as: 
  \bea \lim_{|\nu|>>1, f_p\rightarrow 0}{\bf c_6}(\alpha,\nu)&\approx&\displaystyle
                       \displaystyle  -\frac{\nu^3}{3\pi}\left[\ln\left(1-e^{2\pi\nu}\left(1+\tan\alpha\right)^2\right)+\frac{\left(2\ln\left(1+\tan\alpha\right)+2\pi\nu\right)~e^{2\pi\nu}\left(1+\tan\alpha\right)^2}{\left(1-e^{2\pi\nu}\left(1+\tan\alpha\right)^2\right)}\right].~~~~~~~~~~\eea 

                     In fig.~(\ref{azgax5}) and fig.~(\ref{azgax6}), we have demonstrated the behaviour of entanglement entropy in $D=4$ de Sitter space in absence ($f^{(\alpha)}_p=0$) and in presence ($f^{(\alpha)}_p=10^{-7}$) of axionic source with respect to the mass parameter $\nu^2$. In both the cases we have normalized the entanglement entropy with the result obtained from conformal mass parameter $\nu=1/2$ in presence of $\alpha$ vacuum. In fig.~(\ref{azgax5}), it is clearly observed that in absence of axionic source in the large mass regime (where $\nu^2<0$) the normalised entanglement entropy $S_{intr}(\alpha)/S_{\nu=1/2}(\alpha)$ asymptotically approaches towards zero. In the large mass regime the measure of long range correlation (or more precisely the entanglement entropy) in presence of $\alpha$ vacuum for axion can be expressed for $\gamma^{(\alpha)}_p=e^{-\pi|m_{axion}/H|}\left(1+\tan\alpha\right)$ as:
                     \bea\label{c1}&& {\bf c_6}\left(\alpha,|\nu|\approx\frac{m_{axion}}{H}\right)= S_{\bf intr}\left(\alpha,|\nu|\approx\frac{m_{axion}}{H}\right)\nonumber\\
                     &\approx& \frac{2}{3}\left(\frac{m_{axion}}{H}\right)^4 e^{-\frac{2\pi m_{axion}}{H}}\left(1+\tan\alpha\right)^2\left\{1-\frac{H}{\pi m_{axion}}\ln\left(1+\tan\alpha\right)\right\}\left[1+\left(1+\tan\alpha\right)^2{\cal O}\left(\frac{H}{m_{axion}}\right)\right],~~~~~~~~~~\eea
                                          If we further use Eq~(\ref{c1}) then it is clearly observed that in presence of $\alpha$ vacuum one is able to get considerably large entanglement compared to the result obtained for Bunch Davies vacuum ($\alpha=0$) for large mass regime ($\nu^2<0$). To demonstrate this clearly we have depicted the numerical values of the entanglement entropy for $\alpha=0$ (\textcolor{red}{\bf red}), $\alpha=0.03$ (\textcolor{blue}{\bf blue}), $\alpha=0.1$ (\textcolor{green}{\bf green}) and $\alpha=0.3$ (\textcolor{violet}{\bf violet}). Now from the fig.~(\ref{azgax5}) it is observed that in $\nu^2>0$ region $S_{intr}(\alpha)/S_{\nu=1/2}(\alpha)$ reaches its maximum value at $\alpha=0.1$ (\textcolor{green}{\bf green}) and $\alpha=0.3$ (\textcolor{violet}{\bf violet}) with $\nu=0$ (or $m_{ axion}=3H/2)$, as given by, $ \left(S_{intr}(0.1)/S_{\nu=1/2}(0.1)\right)_{\rm max, \nu=0}\sim 1.2$ and $ \left(S_{intr}(0.3)/S_{\nu=1/2}(0.3)\right)_{\rm max, \nu=0}\sim 2.1.$
                                          On the other hand, at $\alpha=0.03$ (\textcolor{blue}{\bf blue})  and $\alpha=0$ and $\alpha=0.3$ (\textcolor{red}{\bf red}) with $\nu=1/2$ (or $m_{ axion}=\sqrt{2}H)$ the maximum value of $S_{intr}(\alpha)/S_{\nu=1/2}(\alpha)$ is given by, $ \left(S_{intr}(0.03)/S_{\nu=1/2}(0.03)\right)_{\rm max, \nu=1/2}\sim  \left(S_{intr}(0)/S_{\nu=1/2}(0)\right)_{\rm max, \nu=1/2}\sim 1.$ 
                   Further if we consider the interval $3/2<\nu<5/2$ then 
                      $S_{intr}(\alpha)/S_{\nu=1/2}(\alpha)$ show one oscillation with different amplitude for all values of the parameter $\alpha$. After that it reaches its maximum value for $\alpha=0$ and $\alpha=0.03$, as given by,
                      $\left(S_{intr}(0.03)/S_{\nu=1/2}(0.03)\right)_{\rm max, 3/2<\nu<5/2}\sim \left(S_{intr}(0)/S_{\nu=1/2}(0)\right)_{\rm max, 3/2<\nu<5/2}\sim 1.$
                      On the other hand, $S_{intr}(\alpha)/S_{\nu=1/2}(\alpha)$ reaches its minimum value for $\alpha=0.1$ and $\alpha=0.3$, as given by,
                       $\left(S_{intr}(0.1)/S_{\nu=1/2}(0.1)\right)_{\rm min, 3/2<\nu<5/2}\sim 1$ and  $\left(S_{intr}(0.3)/S_{\nu=1/2}(0.3)\right)_{\rm min, 3/2<\nu<5/2}\sim 1.$
                      Similarly in the interval $5/2<\nu<7/2$ we can observe the same feature for the same values of $\alpha$ with larger period of oscillation. In fig.~(\ref{azgax6}), the significant role of axionic source term is explicitly shown. In both $\nu^2<0$ and $\nu^2>0$ regime the behaviour of $S_{intr}(\alpha)/S_{\nu=1/2}(\alpha)$ is exactly same as depicted in fig.~(\ref{azgax5}). But in presence of axionic source term the amount of $S_{intr}(\alpha)/S_{\nu=1/2}(\alpha)$ increase for $\alpha=0$, $\alpha=0.03$ and decrease for $\alpha=0.1$, $\alpha=0.3$ compared to fig.~(\ref{azgax5}). Also it is important to note that, the amplitude of the maximum and minimum of the oscillations change in presence of axionic source term.  
                      
                      On the other hand, for $\gamma_p=e^{\pi|m_{axion}/H|}\left(1+\tan\alpha\right)$ the entanglement entropy for axion in the large mass limiting range is given by the following expression:
                        \bea\label{c4} &&{\bf c_6}\left(\alpha,|\nu|\approx\frac{m_{axion}}{H}\right)= S_{\bf intr}\left(\alpha,|\nu|\approx\frac{m_{axion}}{H}\right)\nonumber\\
                                             &\approx& -\frac{1}{3\pi}\left(\frac{m_{axion}}{H}\right)^3\left[\ln\left(1-e^{\frac{2\pi m_{axion}}{H}}\left(1+\tan\alpha\right)^2\right)+\frac{\left(2\ln\left(1+\tan\alpha\right)+\frac{2\pi m_{axion}}{H}\right)~e^{\frac{2\pi m_{axion}}{H}}\left(1+\tan\alpha\right)^2}{\left(1-e^{\frac{2\pi m_{axion}}{H}}\left(1+\tan\alpha\right)^2\right)}\right],~~~~~~~~\eea
  Next, in fig.~(\ref{zpxxgg2}), we have depicted the behaviour of entanglement entropy $S_{\bf intr}(\alpha)$ with respect to the parameter $\alpha$ in absence ($f^{(\alpha)}_{p}=0$) and presence ($f^{(\alpha)}_{p}=10^{-7}$) of axionic source for the mass parameter $\nu^2<0$ and $\nu^2>0$  respectively. In fig.~(\ref{zgax5}) and fig.~(\ref{zgax6}) it is observed that a crossover takes place for $\nu^2=1/4,9/4,25/4$ (\textcolor{green}{\bf green}), $\nu^2=1/16,9/16,25/16$ (\textcolor{blue}{\bf blue}) and $\nu^2=0$ (\textcolor{red}{\bf red}) with small values of the parameter $\alpha$. We also observe that for $\nu^2=1/4,9/4,25/4$ (\textcolor{green}{\bf green}) entanglement entropy decreases with increasing value of the parameter $\alpha$. On the other hand, for $\nu^2=1/16,9/16,25/16$ (\textcolor{blue}{\bf blue}) and $\nu^2=0$ (\textcolor{red}{\bf red}) entanglement entropy increases with increasing value of the parameter $\alpha$. Additionally, we observe that, in presence of axionic source the entanglement entropy is significantly larger compared to the result obtained in absence of source contribution. In fig.~(\ref{zgax7}) and fig.~(\ref{zgax8}) it is observed that no crossover takes place for $\nu^2=-1/2$ (\textcolor{green}{\bf green}), $\nu^2=-1/4,-9/4,-25/4$ (\textcolor{blue}{\bf blue}) and $\nu^2=-1/16,-9/16,-25/16$ (\textcolor{red}{\bf red}) with all values of the parameter $\alpha$. Also it is important to note that, for all values of $\nu^2<0$ entanglement entropy increases with increasing value of the parameter $\alpha$.

 \end{itemize}

     \subsection{Computation of R$\acute{e}$nyi entropy using $\alpha$ vacua}
     \label{3d}
      In this context one can further use the density matrix to compute the R$\acute{e}$nyi entropy for $\alpha$ vacuum, which is defined as:
         \bea\label{ooaap1} S_{q}(p,\nu,\alpha)&=& \frac{1}{1-q}\ln~{\rm \bf Tr}~\left[\rho_{\bf L}(p,\alpha)\right]^q.~~~~~~~~~~{\rm with}~~q>0.\eea
The obtained solution for $\alpha$ vacuum with a given ${\bf SO(1,3)}$ principal quantum number $p$ can be written as:
            \bea S_{q}(p,\nu,\alpha)&=&\frac{1}{1-q}\left[q\ln\left(\frac{1-|\gamma^{(\alpha)}_{p}|^2}{1+f^{(\alpha)}_p}\right)-\ln\left(1-|\gamma^{(\alpha)}_{p}|^{2q}\right)\right]+\ln\left[1+\sum^{q}_{k=1}{}^{q}{\bf C}_{k} (f^{(\alpha)}_p)^{k
               }\frac{\left(1-|\gamma^{(\alpha)}_{p}|^2\right)^{-k}}{\left(1-|\gamma^{(\alpha)}_{p}|^{-2k}\right)}\right],~~~~~~~~\eea
               using which the interesting part of the R$\acute{e}$nyi entropy in de Sitter space for $\alpha$ vacuum can be written as:
                                    \bea  S_{q,\bf intr}(\alpha,\nu)= \frac{1}{\pi}\int^{\infty}_{p=0}~dp~p^2~S_{q}(p,\nu,\alpha).\eea

               Now to study the properties of the derived result we check the following physical limiting situations as given by:
               \begin{itemize}
               
               \item If we take the limit $q\rightarrow 1$ limit then from the R$\acute{e}$nyi entropy in $\alpha$ vacuum we get, 
               $\lim_{q\rightarrow 1}S_{q}(p,\nu,\alpha)
               \neq S(p,\nu,\alpha).$
            which shows that in presence of axionic source, the entanglement entropy and R$\acute{e}$nyi entropy are not same in the limit $q\rightarrow 1$. Now if we take further $f_p\rightarrow 0$ then entanglement entropy and R$\acute{e}$nyi entropy both are same.

               \item Further if we take the limit $q\rightarrow \infty$ limit then from the R$\acute{e}$nyi entropy in $\alpha$ vacuum we get:
                           \bea \label{df1xzz} \lim_{q\rightarrow \infty}S_{q}(p,\nu,\alpha)&=&-\ln[\rho_{\bf L}]_{\bf max}\approx\ln\left(\frac{1+f^{(\alpha)}_p}{1-|\gamma^{(\alpha)}_{p}|^2}\right).~~~~~~~~~~~~\eea
                           which directly implies the largest eigenvalue of density matrix. Now if we take further $f_p\rightarrow 0$ in Eqn~(\ref{df1xzz}) then entanglement entropy and R$\acute{e}$nyi entropy both are same. 
               \end{itemize}                                                                                      
 Further substituting the expression for entanglement entropy $S(p,\nu,\alpha)$ computed in presence of axion for $\alpha$ vacuum and integrating over all possible ${\bf SO(1,3)}$ principal quantum number, lying within the window $0<p<\infty$, we get:
                     \bea S_{q,\bf intr}(\alpha,\nu)= \left[{\cal M}_{1,q}+\ln\left(1+f^{(\alpha)}_p\right){\cal M}^{(\alpha)}_{2,q}+{\cal M}^{(\alpha)}_{3,q}\right],\eea
                     where the integrals ${\cal M}_{1,q}$, ${\cal M}^{(\alpha)}_{2,q}$ and ${\cal M}^{(\alpha)}_{3,q}$ can be written as:
                     \bea {\cal M}_{1,q}&=&\frac{1}{\pi}\int^{\infty}_{p=0}~dp~p^2~\left[\frac{q}{1-q}\ln\left(1-|\gamma^{(\alpha)}_{p}|^2\right)-\frac{1}{1-q}\ln\left(1-|\gamma^{(\alpha)}_{p}|^{2q}\right)\right],\\
                     {\cal M}^{(\alpha)}_{2,q}&=&-\frac{1}{\pi}\frac{q}{1-q}\int^{\infty}_{p=0}~dp~p^2,\\
                                          {\cal M}^{(\alpha)}_{3,q}&=&\frac{1}{\pi}\frac{1}{1-q}\int^{\infty}_{p=0}~dp~p^2~\ln\left[1+\sum^{q}_{k=1}{}^{q}{\bf C}_{k} (f^{(\alpha)}_p)^{k
                                                }\frac{\left(1-|\gamma^{(\alpha)}_{p}|^2\right)^{-k}}{\left(1-|\gamma^{(\alpha)}_{p}|^{-2k}\right)}\right].\eea
                     Here it is important to note that:
                     \begin{itemize}
                     \item Here the integral ${\cal M}^{(\alpha)}_{2,q}$ diverges. Further introducing a change in variable to $x=2\pi p$ along with a cut-off $\Lambda_{\bf C}$  the regularised version of this integral can be written as:
                     \bea  {\cal M}^{(\alpha)}_{2,q}&=&-\frac{1}{8\pi^4}\frac{q}{1-q}\int^{\Lambda_{\bf C}}_{x=0}~dx~x^2=-\frac{\Lambda^3_{\bf C}}{24\pi^4}\frac{q}{1-q}.\eea

                     \item On the other hand for arbitrary $\nu$ and $\alpha$  we get:
                     \bea {\cal M}^{(\alpha)}_{1,q}&=&\frac{1}{8\pi^4}\int^{\Lambda_{\bf C}}_{x=0}~dx~x^2~\left[\frac{q}{1-q}\ln\left(1-2G_{\pm}(x,\nu,\alpha)\right)-\frac{1}{1-q}\ln\left(1-(2G_{\pm}(x,\nu,\alpha)\right)^q)\right],~~~~~~~~~~~~\\ {\cal M}^{(\alpha)}_{3,q}&=&\frac{1}{8\pi^4}\frac{1}{1-q}\int^{\Lambda_{\bf C}}_{x=0}~dx~x^2~\ln\left[1+\sum^{q}_{k=1}{}^{q}{\bf C}_{k} (f^{(\alpha)}_p)^{k
                }\frac{\left(1-2G_{\pm}(x,\nu,\alpha)\right)^{-k}}{\left(1-(2G_{\pm}(x,\nu,\alpha))^{-k}\right)}\right],~~~~~~~~~~\eea
                    
                     where $G_{\pm}(x,\nu,\alpha)$ is defined in Eqn~(\ref{ccx}). We consider large axion mass ($\nu^2<0$ where $\nu\rightarrow -i|\nu|$) limiting situation which is important to study the physics from this case.  In this large axion mass limiting situation we consider the window of ${\bf SO(1,3)}$ principal quantum number is $0<p<|\nu|$.
                     
                      As a result, the regularized integral ${\cal M}^{(\alpha)}_{1,q}$ and ${\cal M}^{(\alpha)}_{3,q}$ for the first solution for $|\gamma^{(\alpha)}_p|$ in presence of $\alpha$ vacuum can be expressed as:
                                          \bea \label{rt1z}
                                 \displaystyle {\cal M}^{(\alpha)}_{1,q} &=&\displaystyle\footnotesize\left\{\begin{array}{ll}
                                \displaystyle \frac{{\cal A}(\nu)}{8\pi^4}\left[\frac{q}{1-q}\ln\left(1-e^{-2\pi\nu}\left(1+\tan\alpha\right)^2\right)\-\frac{1}{1-q}\ln\left(1-e^{-2\pi\nu q}\left(1+\tan\alpha\right)^{2q}\right)\right]~~~~~~ &
                                                                                          \mbox{\small {\textcolor{red}{\bf for $0<x<2\pi|\nu|$}}}  
                                                                                         \\ 
                                         \displaystyle \frac{{\cal D}(\nu,\alpha,\Lambda_{\bf C},q)}{8\pi^4} & \mbox{\small { \textcolor{red}{\bf for $2\pi|\nu|<x<\Lambda_{\bf C}$}}}.~~~~~~~~
                                                                                                   \end{array}
                                                                                         \right.\\ \label{ss}
                                                                                                                    \displaystyle {\cal M}^{(\alpha)}_{3,q} &=&\displaystyle\footnotesize\left\{\begin{array}{ll}
                                                                                                                   \displaystyle \frac{{\cal A}(\nu)}{8\pi^4}\frac{1}{1-q}~\ln\left[1+\sum^{q}_{k=1}{}^{q}{\bf C}_{k} (f^{(\alpha)}_p)^{k                                                                                                             }\frac{\left(1-e^{-2\pi\nu}\left(1+\tan\alpha\right)^2\right)^{-k}}{\left(1-e^{2\pi \nu k}\left(1+\tan\alpha\right)^{-2k}\right)}\right] &
                                                                                                                                                                             \mbox{\small {\textcolor{red}{\bf for $0<x<2\pi|\nu|$}}}  
                                                                                                                                                                            \\ 
                                                                                                                            \displaystyle \frac{1}{8\pi^4}\frac{1}{1-q}\int^{\Lambda_{\bf C}}_{x=2\pi\nu}dx~x^2\ln\left[1+\sum^{q}_{k=1}{}^{q}{\bf C}_{k} (f^{(\alpha)}_p)^{k                                                                                                             }\frac{\left(1-\frac{e^{-x}\left(1+\tan\alpha\right)^2\left(1+\tan\alpha~e^{2\pi\nu}\right)^2}{\left(1+\tan^2\alpha~e^{-x}\right)^2}\right)^{-k}}{\left(1-\frac{e^{xk}\left(1+\tan\alpha\right)^{-2k}\left(1+\tan\alpha~e^{2\pi\nu}\right)^{-2k}}{\left(1+\tan^2\alpha~e^{-x}\right)^{-2k}}\right)}\right] & \mbox{\small { \textcolor{red}{\bf for $2\pi|\nu|<x<\Lambda_{\bf C}$}}}.~~~
                                                                                                                                                                                      \end{array}
                                                                                                                                                                            \right.\eea
                              and for the second solution for $|\gamma^{(\alpha)}_p|$ in presence of $\alpha$ vacuum we get:
                             \bea \label{zs}
                                   \displaystyle {\cal M}^{(\alpha)}_{1,q}  &=&\displaystyle\left\{\begin{array}{ll}
                                  \displaystyle \frac{{\cal A}(\nu)}{8\pi^4}\left[\frac{q}{1-q}\ln\left(1-e^{2\pi\nu}\left(1+\tan\alpha\right)^2\right)-\frac{1}{1-q}\ln\left(1-e^{2\pi\nu q}\left(1+\tan\alpha\right)^{2q}\right)\right]~~~ &
                                                                                            \mbox{\small {\textcolor{red}{\bf for $0<x<2\pi|\nu|$}}}  
                                                                                           \\ 
                                           \displaystyle \frac{{\cal W}(\nu,\alpha,\Lambda_{\bf C},q)}{8\pi^4} & \mbox{\small { \textcolor{red}{\bf for $2\pi|\nu|<x<\Lambda_{\bf C}$}}}.~~~
                                                                                                     \end{array}
                                                                                           \right.\\ \label{s1}
                                                                                                                      \displaystyle {\cal M}^{(\alpha)}_{3,q} &=&\displaystyle\footnotesize\left\{\begin{array}{ll}
                                                                                                                                                                                                                                         \displaystyle \frac{{\cal A}(\nu)}{8\pi^4}\frac{1}{1-q}~\ln\left[1+\sum^{q}_{k=1}{}^{q}{\bf C}_{k} (f^{(\alpha)}_p)^{k                                                                                                             }\frac{\left(1-e^{2\pi\nu}\left(1+\tan\alpha\right)^2\right)^{-k}}{\left(1-e^{-2\pi \nu k}\left(1+\tan\alpha\right)^{-2k}\right)}\right] &
                                                                                                                                                                                                                                                                                                   \mbox{\small {\textcolor{red}{\bf for $0<x<2\pi|\nu|$}}}  
                                                                                                                                                                                                                                                                                                  \\ 
                                                                                                                                                                                                                                                  \displaystyle \frac{1}{8\pi^4}\frac{1}{1-q}\int^{\Lambda_{\bf C}}_{x=2\pi\nu}~dx~x^2~\ln\left[1+\sum^{q}_{k=1}{}^{q}{\bf C}_{k} (f^{(\alpha)}_p)^{k                                                                                                             }\frac{\left(1-\frac{e^{x}\left(1+\tan\alpha\right)^2\left(1+\tan\alpha~e^{2\pi\nu}\right)^2}{\left(1+\tan^2\alpha~e^{-x}\right)^2}\right)^{-k}}{\left(1-\frac{e^{-xk}\left(1+\tan\alpha\right)^{-2k}\left(1+\tan\alpha~e^{2\pi\nu}\right)^{-2k}}{\left(1+\tan^2\alpha~e^{-x}\right)^{-2k}}\right)}\right] & \mbox{\small { \textcolor{red}{\bf for $2\pi|\nu|<x<\Lambda_{\bf C}$}}}.~~~
                                                                                                                                                                                                                                                                                                            \end{array}
                                                                                                                                                                                                                                                                                                  \right.\eea                                            
     Here the coefficient function ${\cal A}(\nu)$ is defined in Eq~(\ref{p2}) and other $\alpha$ parameter dependent functions ${\cal D}(\nu,\alpha,\Lambda_{\bf C})$ and ${\cal W}(\nu,\alpha,\Lambda_{\bf C})$ are defined as:
                                \bea 
                                \label{f1}{\cal D}(\nu,\alpha,\Lambda_{\bf C},q)&=&\int^{\Lambda_{\bf C}}_{x=2\pi\nu}~dx~x^2~\left[\frac{q}{1-q}\ln\left(1-\frac{e^{-x}\left(1+\tan\alpha\right)^2\left(1+\tan\alpha~e^{2\pi\nu}\right)^2}{\left(1+\tan^2\alpha~e^{-x}\right)^2}\right)\right.\nonumber\\
                                &&\left.~~~~~~~~~-\frac{1}{1-q}\ln\left(1-\frac{e^{-xq}\left(1+\tan\alpha\right)^{2q}\left(1+\tan\alpha~e^{2\pi\nu}\right)^{2q}}{\left(1+\tan^2\alpha~e^{-x}\right)^{2q}}\right)\right]                                                                   ,~~~~~~~~~~~~\\
                          \label{f2} {\cal W}(\nu,\alpha,\Lambda_{\bf C},q)&=&\int^{\Lambda_{\bf C}}_{x=2\pi\nu}~dx~x^2~\left[\frac{q}{1-q}\ln\left(1-\frac{e^{x}\left(1+\tan\alpha\right)^2\left(1+\tan\alpha~e^{2\pi\nu}\right)^2}{\left(1+\tan^2\alpha~e^{-x}\right)^2}\right)\right.\nonumber\\
                                                           &&\left.~~~~~~~~~-\frac{1}{1-q}\ln\left(1-\frac{e^{xq}\left(1+\tan\alpha\right)^{2q}\left(1+\tan\alpha~e^{2\pi\nu}\right)^{2q}}{\left(1+\tan^2\alpha~e^{-x}\right)^{2q}}\right)\right].                                                   \eea
    \begin{figure*}[htb]
    	\centering
    	\subfigure[Normalized R$\acute{e}$nyi entropy vs $\nu^2$ in $3+1$ D de Sitter space without axionic source ($f^{(\alpha)}_p=0$).]{
    		\includegraphics[width=7.2cm,height=5cm] {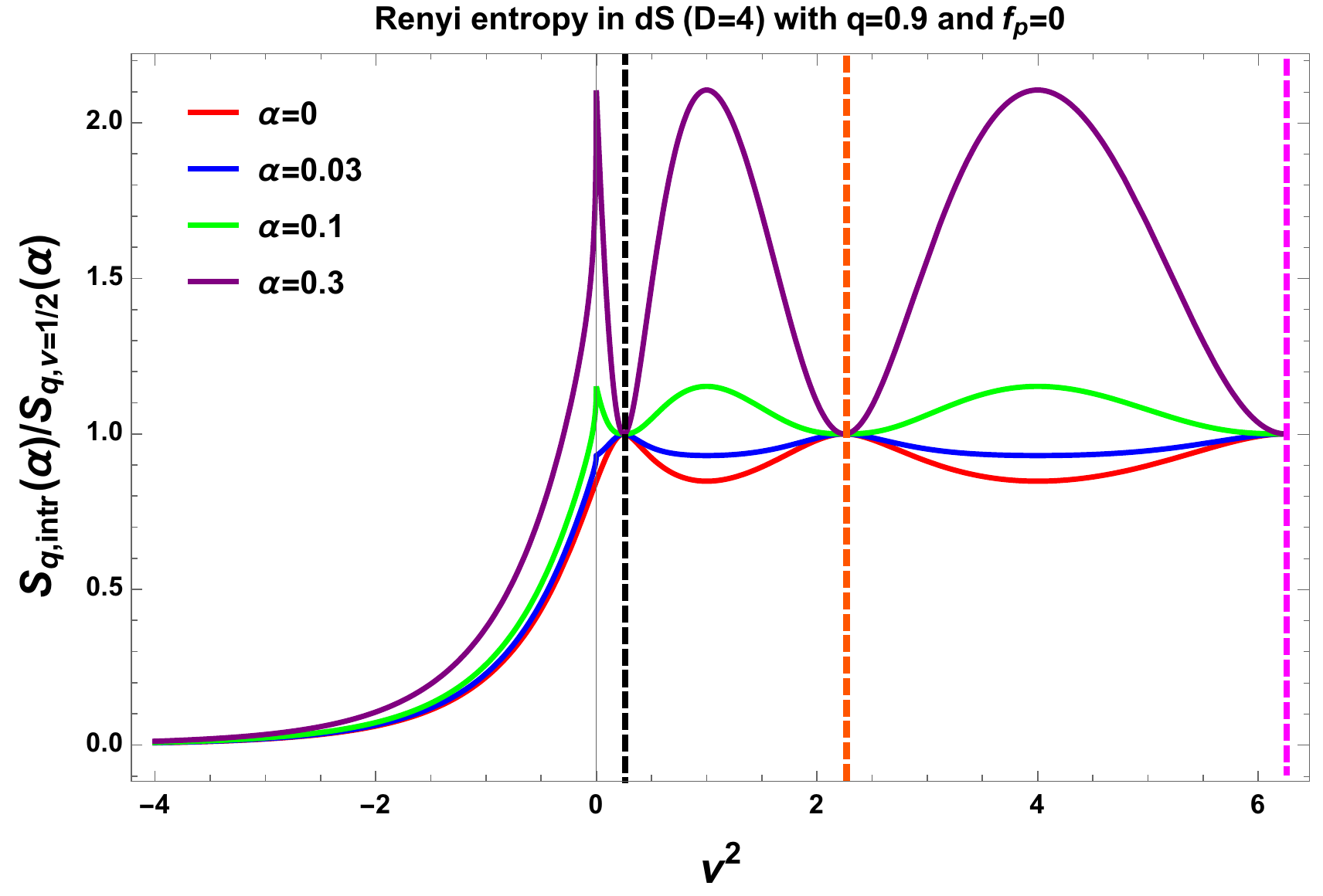}
    		\label{ix5c}}
    	\subfigure[Normalized R$\acute{e}$nyi entropy vs $\nu^2$ in $3+1$ D de Sitter space without axionic source ($f^{(\alpha)}_p=10^{-7}$).]{
    		\includegraphics[width=7.2cm,height=5cm] {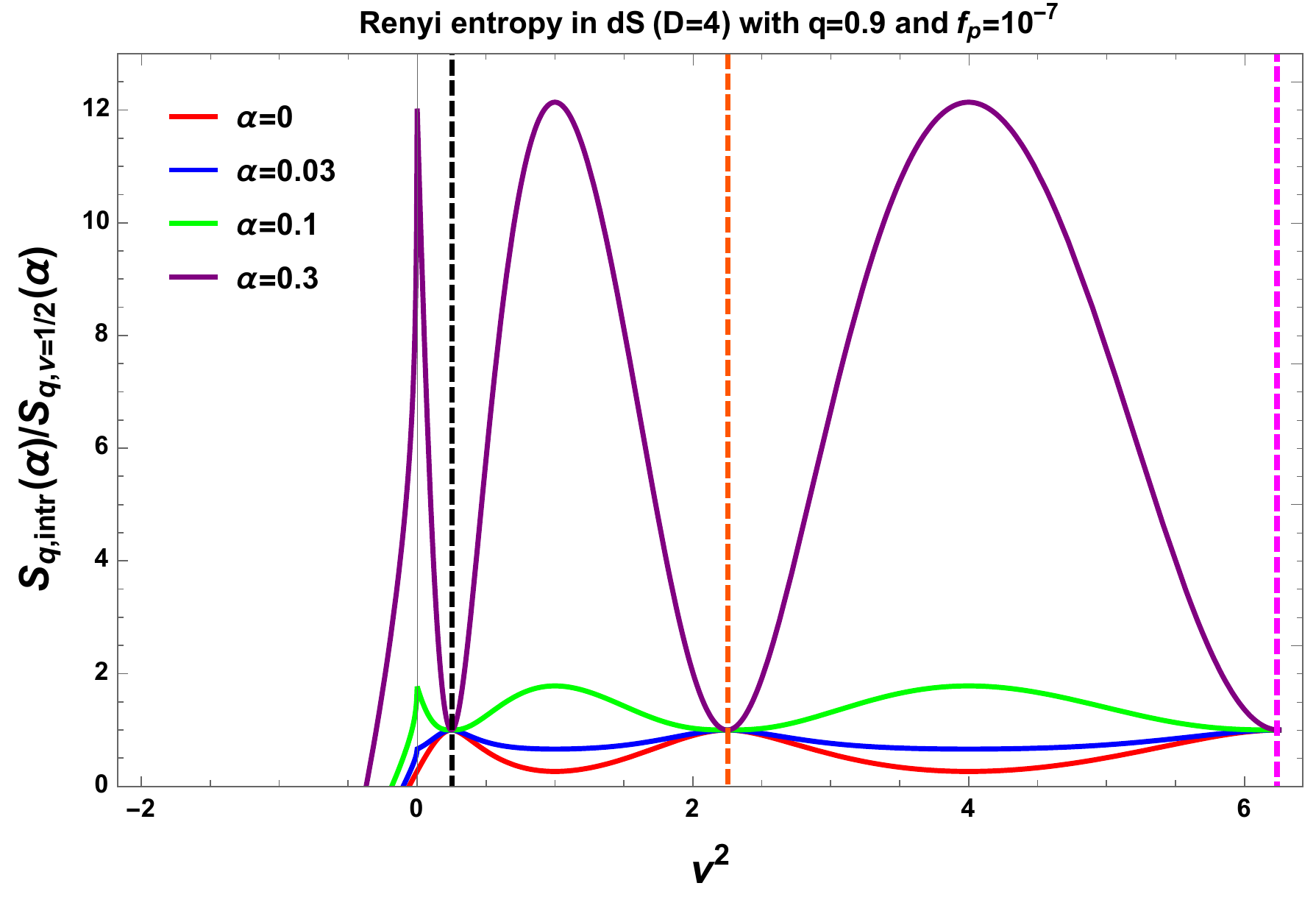}
    		\label{ix5a}}                            
    	\caption[Optional caption for list of figures]{Normalized R$\acute{e}$nyi entropy $S_{q,intr}(\alpha)/S_{q,\nu=1/2}(\alpha)$ vs mass parameter $\nu^2$ in $3+1$ D de Sitter space in absence of axionic source ($f^{(\alpha)}_p=0$) and in presence of axionic source ($f^{(\alpha)}_p=10^{-7}$) for $q=0.9$ and $\alpha=0$ (\textcolor{red}{\bf red}),$\alpha=0.03$ (\textcolor{blue}{\bf blue}),$\alpha=0.1$ (\textcolor{green}{\bf green}),$\alpha=0.3$ (\textcolor{violet}{\bf violet}) with $`+'$ branch of solution of $|\gamma^{(\alpha)}_p|$ and $|\Gamma^{(\alpha)}_{p,n}|$. Here we set the cut-off $\Lambda_{\bf C}=300$ for numerical computation.} 
    	\label{zbc1i}
    \end{figure*}   
    
    \begin{figure*}[htb]
    	\centering
    	\subfigure[Normalized R$\acute{e}$nyi entropy vs $\nu^2$ in $3+1$ D de Sitter space without axionic source ($f^{(\alpha)}_p=0$).]{
    		\includegraphics[width=7.2cm,height=5cm] {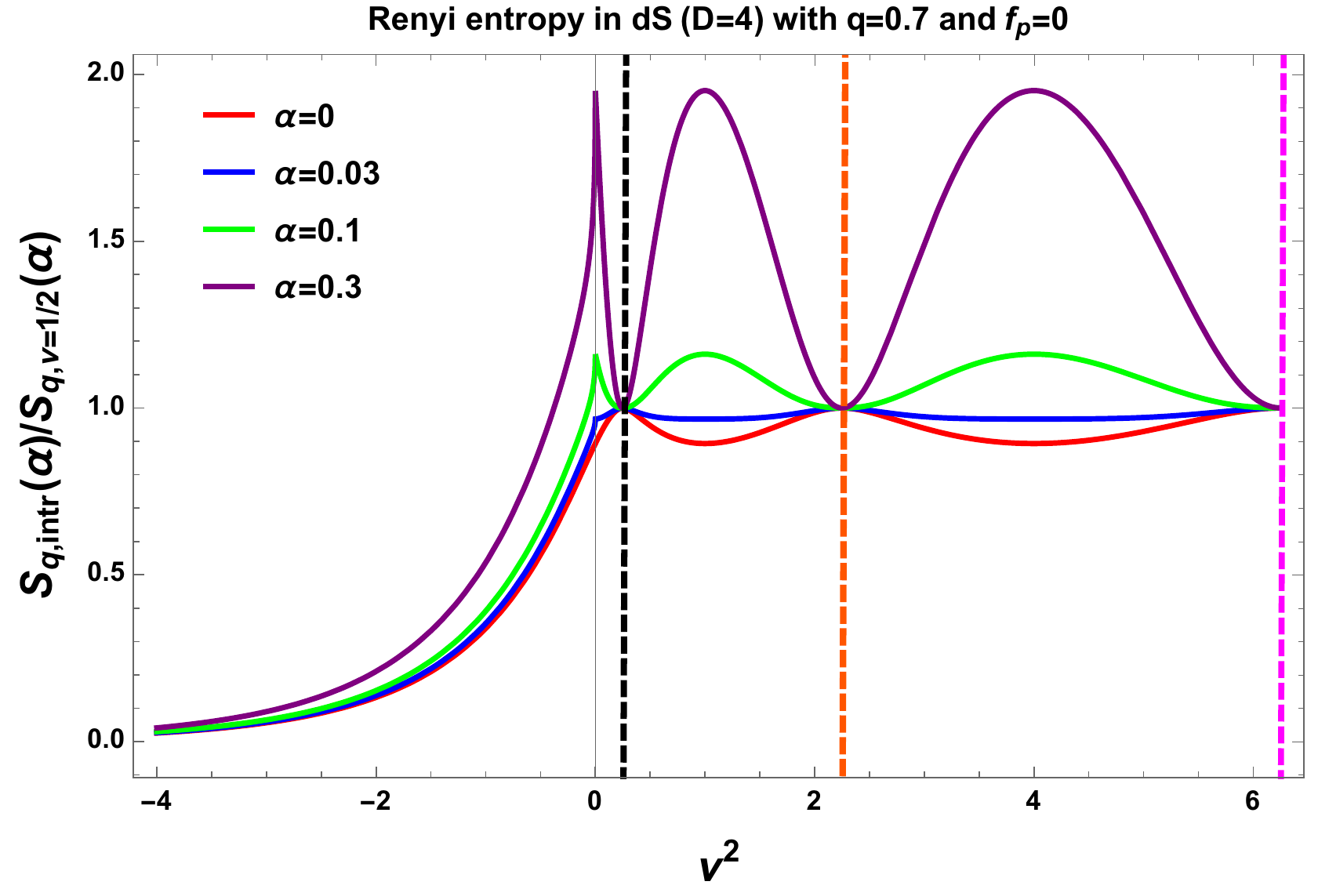}
    		\label{i5xc}}
    	\subfigure[Normalized R$\acute{e}$nyi entropy vs $\nu^2$ in $3+1$ D de Sitter space without axionic source ($f^{(\alpha)}_p=10^{-7}$).]{
    		\includegraphics[width=7.2cm,height=5cm] {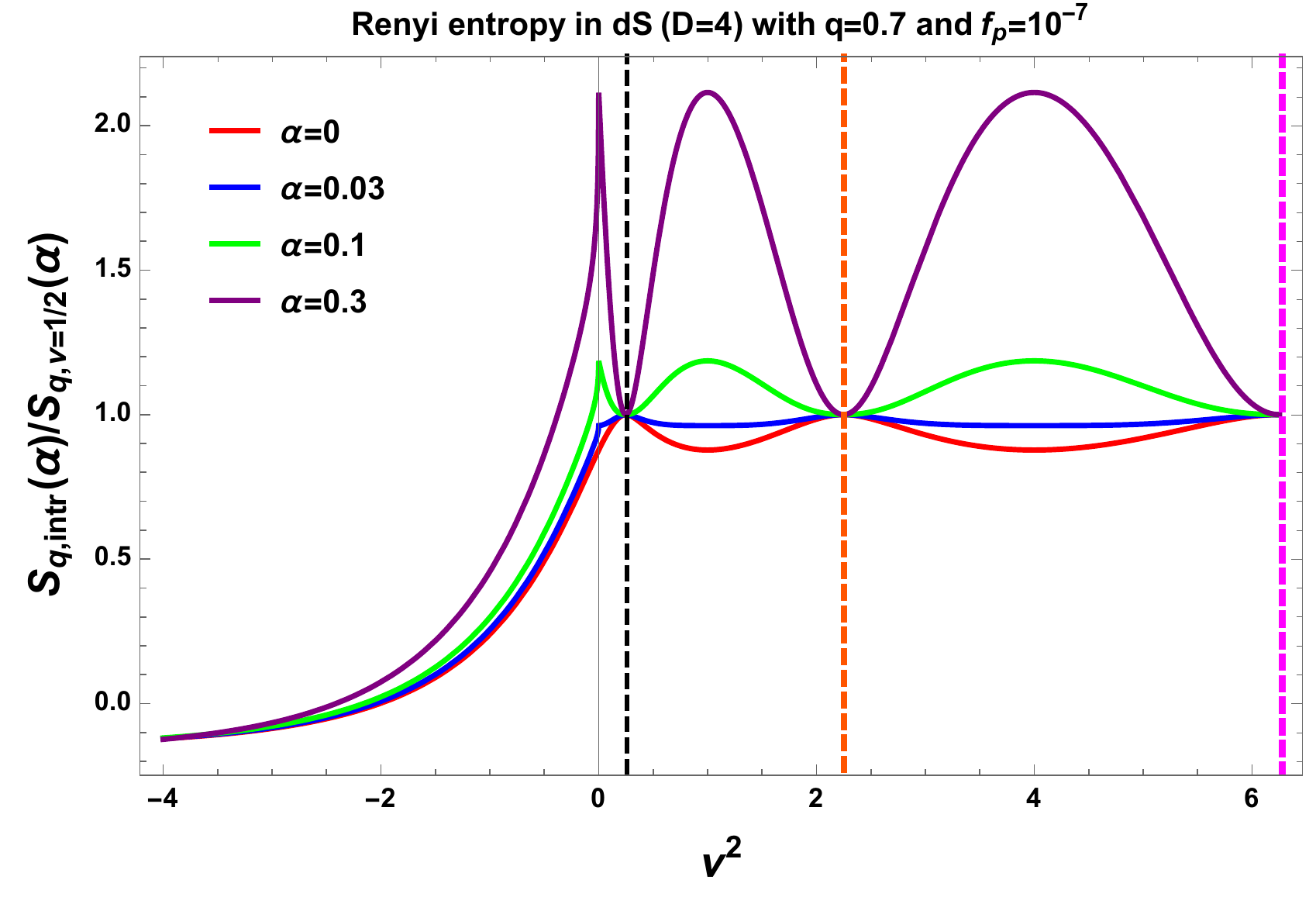}
    		\label{i5xa}}                            
    	\caption[Optional caption for list of figures]{Normalised R$\acute{e}$nyi entropy $S_{q,intr}(\alpha)/S_{q,\nu=1/2}(\alpha)$ vs mass parameter $\nu^2$ in $3+1$ D de Sitter space in absence of axionic source ($f^{(\alpha)}_p=0$) and in presence of axionic source ($f^{(\alpha)}_p=10^{-7}$) for $q=0.7$ and $\alpha=0$ (\textcolor{red}{\bf red}),$\alpha=0.03$ (\textcolor{blue}{\bf blue}),$\alpha=0.1$ (\textcolor{green}{\bf green}),$\alpha=0.3$ (\textcolor{violet}{\bf violet}) with $`+'$ branch of solution of $|\gamma^{(\alpha)}_p|$ and $|\Gamma^{(\alpha)}_{p,n}|$. Here we set the cut-off $\Lambda_{\bf C}=300$ for numerical computation.} 
    	\label{zbc2i}
    \end{figure*}   
    \begin{figure*}[htb]
    	\centering
    	\subfigure[Normalized R$\acute{e}$nyi entropy vs $\nu^2$ in $3+1$ D de Sitter space without axionic source ($f^{(\alpha)}_p=0$).]{
    		\includegraphics[width=7.2cm,height=5cm] {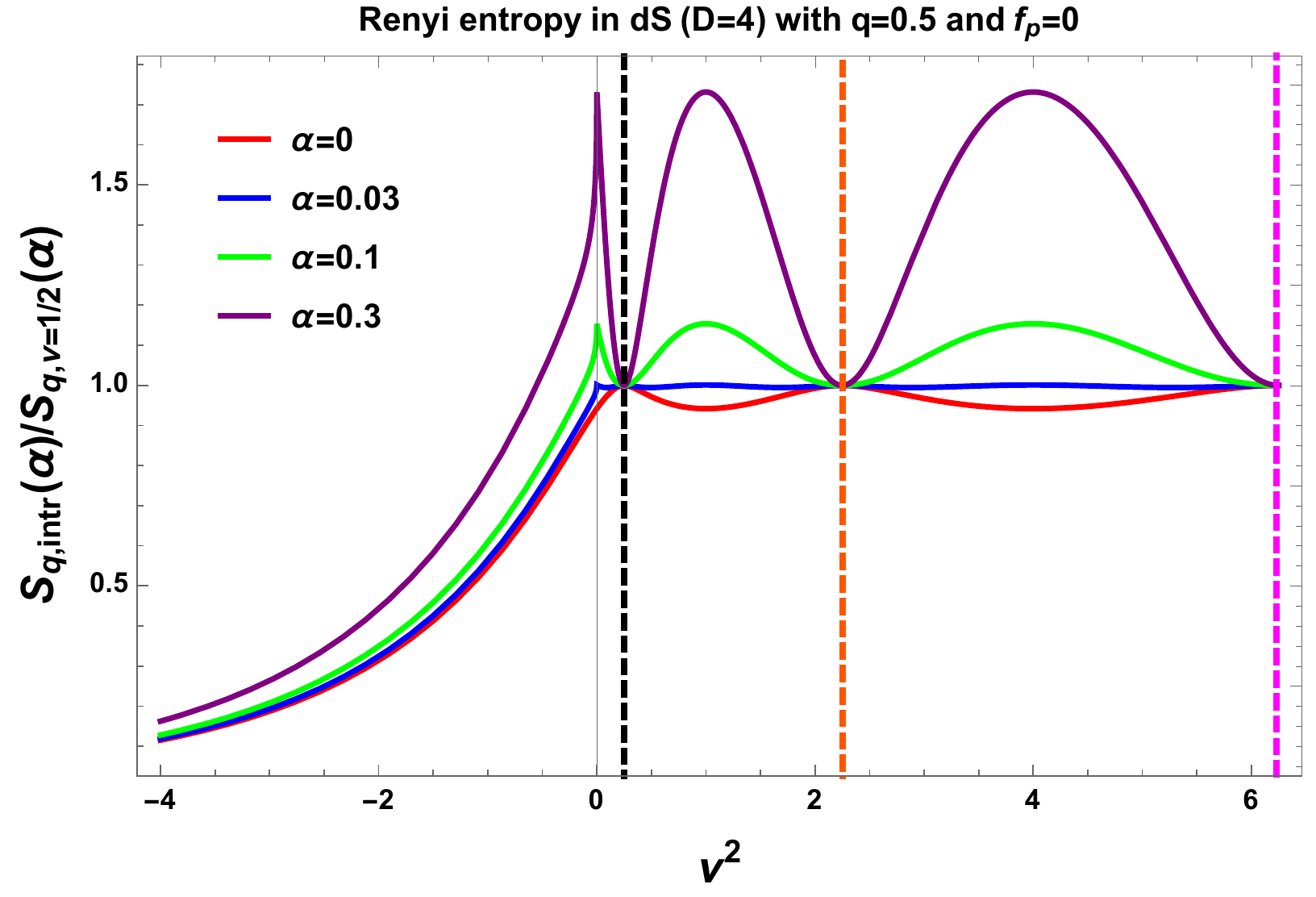}
    		\label{is5xc}}
    	\subfigure[Normalized R$\acute{e}$nyi entropy vs $\nu^2$ in $3+1$ D de Sitter space without axionic source ($f^{(\alpha)}_p=10^{-7}$).]{
    		\includegraphics[width=7.2cm,height=5cm] {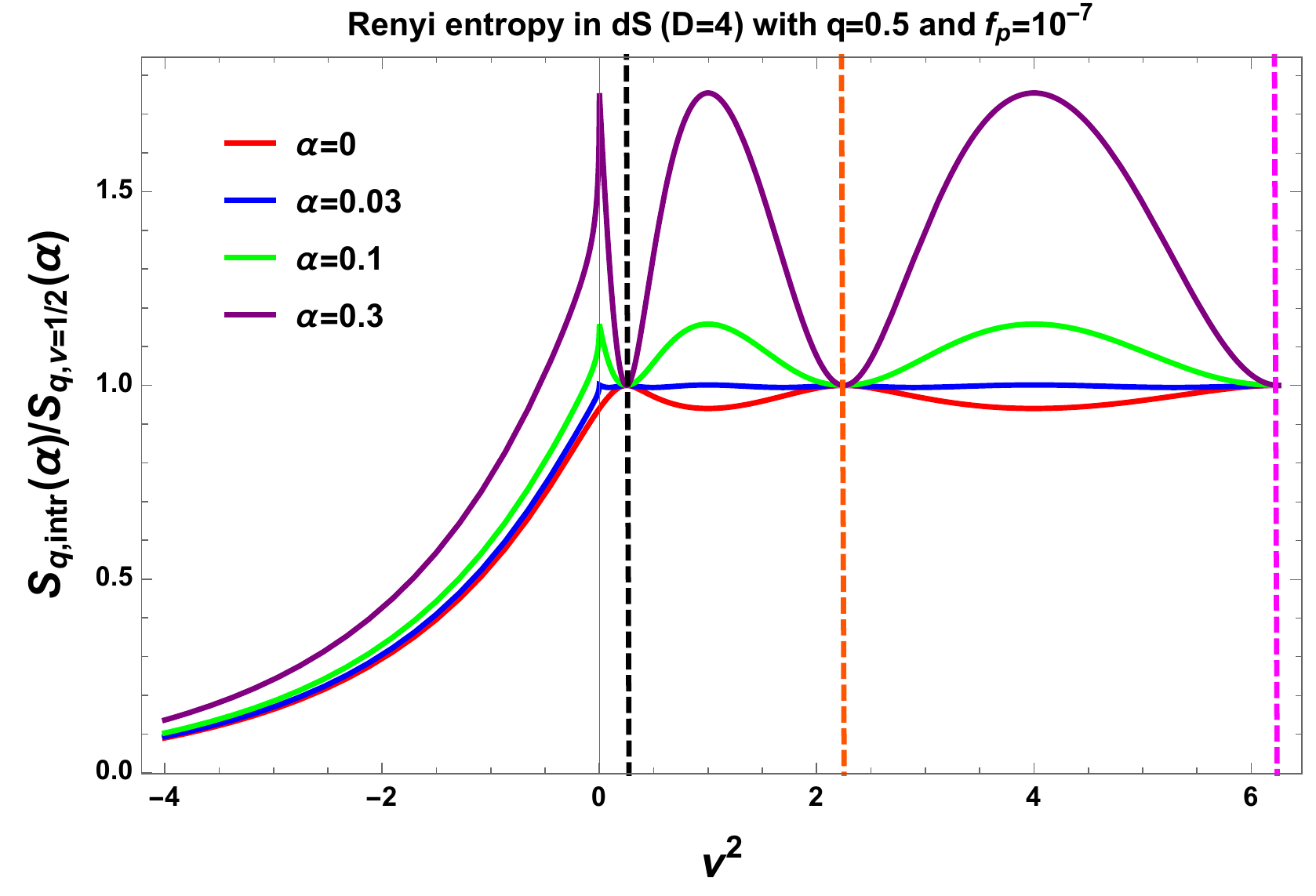}
    		\label{is5xa}}                            
    	\caption[Optional caption for list of figures]{Normalised R$\acute{e}$nyi entropy $S_{q,intr}(\alpha)/S_{q,\nu=1/2}(\alpha)$ vs mass parameter $\nu^2$ in $3+1$ D de Sitter space in absence of axionic source ($f^{(\alpha)}_p=0$) and in presence of axionic source ($f^{(\alpha)}_p=10^{-7}$) for $q=0.5$ and $\alpha=0$ (\textcolor{red}{\bf red}),$\alpha=0.03$ (\textcolor{blue}{\bf blue}),$\alpha=0.1$ (\textcolor{green}{\bf green}),$\alpha=0.3$ (\textcolor{violet}{\bf violet}) with $`+'$ branch of solution of $|\gamma^{(\alpha)}_p|$ and $|\Gamma^{(\alpha)}_{p,n}|$. Here we set the cut-off $\Lambda_{\bf C}=300$ for numerical computation.} 
    	\label{bcf1i}
    \end{figure*}   
    \begin{figure*}[htb]
    	\centering
    	\subfigure[Normalized R$\acute{e}$nyi entropy vs $\nu^2$ in $3+1$ D de Sitter space without axionic source ($f^{(\alpha)}_p=0$).]{
    		\includegraphics[width=7.2cm,height=5cm] {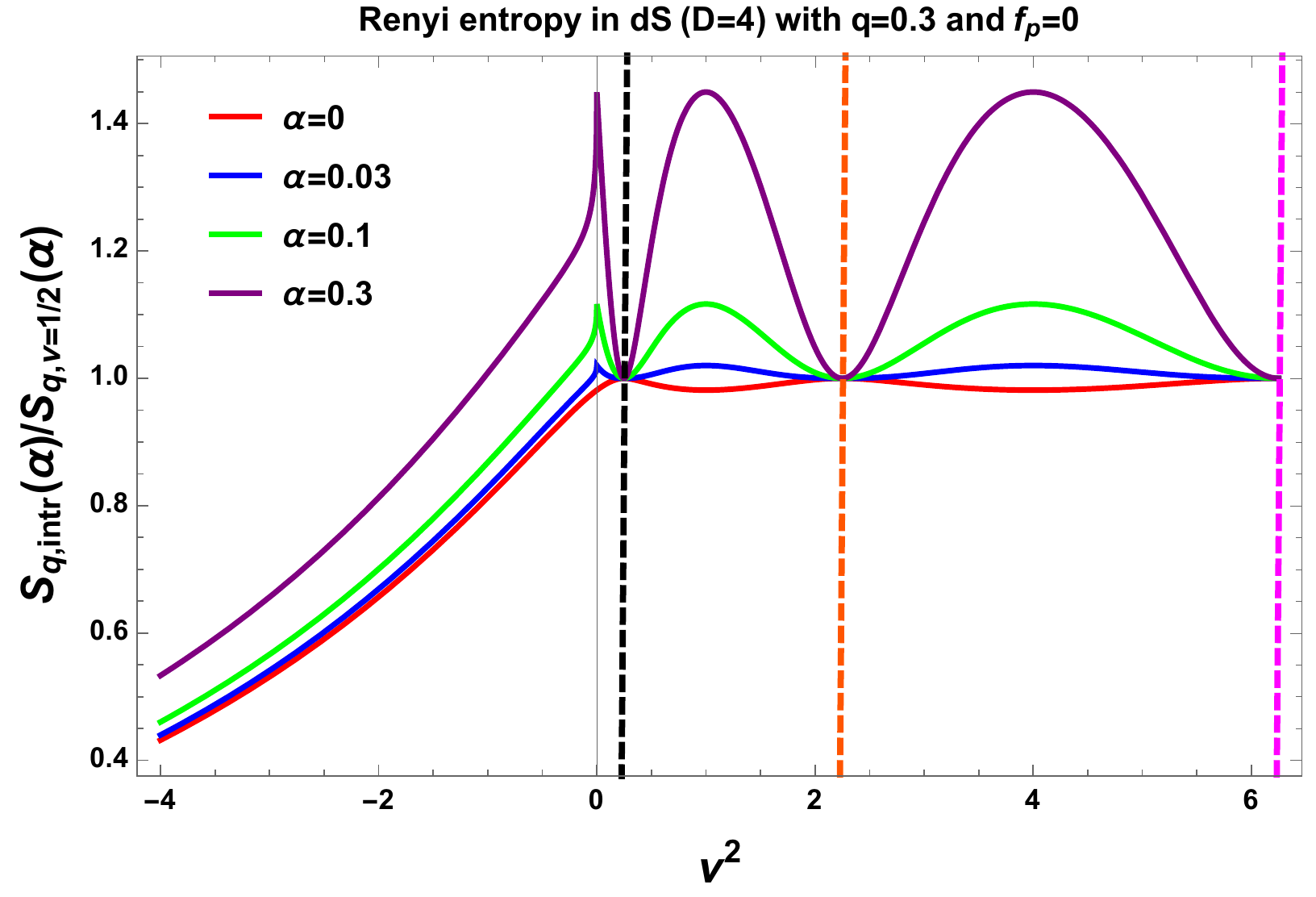}
    		\label{i5xcv}}
    	\subfigure[Normalized R$\acute{e}$nyi entropy vs $\nu^2$ in $3+1$ D de Sitter space without axionic source ($f^{(\alpha)}_p=10^{-7}$).]{
    		\includegraphics[width=7.2cm,height=5cm] {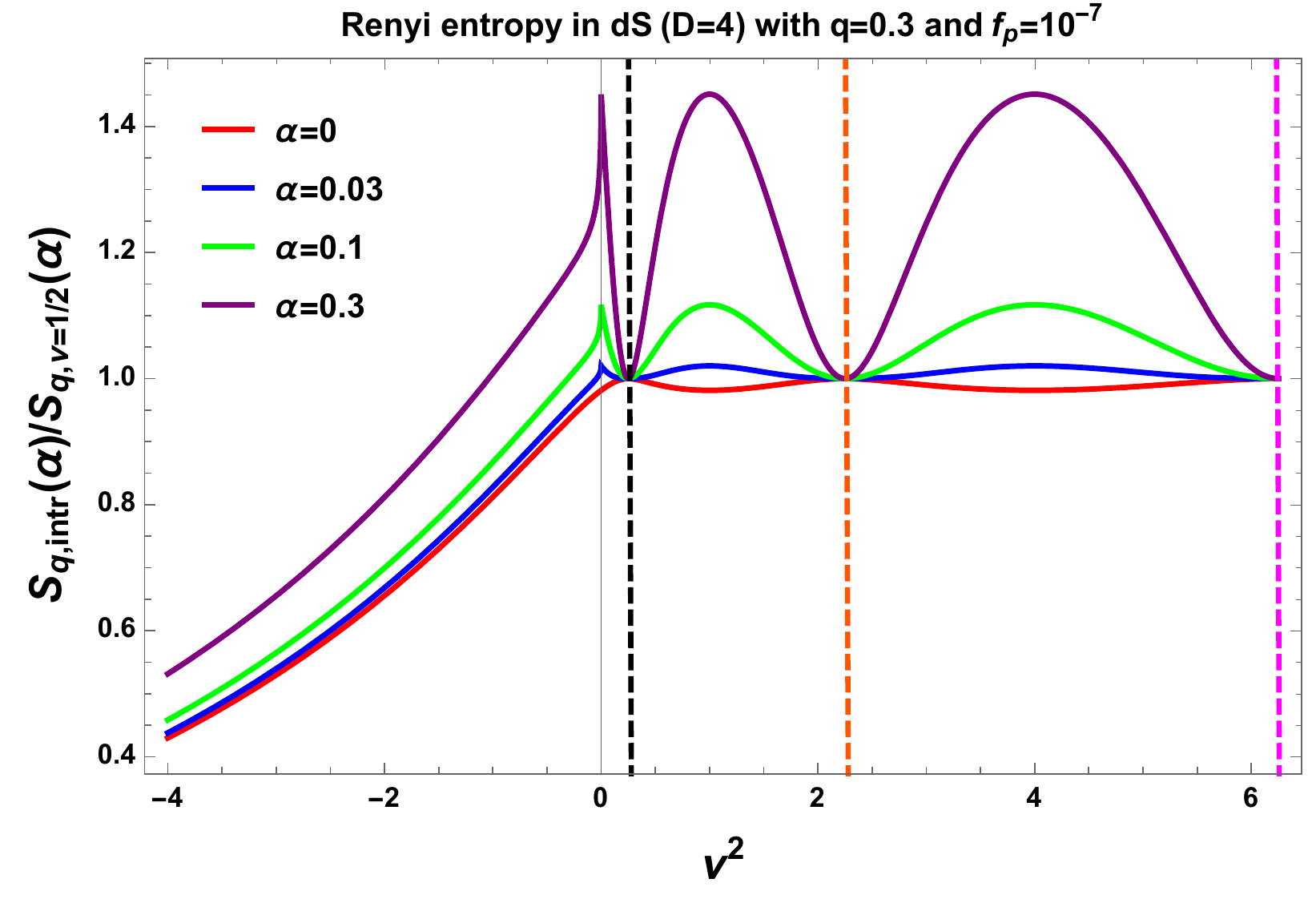}
    		\label{i5xav}}                            
    	\caption[Optional caption for list of figures]{Normalised R$\acute{e}$nyi entropy $S_{q,intr}(\alpha)/S_{q,\nu=1/2}(\alpha)$ vs mass parameter $\nu^2$ in $3+1$ D de Sitter space in absence of axionic source ($f^{(\alpha)}_p=0$) and in presence of axionic source ($f^{(\alpha)}_p=10^{-7}$) for $q=0.3$ and $\alpha=0$ (\textcolor{red}{\bf red}),$\alpha=0.03$ (\textcolor{blue}{\bf blue}),$\alpha=0.1$ (\textcolor{green}{\bf green}),$\alpha=0.3$ (\textcolor{violet}{\bf violet}) with $`+'$ branch of solution of $|\gamma^{(\alpha)}_p|$ and $|\Gamma^{(\alpha)}_{p,n}|$. Here we set the cut-off $\Lambda_{\bf C}=300$ for numerical computation.} 
    	\label{zbc2iv}
    \end{figure*}   
    \begin{figure*}[htb]
    	\centering
    	\subfigure[Normalized R$\acute{e}$nyi entropy vs $\nu^2$ in $3+1$ D de Sitter space without axionic source ($f^{(\alpha)}_p=0$).]{
    		\includegraphics[width=7.2cm,height=5cm] {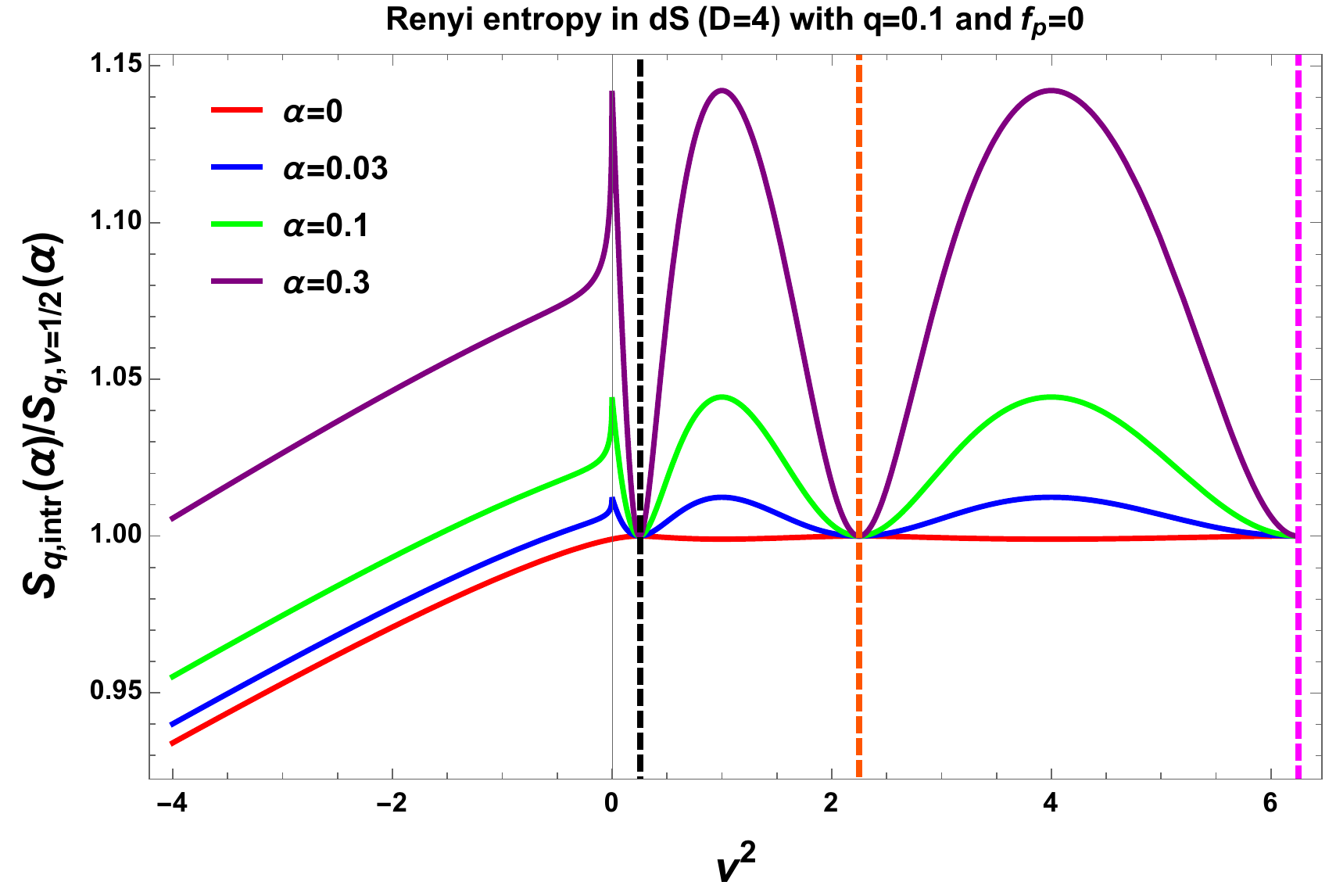}
    		\label{ci5xcv}}
    	\subfigure[Normalized R$\acute{e}$nyi entropy vs $\nu^2$ in $3+1$ D de Sitter space without axionic source ($f^{(\alpha)}_p=10^{-7}$).]{
    		\includegraphics[width=7.2cm,height=5cm] {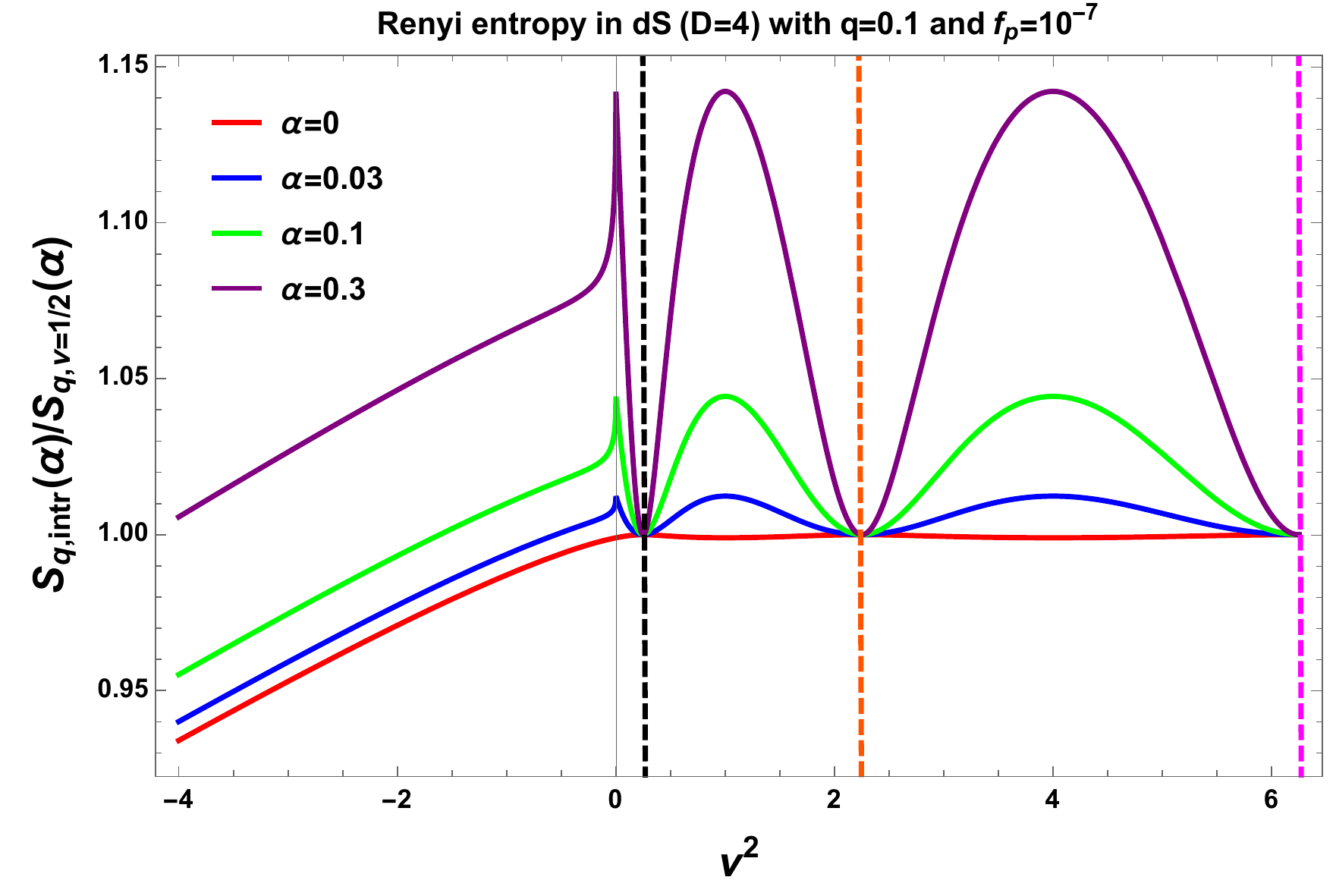}
    		\label{ci5xav}}                            
    	\caption[Optional caption for list of figures]{Normalised R$\acute{e}$nyi entropy $S_{q,intr}(\alpha)/S_{q,\nu=1/2}(\alpha)$ vs mass parameter $\nu^2$ in $3+1$ D de Sitter space in absence of axionic source ($f^{(\alpha)}_p=0$) and in presence of axionic source ($f^{(\alpha)}_p=10^{-7}$) for $q=0.1$ and $\alpha=0$ (\textcolor{red}{\bf red}),$\alpha=0.03$ (\textcolor{blue}{\bf blue}),$\alpha=0.1$ (\textcolor{green}{\bf green}),$\alpha=0.3$ (\textcolor{violet}{\bf violet}) with $`+'$ branch of solution of $|\gamma^{(\alpha)}_p|$ and $|\Gamma^{(\alpha)}_{p,n}|$.Here we set the cut-off $\Lambda_{\bf C}=300$ for numerical computation.} 
    	\label{zcbc2iv}
    \end{figure*}   
    \begin{figure*}[htb]
    	\centering
    	\subfigure[Normalized R$\acute{e}$nyi entropy vs $\nu^2$ in $3+1$ D de Sitter space without axionic source ($f_p=0$).]{
    		\includegraphics[width=7.2cm,height=5cm] {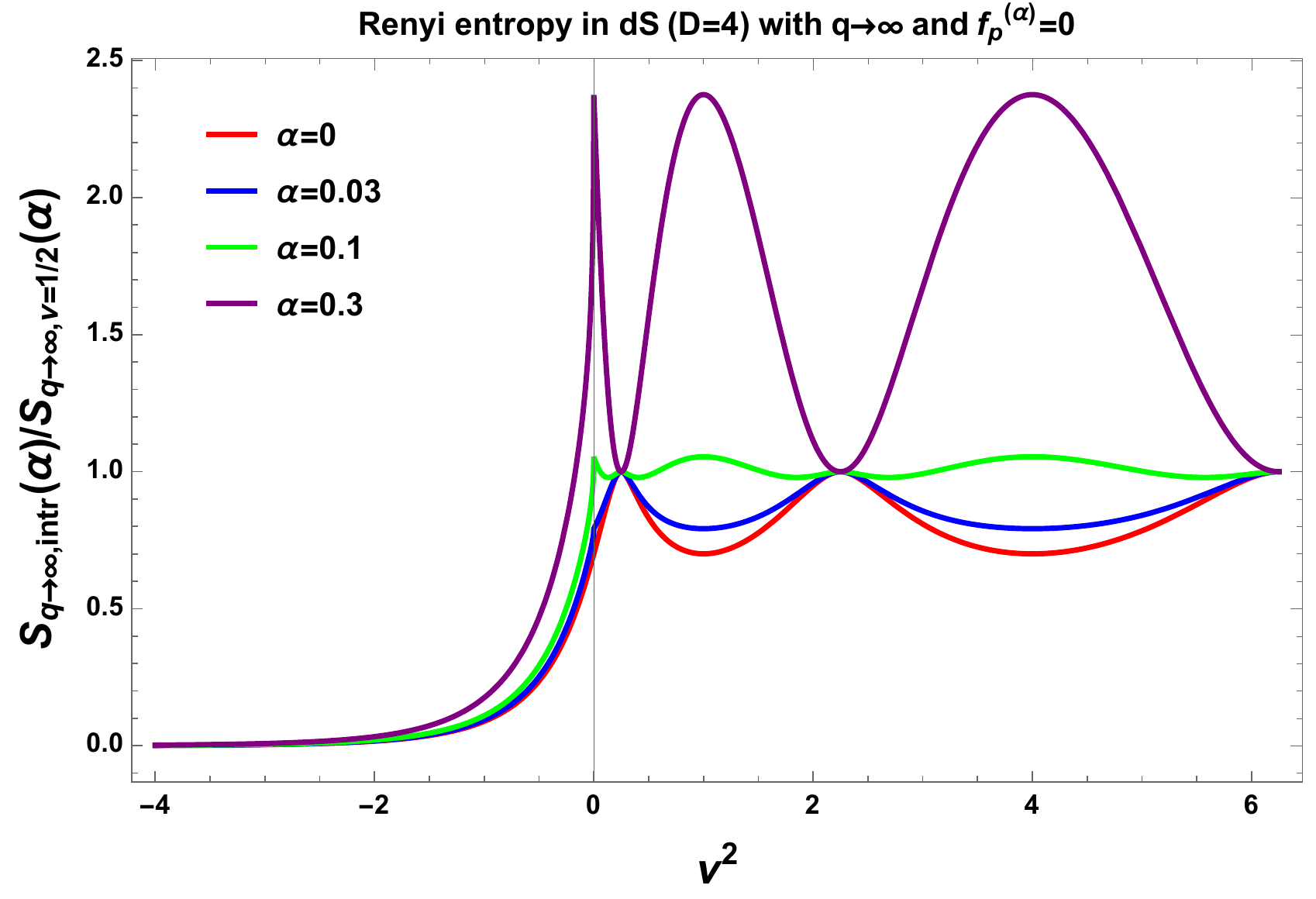}
    		\label{axz}
    	}
    	\subfigure[Normalised R$\acute{e}$nyi entropy vs $\nu^2$ in de Sitter space with axionic source ($f_p=10^{-7}$).]{                             \includegraphics[width=7.2cm,height=5cm] {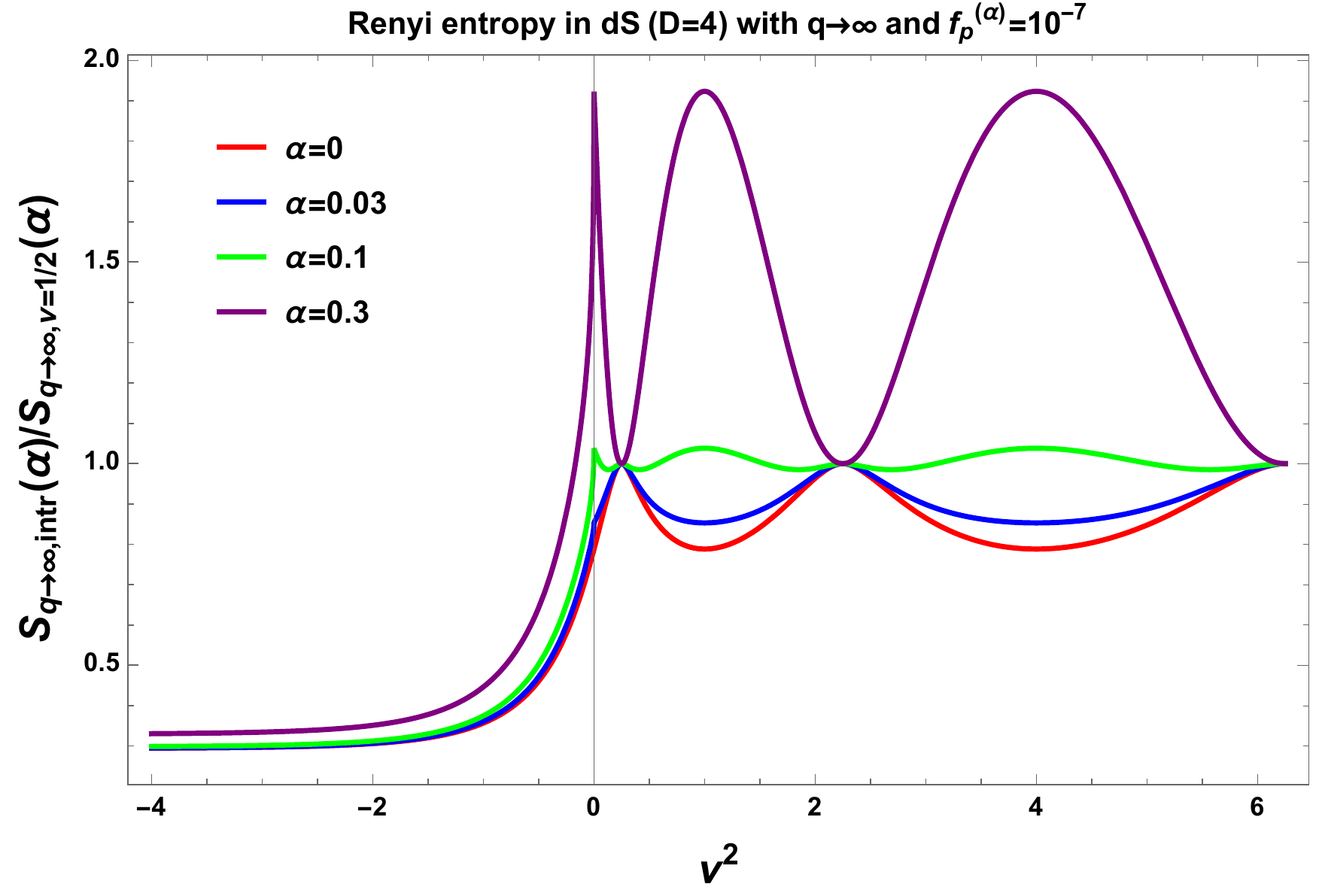}\label{axza}
    	}\caption[Optional caption for list of figures]{Normalized R$\acute{e}$nyi entropy $S_{q\rightarrow \infty,intr}(\alpha)/S_{q\rightarrow \infty,\nu=1/2}(\alpha)$ vs mass parameter $\nu^2$ in $3+1$ D de Sitter space in absence of axionic source ($f^{(\alpha)}_p=0$) and in presence of axionic source ($f^{(\alpha)}_p=10^{-7}$) for $q\rightarrow \infty$ and $\alpha=0$ (\textcolor{red}{\bf red}),$\alpha=0.03$ (\textcolor{blue}{\bf blue}),$\alpha=0.1$ (\textcolor{green}{\bf green}),$\alpha=0.3$ (\textcolor{violet}{\bf violet}) with $`+'$ branch of solution of $|\gamma^{(\alpha)}_p|$ and $|\Gamma^{(\alpha)}_{p,n}|$, which quantifies largest eigenvalue of density matrix. Here we set the cut-off $\Lambda_{\bf C}=300$ for numerical computation. } 
    	\label{cc}
    \end{figure*}

    \begin{figure*}[htb]
    	\centering
    	\subfigure[For $q=0.9$ and $\nu^2>0$.]{
    		\includegraphics[width=7.8cm,height=4cm] {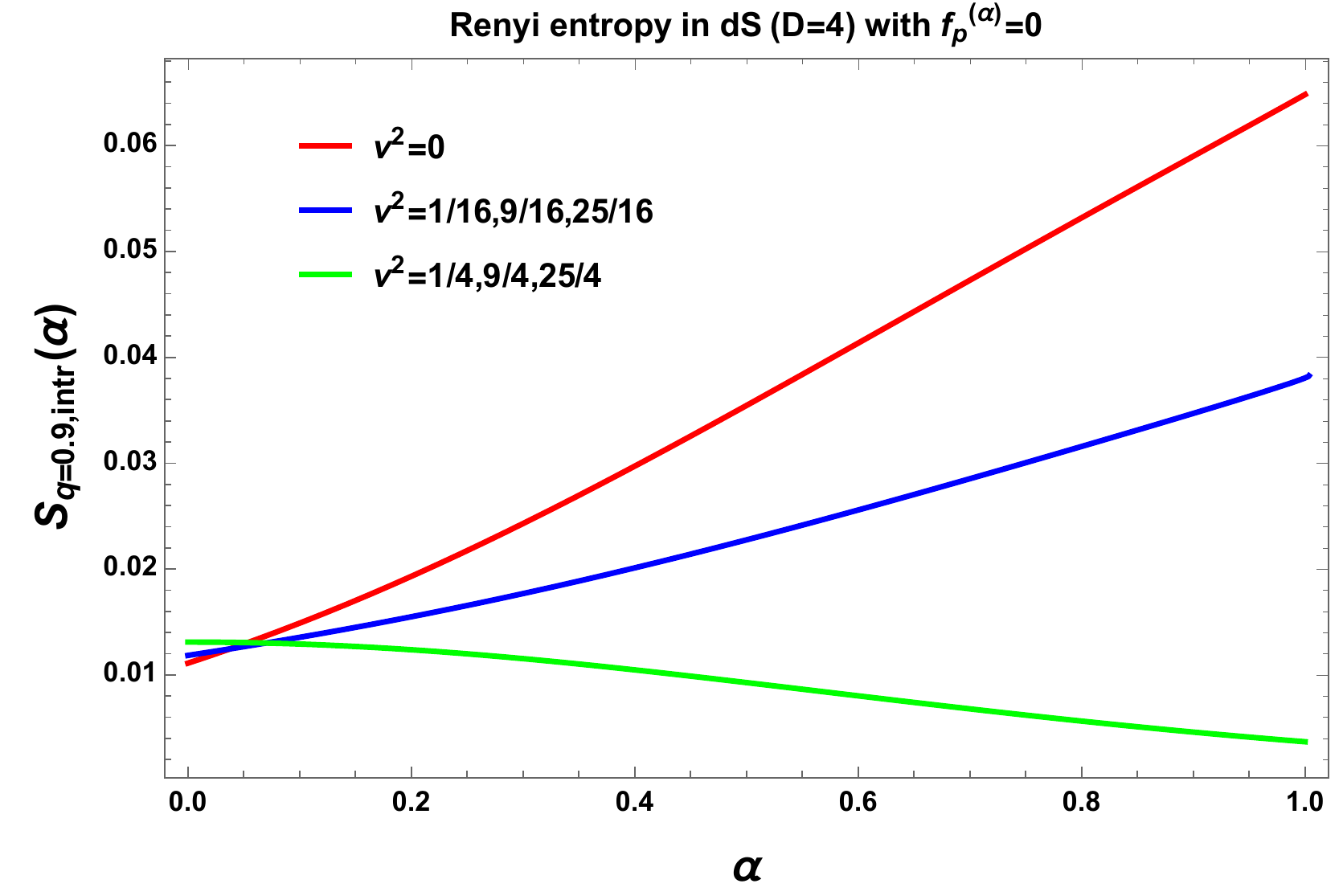}
    		\label{zxx5c}}
    	\subfigure[For $q=0.7$ and $\nu^2>0$.]{
    		\includegraphics[width=7.8cm,height=4cm] {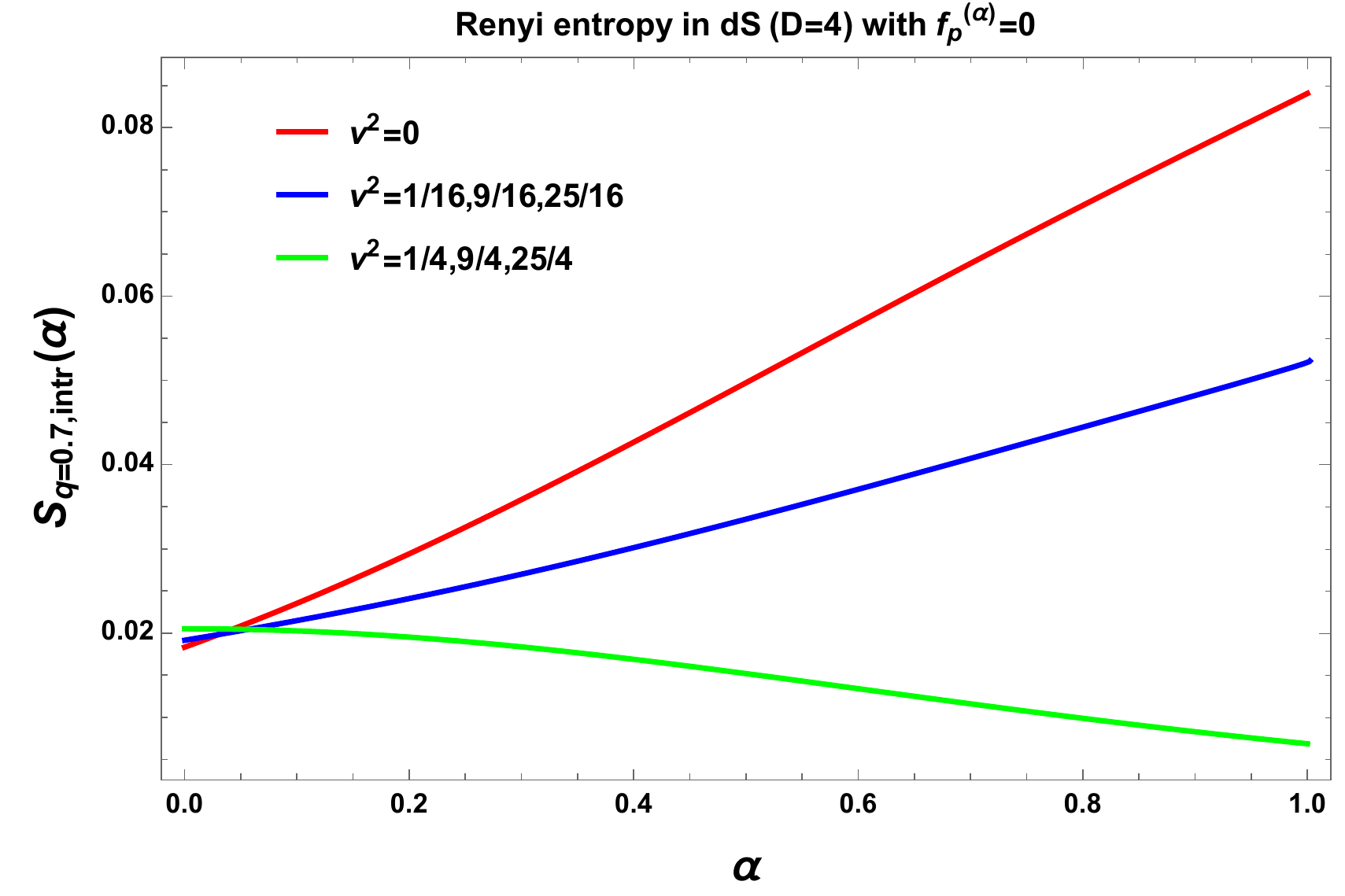}
    		\label{px5a}}
    	\subfigure[For $q=0.5$ and $\nu^2>0$.]{
    		\includegraphics[width=7.8cm,height=4cm] {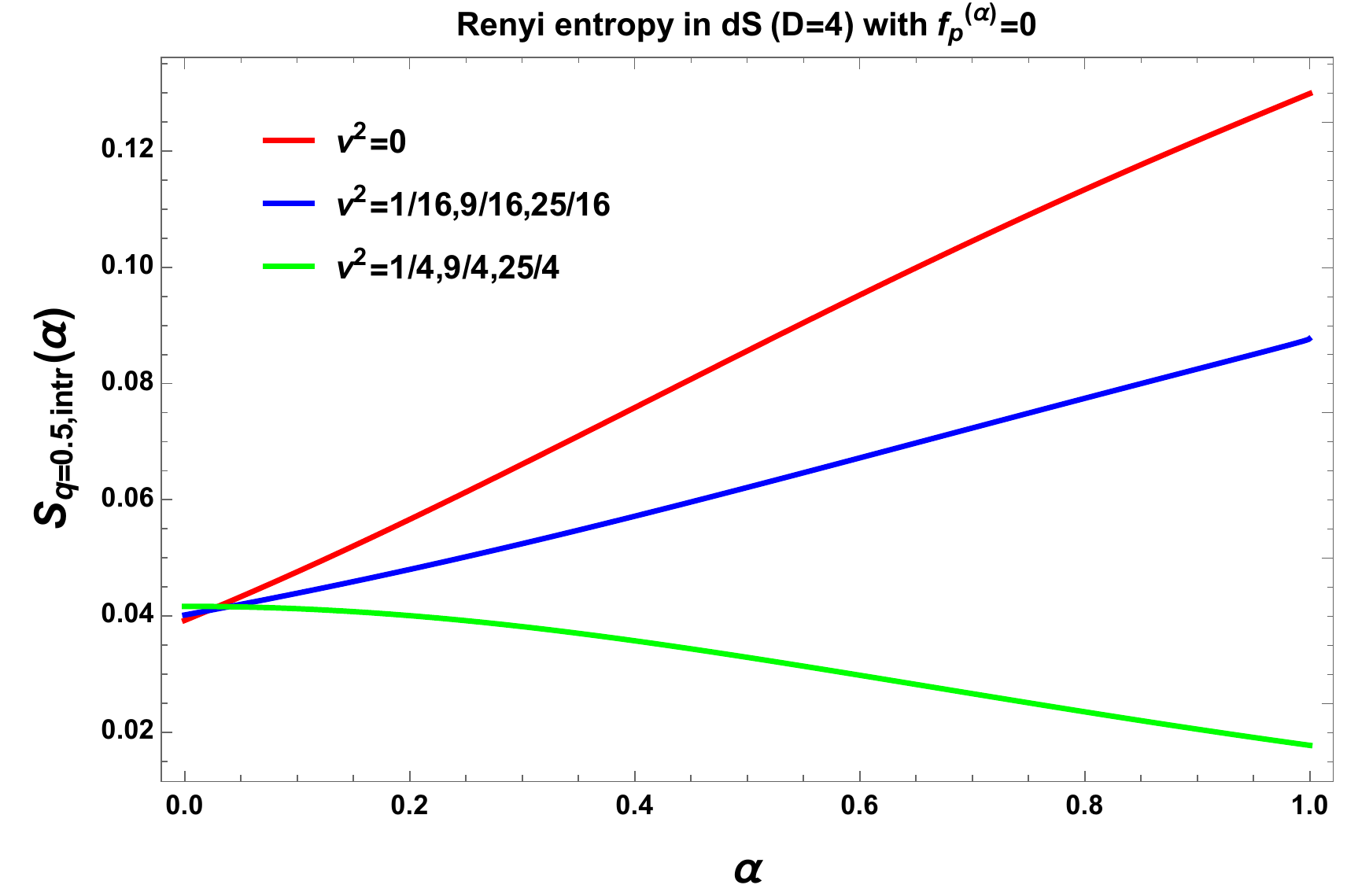}
    		\label{pcx5aa}}  
    	\subfigure[For $q=0.3$ and $\nu^2>0$.]{
    		\includegraphics[width=7.8cm,height=4cm] {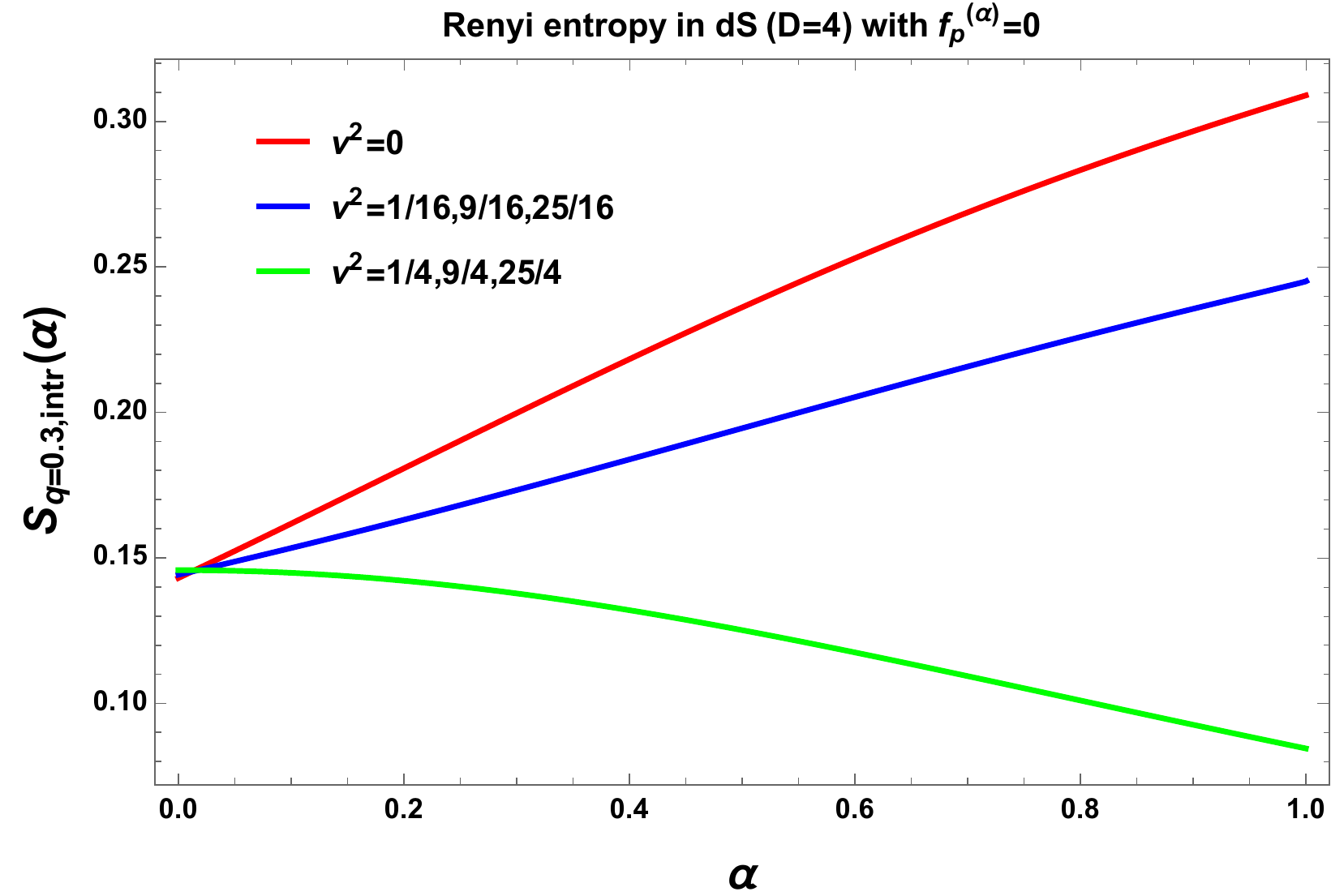}
    		\label{px5aaa} }
    	\caption[Optional caption for list of figures]{R$\acute{e}$nyi entropy $S_{q,intr}(\alpha)$ vs parameter $\alpha$ plot in $3+1$ D de Sitter space in absence of axionic source ($f^{(\alpha)}_p=0$) for $q=0.1$, $q=0.3$, $q=0.5$, $q=0.7$, $q=0.9$ with $`+'$ branch of solution of $|\gamma^{(\alpha)}_p|$ and $|\Gamma^{(\alpha)}_{p,n}|$.} 
    	\label{bc1}
    \end{figure*}               
    \begin{figure*}[htb]
    	\centering
    	\subfigure[For $q=0.9$ and $\nu^2>0$.]{
    		\includegraphics[width=7.8cm,height=4cm] {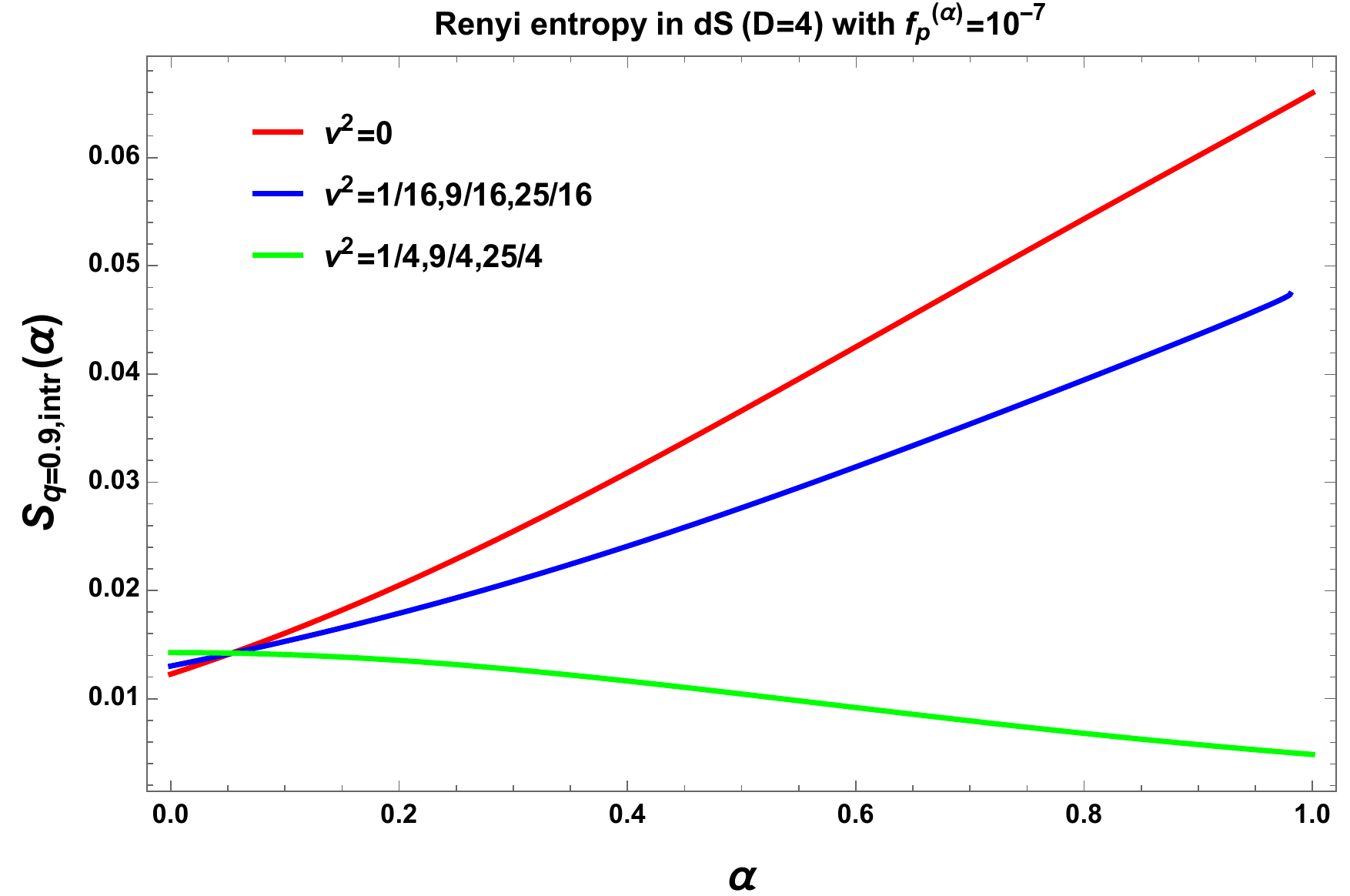}
    		\label{z5c}}
    	\subfigure[For $q=0.7$ and $\nu^2>0$.]{
    		\includegraphics[width=7.8cm,height=4cm] {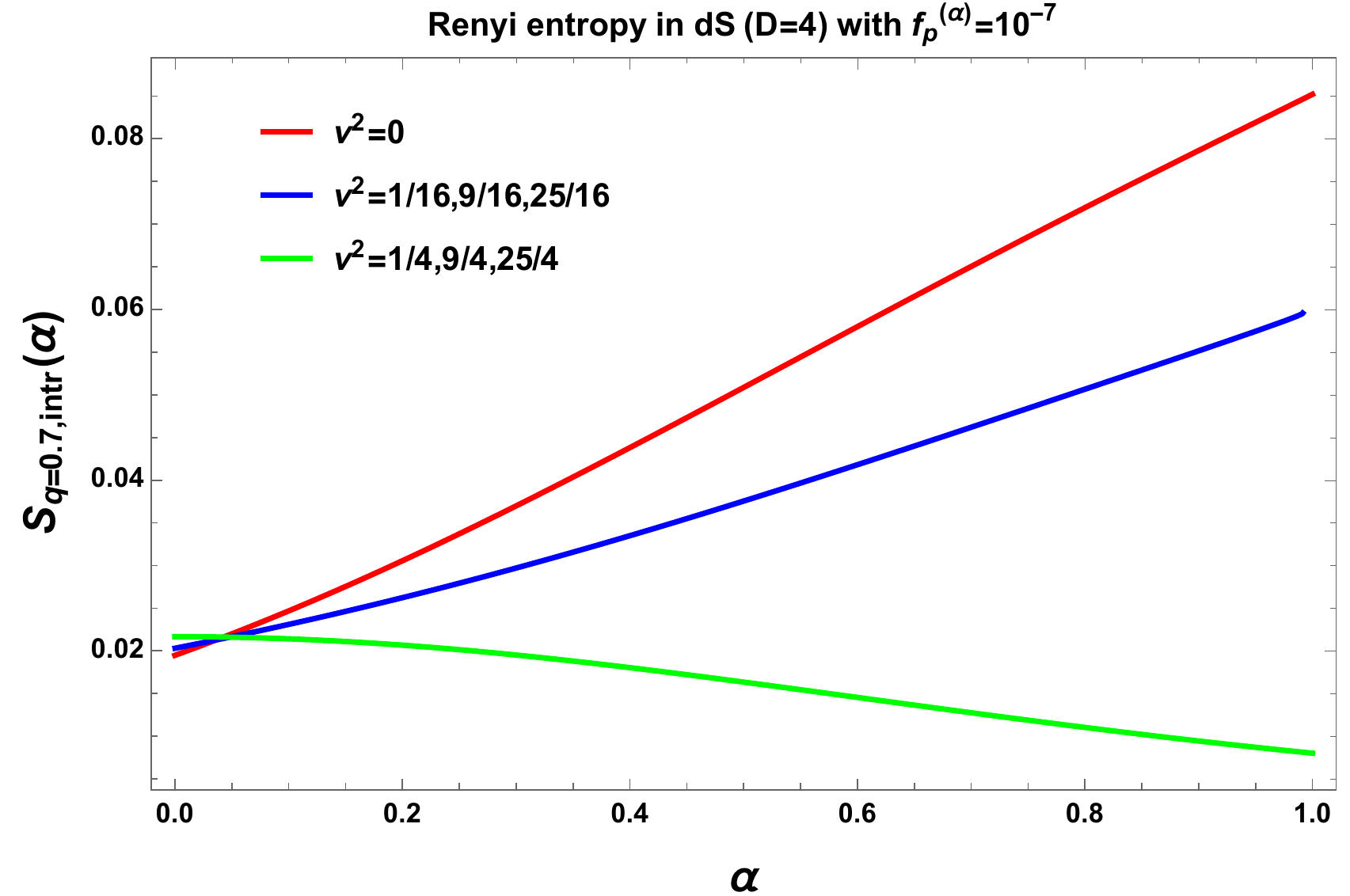}
    		\label{p5a}}
    	\subfigure[For $q=0.5$ and $\nu^2>0$.]{
    		\includegraphics[width=7.8cm,height=4cm] {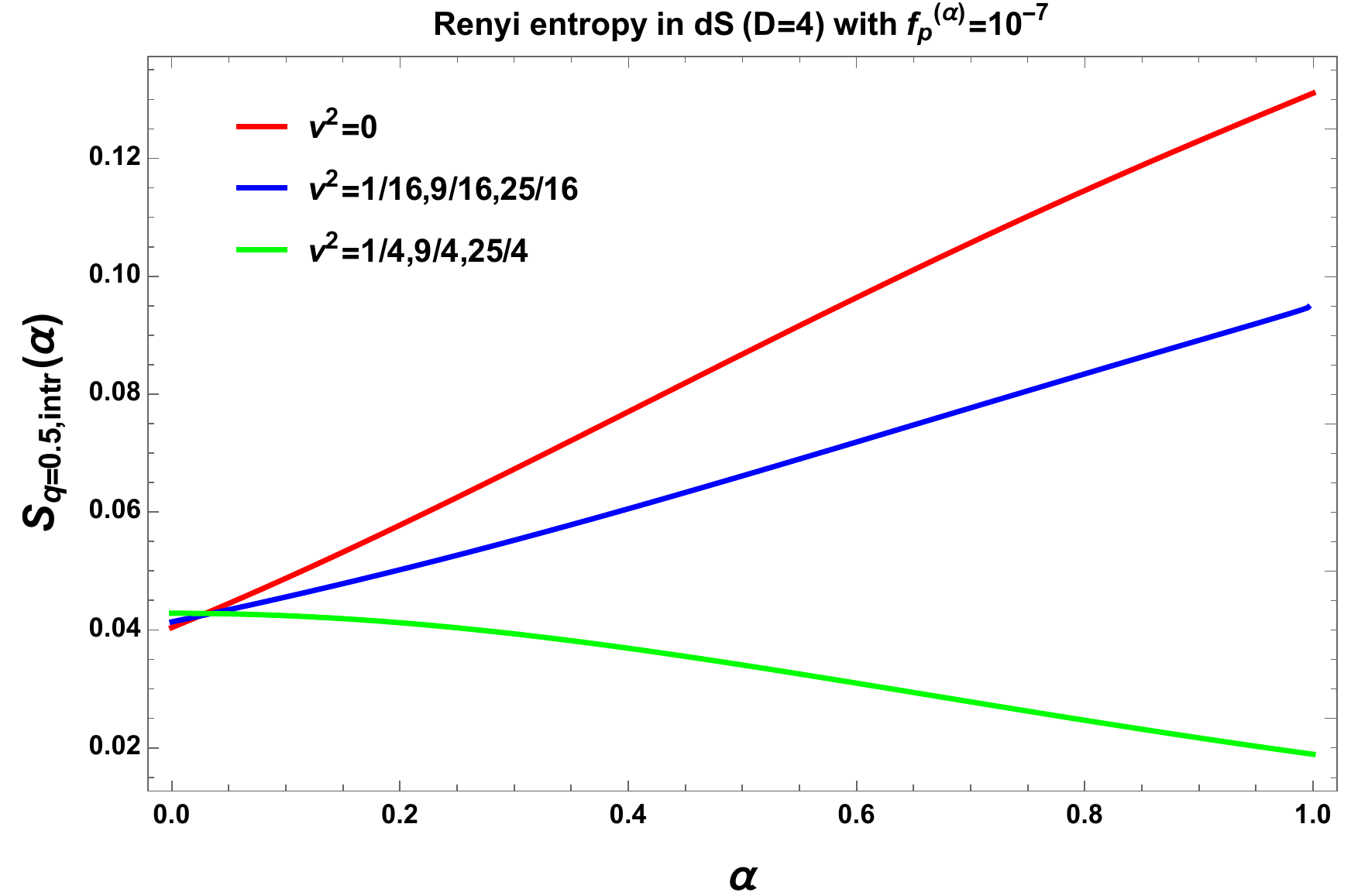}
    		\label{pc5aa}}  
    	\subfigure[For $q=0.3$ and $\nu^2>0$.]{
    		\includegraphics[width=7.8cm,height=4cm] {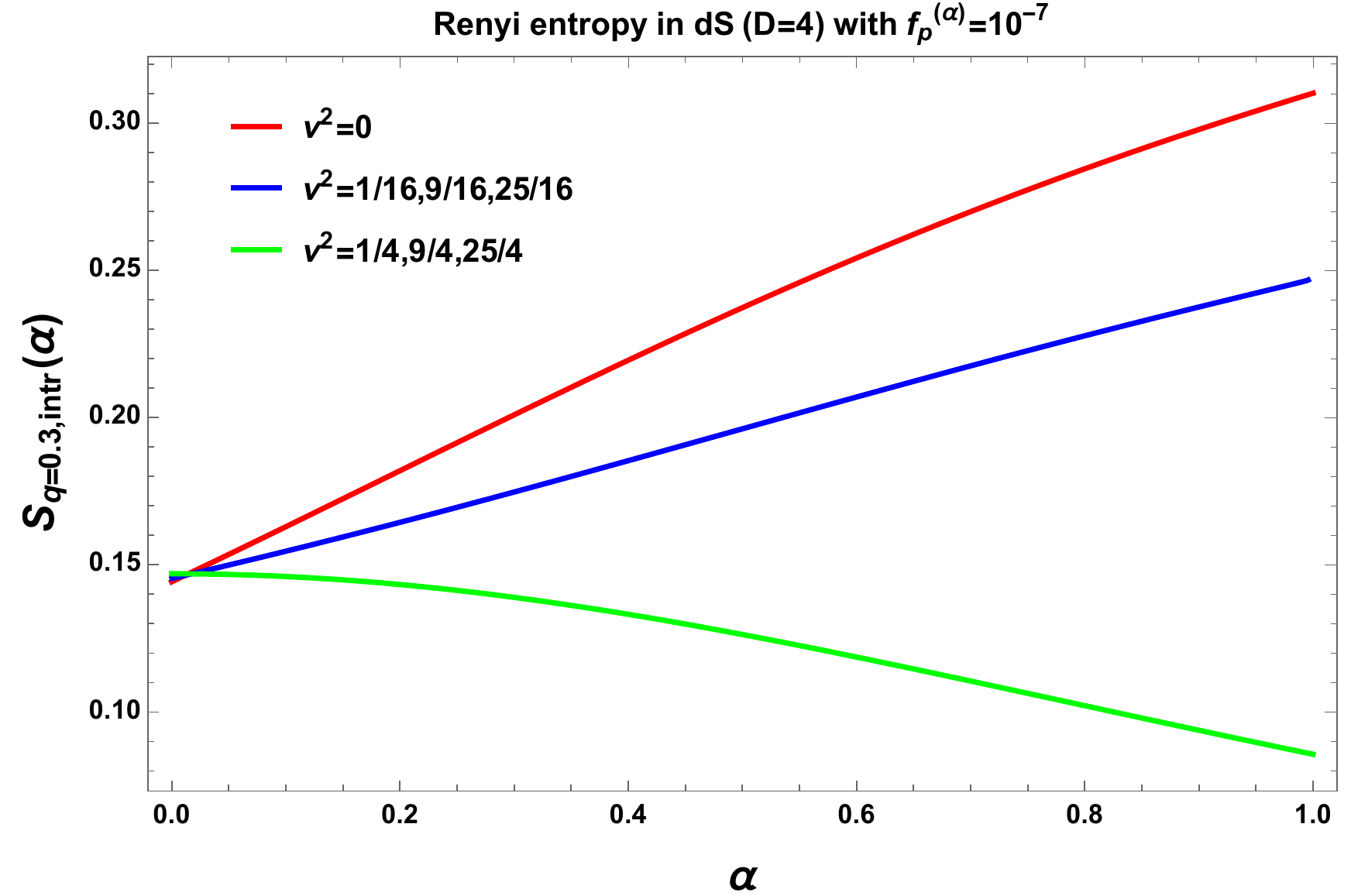}
    		\label{p5aaa} }
    	\caption[Optional caption for list of figures]{R$\acute{e}$nyi entropy $S_{q,intr}(\alpha)$ vs parameter $\alpha$ plot in $3+1$ D de Sitter space in presence of axionic source ($f^{(\alpha)}_p=10^{-7}$) for $q=0.1$, $q=0.3$, $q=0.5$, $q=0.7$, $q=0.9$ with $`+'$ branch of solution of $|\gamma^{(\alpha)}_p|$ and $|\Gamma^{(\alpha)}_{p,n}|$.} 
    	\label{b1}
    \end{figure*}   
    
    \begin{figure*}[htb]
    	\centering
    	\subfigure[For $q=0.9$ and $\nu^2<0$.]{
    		\includegraphics[width=7.8cm,height=4cm] {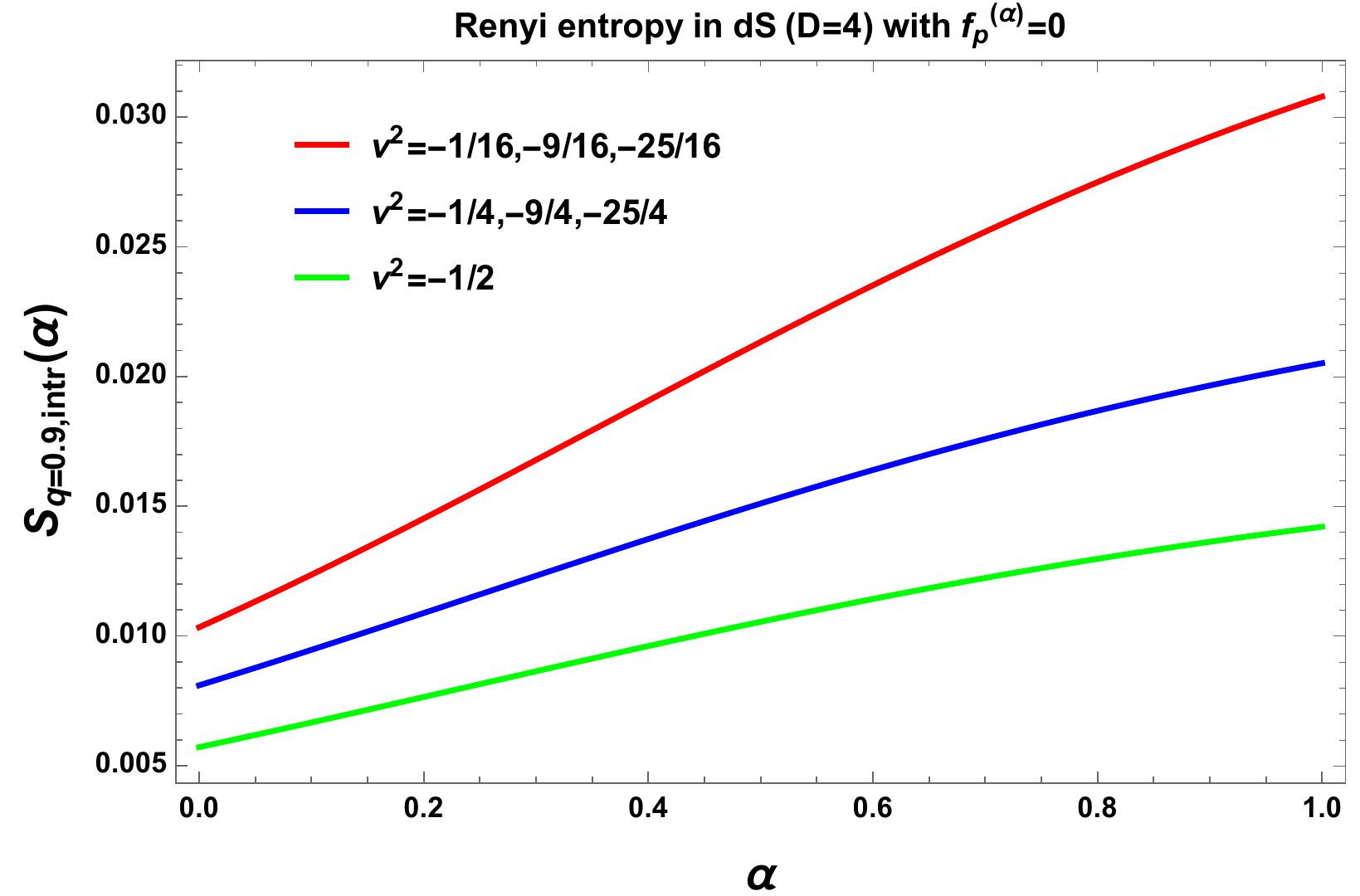}
    		\label{x5cg}}
    	\subfigure[For $q=0.7$ and $\nu^2<0$.]{
    		\includegraphics[width=7.8cm,height=4cm] {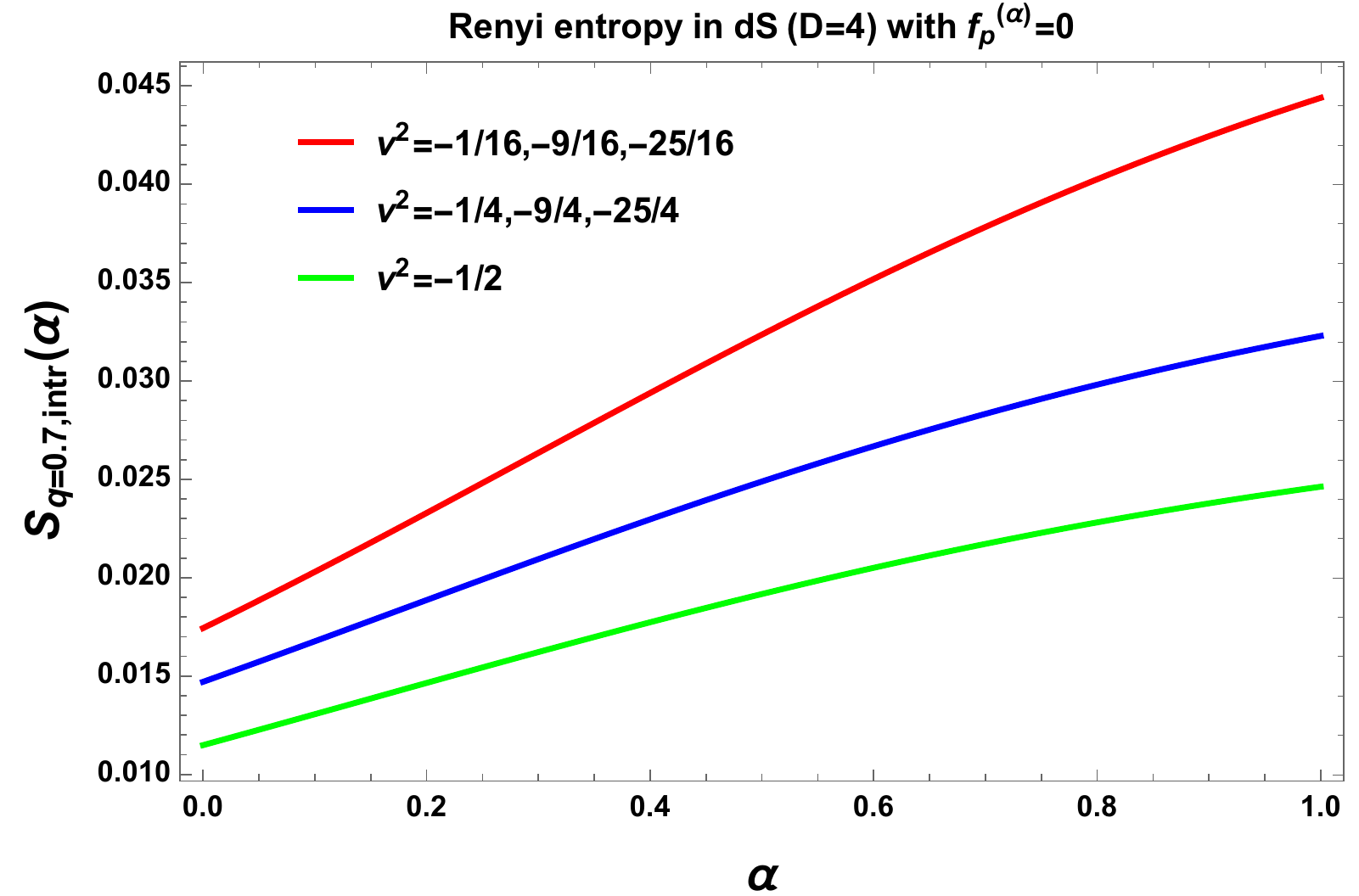}
    		\label{x5ag}}
    	\subfigure[For $q=0.5$ and $\nu^2<0$.]{
    		\includegraphics[width=7.8cm,height=4cm] {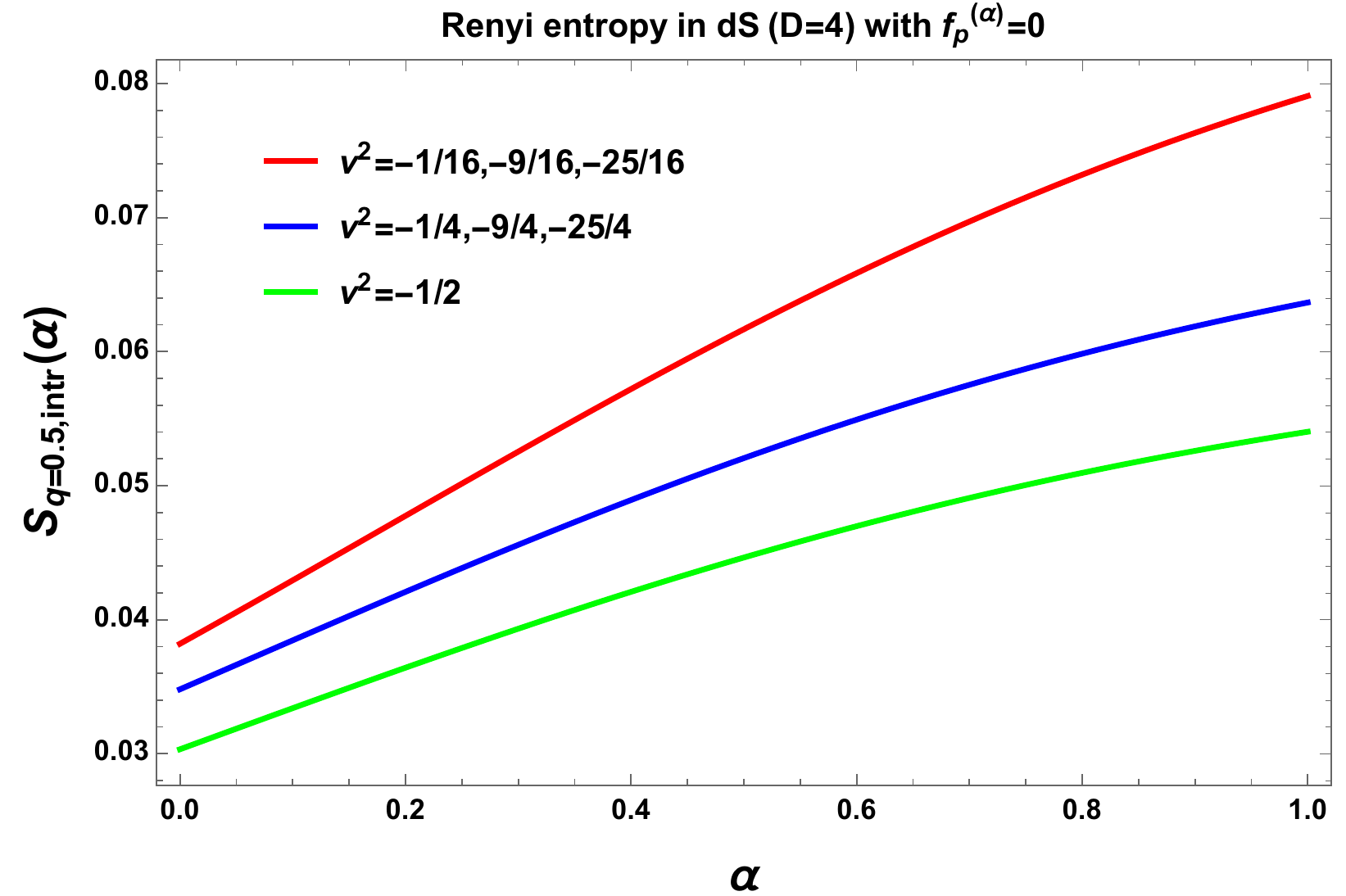}
    		\label{xc5aag}}  
    	\subfigure[For $q=0.3$ and $\nu^2<0$.]{
    		\includegraphics[width=7.8cm,height=4cm] {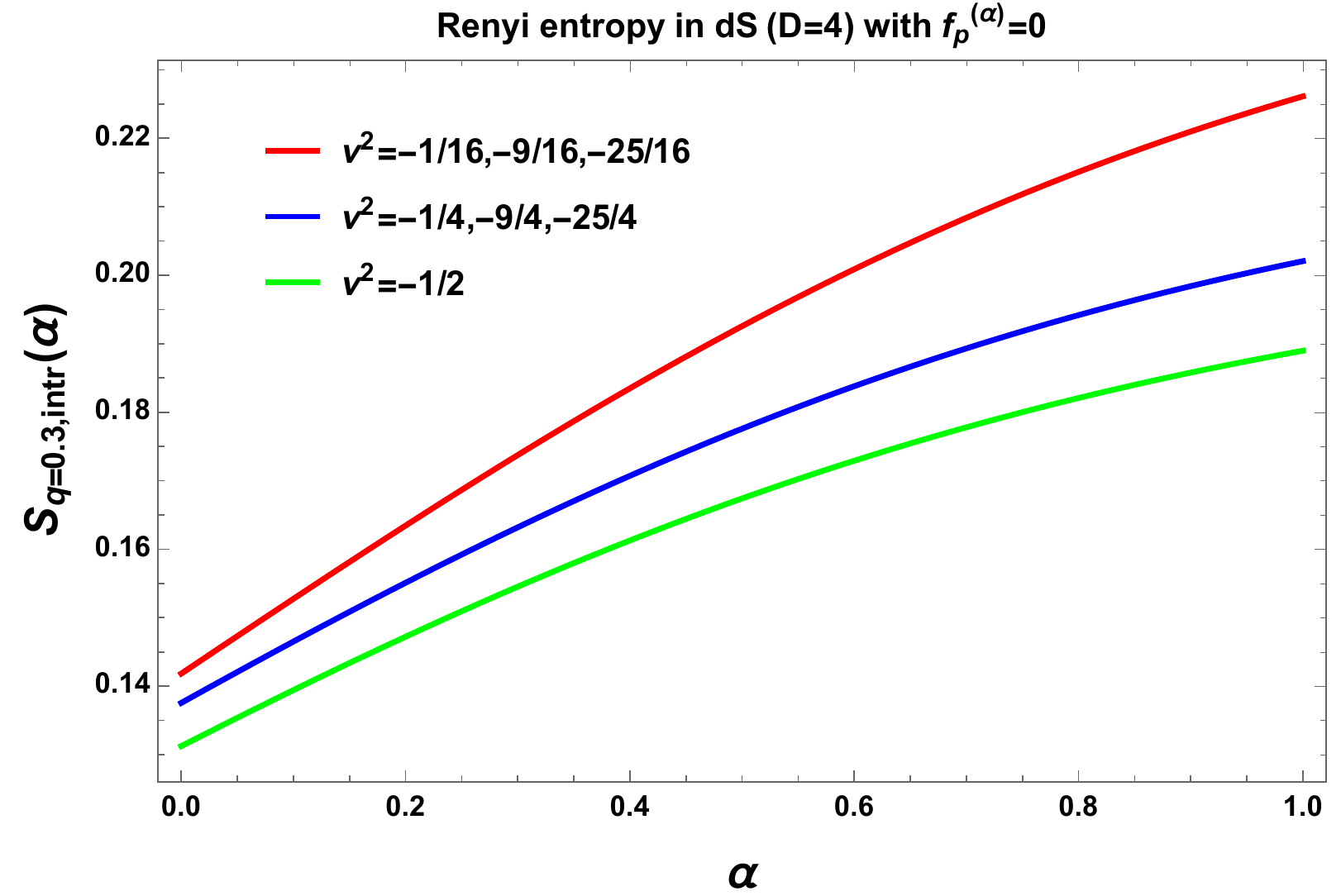}
    		\label{x5aaag} }
    	\caption[Optional caption for list of figures]{R$\acute{e}$nyi entropy $S_{q,intr}(\alpha)$ vs parameter $\alpha$ plot in $3+1$ D de Sitter space in absence of axionic source ($f^{(\alpha)}_p=0$) for $q=0.1$, $q=0.3$, $q=0.5$, $q=0.7$, $q=0.9$ with $`+'$ branch of solution of $|\gamma^{(\alpha)}_p|$ and $|\Gamma^{(\alpha)}_{p,n}|$.} 
    	\label{bcc1g}
    \end{figure*}                        
    \begin{figure*}[htb]
    	\centering
    	\subfigure[For $q=0.9$ and $\nu^2<0$.]{
    		\includegraphics[width=7.8cm,height=4cm] {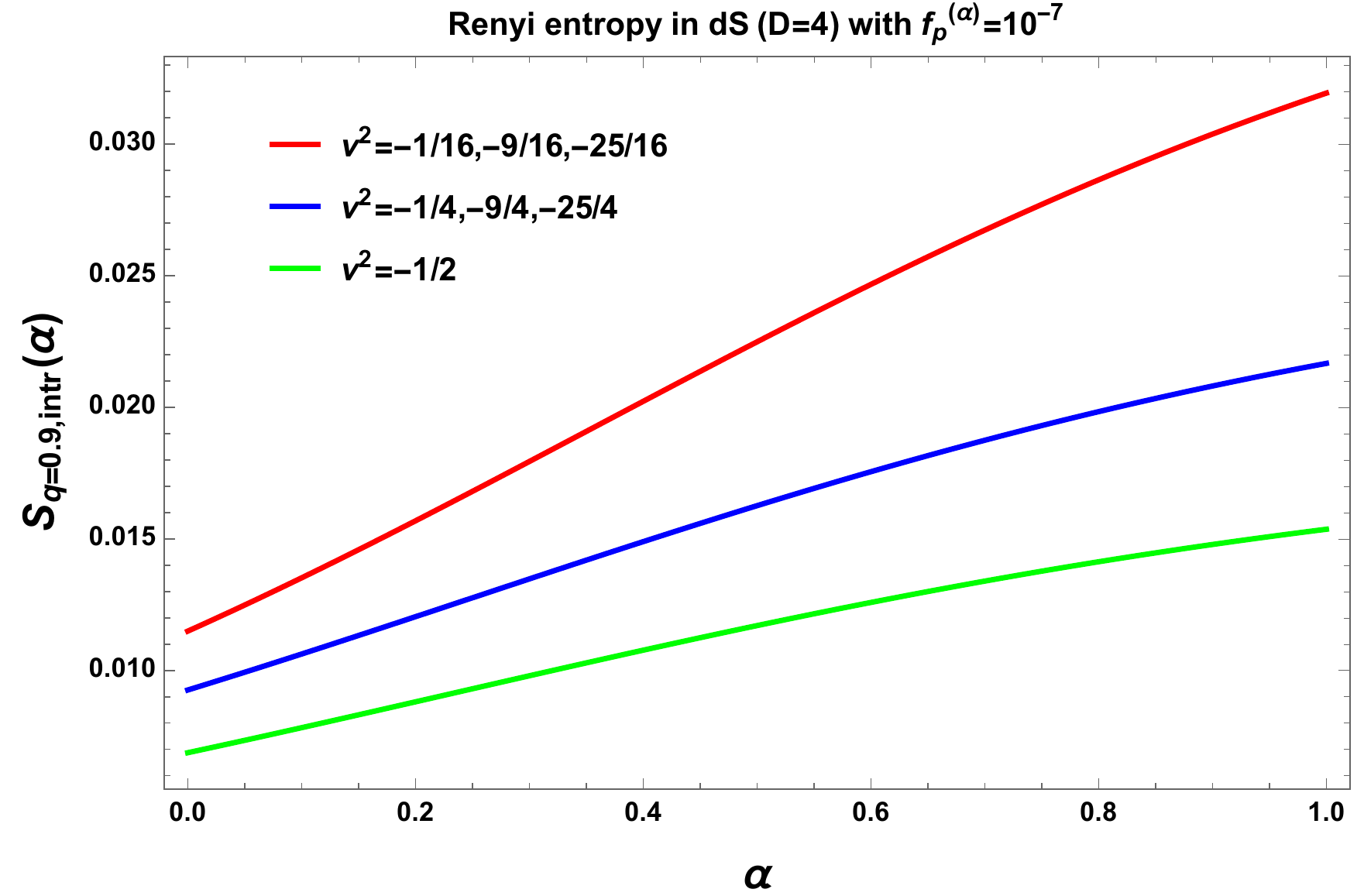}
    		\label{x5bc}}
    	\subfigure[For $q=0.7$ and $\nu^2<0$.]{
    		\includegraphics[width=7.8cm,height=4cm] {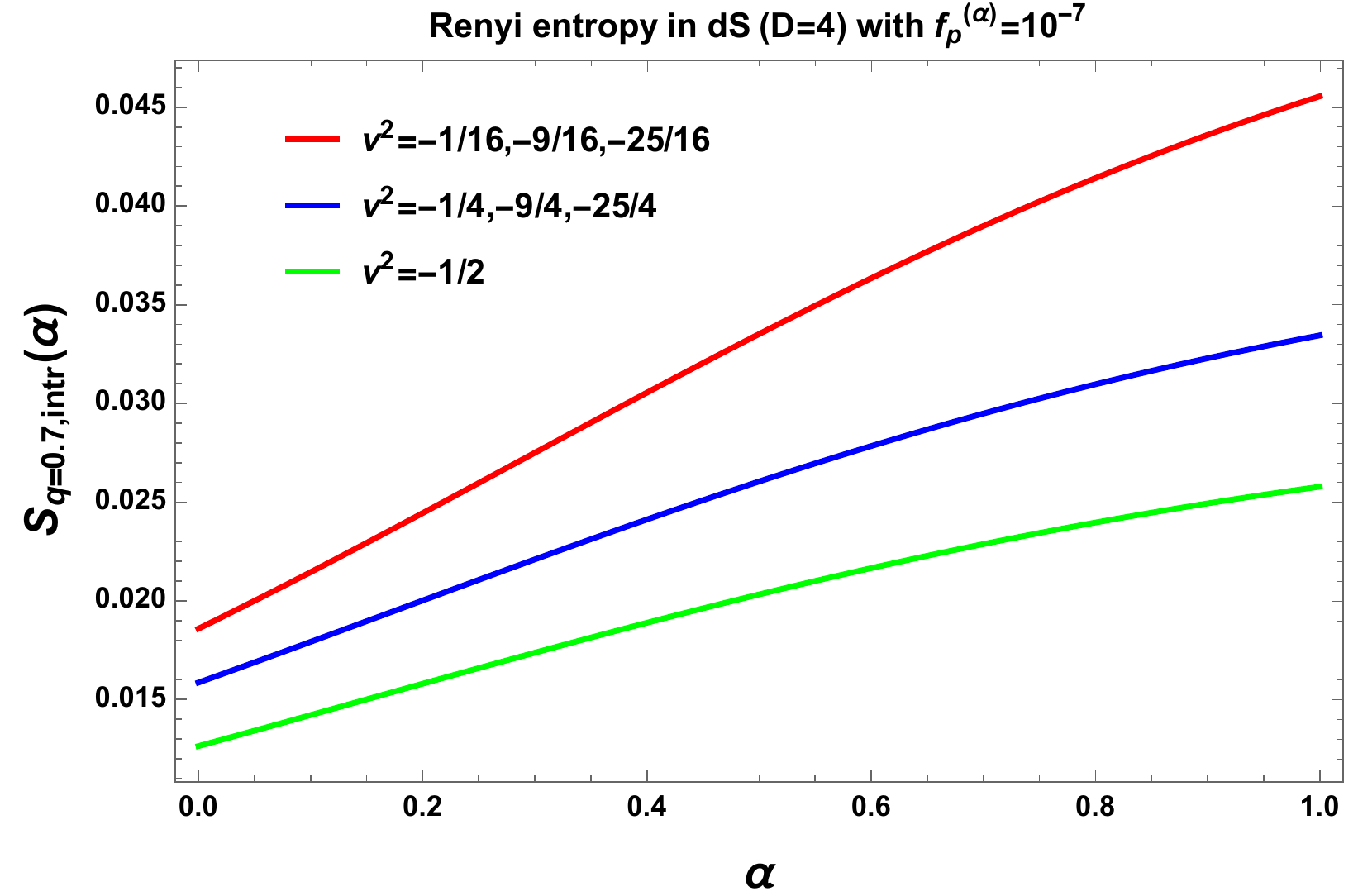}
    		\label{x5ba}}
    	\subfigure[For $q=0.5$ and $\nu^2<0$.]{
    		\includegraphics[width=7.8cm,height=4cm] {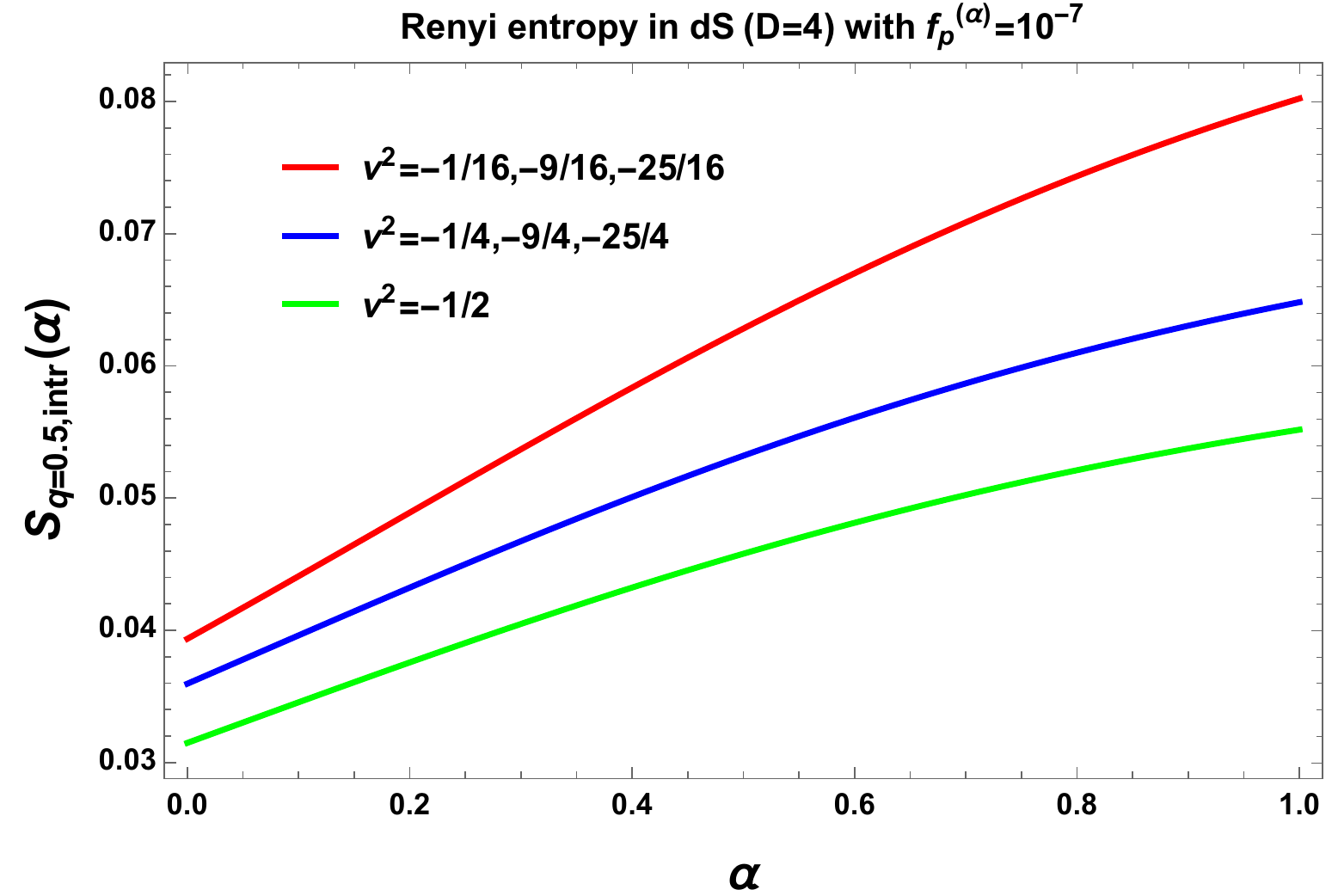}
    		\label{xc5baa}}  
    	\subfigure[For $q=0.3$ and $\nu^2<0$.]{
    		\includegraphics[width=7.8cm,height=4cm] {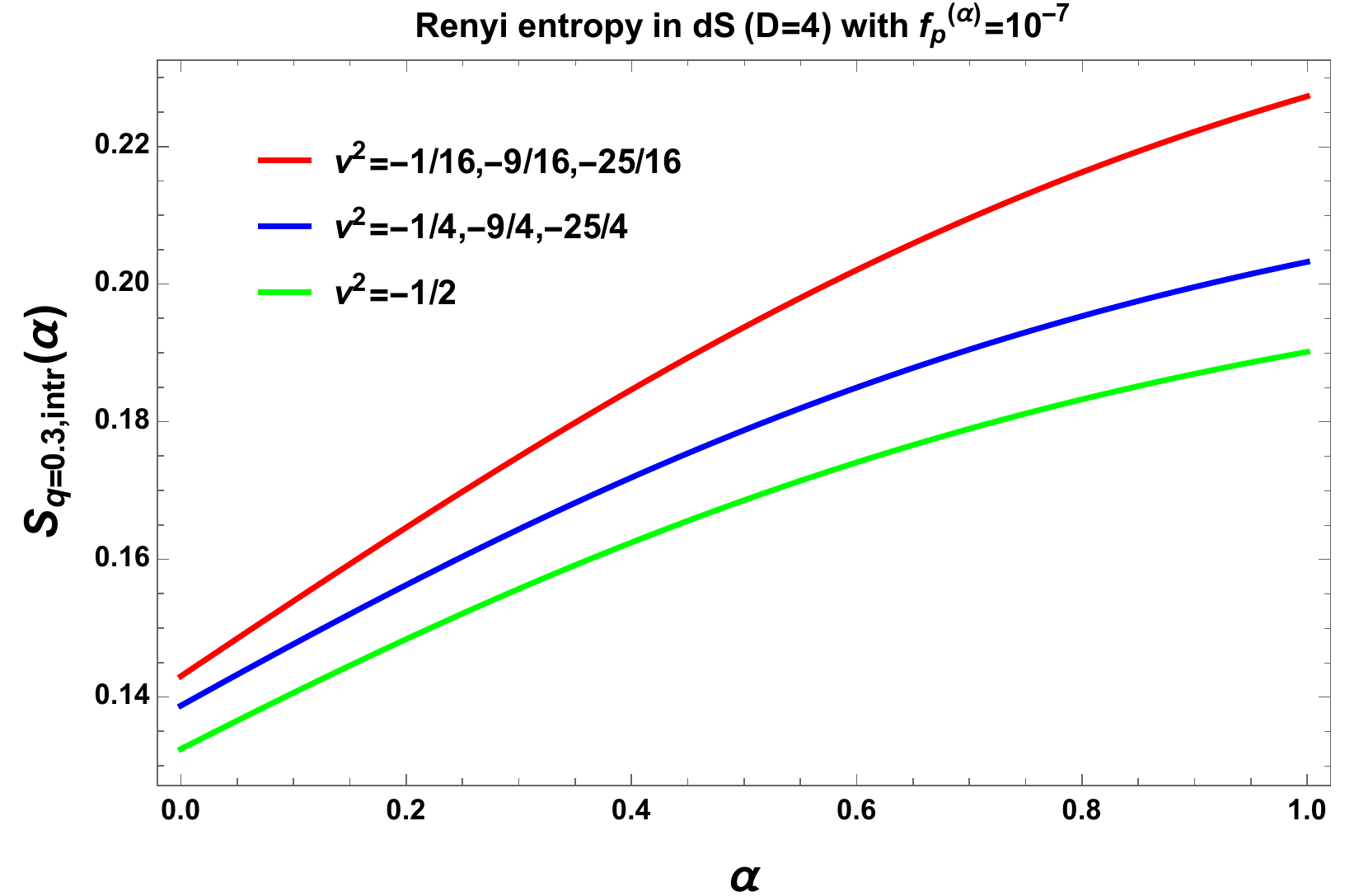}
    		\label{x5baaa} }
    	\caption[Optional caption for list of figures]{R$\acute{e}$nyi entropy $S_{q,intr}(\alpha)$ vs parameter $\alpha$ plot in $3+1$ D de Sitter space in presence of axionic source ($f^{(\alpha)}_p=10^{-7}$) for $q=0.1$, $q=0.3$, $q=0.5$, $q=0.7$, $q=0.9$ with $`+'$ branch of solution of $|\gamma^{(\alpha)}_p|$ and $|\Gamma^{(\alpha)}_{p,n}|$.} 
    	\label{bxcc1}
    \end{figure*}                        

    \begin{figure*}[htb]
    	\centering
    	\subfigure[For $\nu^2=1$ and $f^{(\alpha)}_p=0$.]{
    		\includegraphics[width=7.8cm,height=4cm] {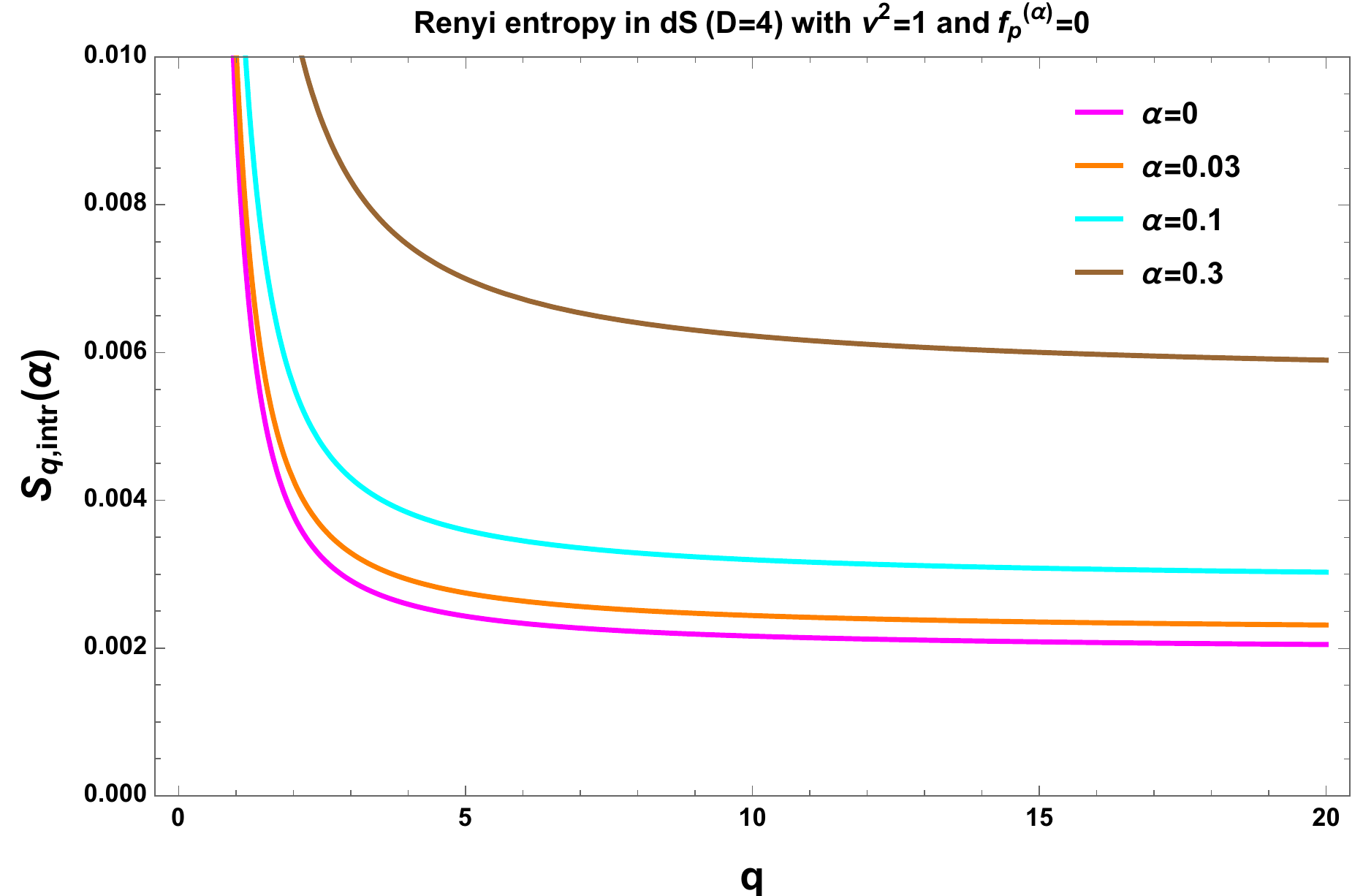}
    		\label{z5}}
    	\subfigure[For $\nu^2=1$ and $f^{(\alpha)}_p=10^{-7}$.]{
    		\includegraphics[width=7.8cm,height=4cm] {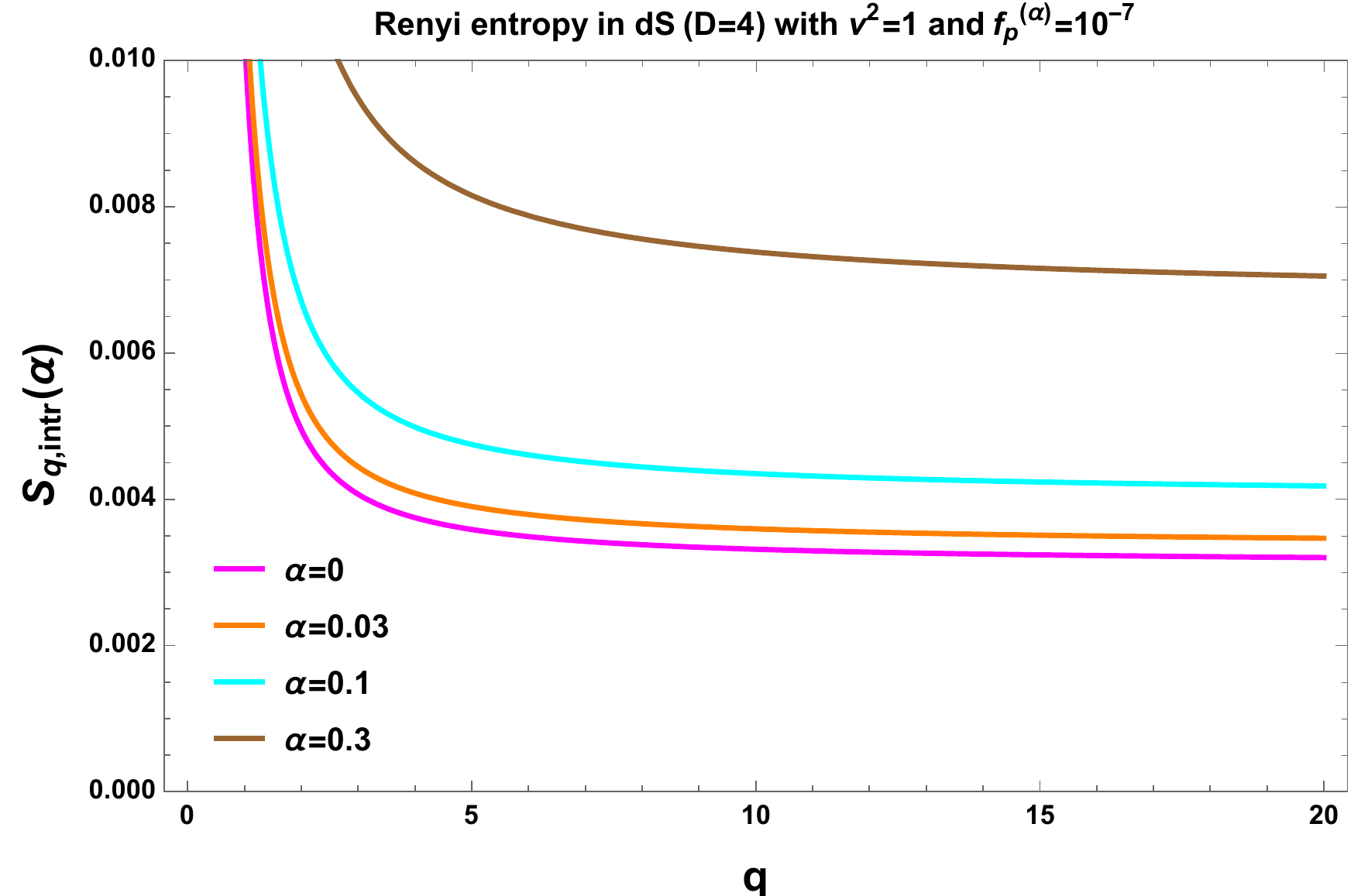}
    		\label{z5a}}
    	\subfigure[For $\nu^2=-1$ and $f^{(\alpha)}_p=0$.]{
    		\includegraphics[width=7.8cm,height=4cm] {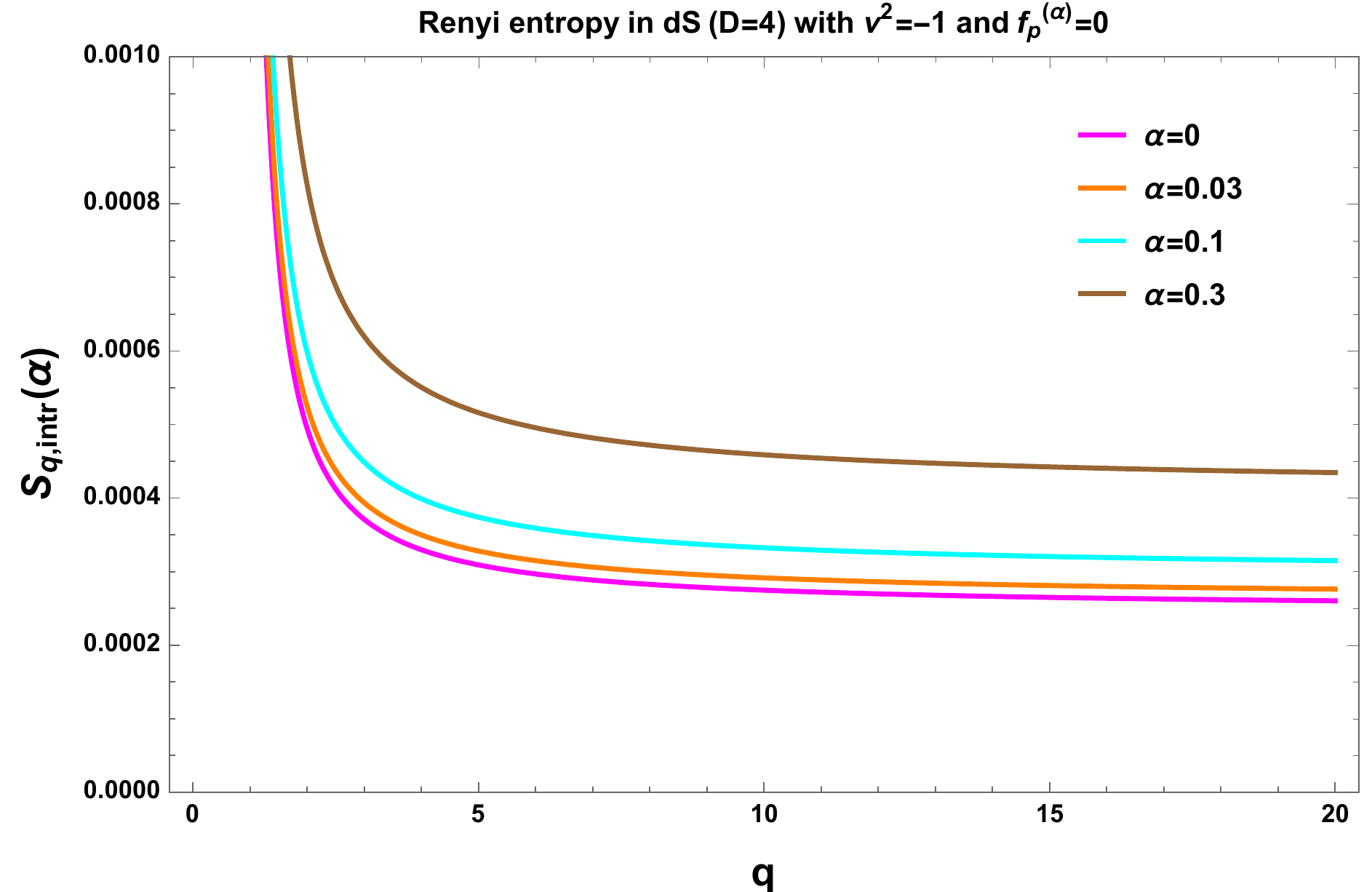}
    		\label{z5aa}}  
    	\subfigure[For $\nu^2=-1$ and $f^{(\alpha)}_p=10^{-7}$.]{
    		\includegraphics[width=7.8cm,height=4cm] {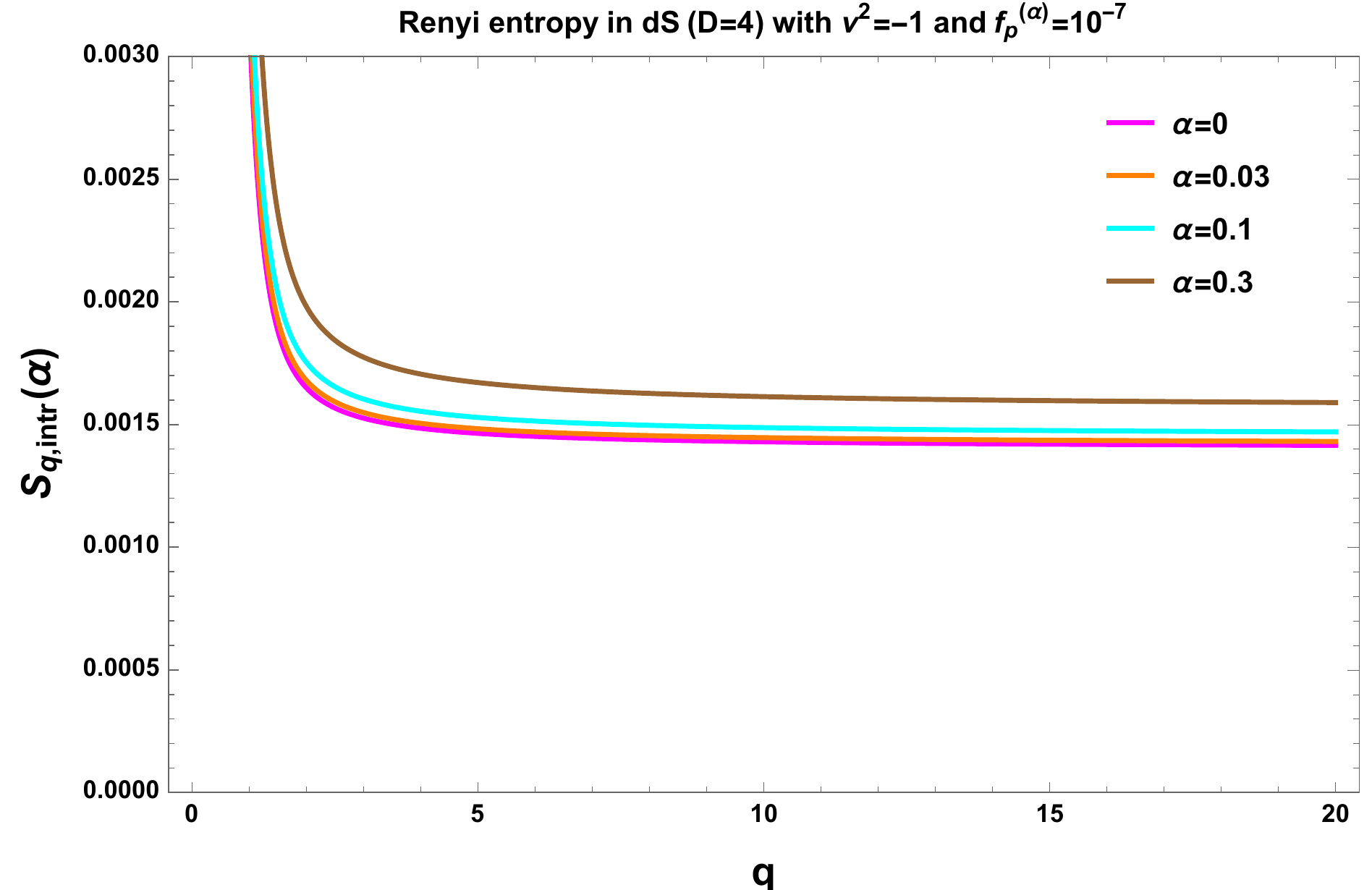}
    		\label{z5aaa} }
    	\caption[Optional caption for list of figures]{R$\acute{e}$nyi entropy $S_{q,intr}(\alpha)$ vs $q$ plot in $3+1$ D de Sitter space in absence and presence of axionic source for $\alpha=0$, $\alpha=0.03$ , $\alpha=0.1$ and $\alpha=0.3$ with $`+'$ branch of solution of $|\gamma^{(\alpha)}_p|$ and $|\Gamma^{(\alpha)}_{p,n}|$. } 
    	\label{b2}
    \end{figure*}                                                           

                       Further, using the results obtained from the first solution for $|\gamma^{(\alpha)}_p|$, within the range  $0<x<2\pi |\nu|$ with $\nu^2<0$, we take $q\rightarrow 1$ limit. This gives the following simplified expression for the integral ${\cal M}^{(\alpha)}_{1,q}$:
                             \bea \label{df1}
                                  \displaystyle \lim_{q\rightarrow 1}{\cal M}^{(\alpha)}_{1,q} &=&\displaystyle
                                 \displaystyle \frac{\nu^3}{3} \left[\frac{2 (1+\tan \alpha)^2 \left\{\nu-\frac{1}{\pi}\ln (1+\tan \alpha)\right\}}{\left(e^{2 \pi  \nu}-(1+\tan \alpha)^2\right)}-\frac{\ln \left(1-e^{-2 \pi  \nu}\left(1+\tan\alpha\right)^2\right)}{\pi}\right].~~~~~~~~~~~\eea
        Now further using $|\nu|>>1$ approximation in Eq~(\ref{df1}) we get:
  \bea \label{df1n}
                                    \displaystyle \lim_{|\nu|>>1,q\rightarrow 1}{\cal M}^{(\alpha)}_{1,q} &=&\displaystyle
                                   \displaystyle \frac{2\nu^4}{3}e^{-2\pi\nu}\left(1+\tan\alpha\right)^2\left\{1-\frac{1}{\pi\nu}\ln\left(1+\tan\alpha\right)\right\}\left[1+\left(1+\tan\alpha\right)^2{\cal O}\left(\nu^{-1}\right)\right],~~~~~~~~~~~\eea                               
 
                               In this context further if we take the source less limit $f^{(\alpha)}_p\rightarrow 0$ then the integral ${\cal M}^{(\alpha)}_{3,q}$ vanishes:
                                  \bea \lim_{q\rightarrow 1,|\nu|>>1,f_p\rightarrow 0}{\cal M}^{(\alpha)}_{3,q} &=& 0.\eea
                                             As a result in the large mass limiting situation with $q\rightarrow 1$ the long range correlation can be expressed in terms of R$\acute{e}$nyi entropy as: 
                                   \bea \lim_{q\rightarrow 1,|\nu|>>1, f^{(\alpha)}_p\rightarrow 0}S_{q,\bf intr}(\alpha)&\approx& \frac{2\nu^4}{3}e^{-2\pi\nu}\left(1+\tan\alpha\right)^2\left\{1-\frac{1}{\pi\nu}\ln\left(1+\tan\alpha\right)\right\}\left[1+\left(1+\tan\alpha\right)^2{\cal O}\left(\nu^{-1}\right)\right]\nonumber\\&=& S_{\bf intr}(\alpha)=\lim_{|\nu|>>1, f^{(\alpha)}_p\rightarrow 0}{\bf c_6}(\alpha,\nu).\eea
        Similarly using the results obtained from the second solution for $|\gamma^{(\alpha)}_p|$, within the range  $0<x<2\pi |\nu|$ with $\nu^2<0$, we take $q\rightarrow 1$ limit. This gives the following simplified expression for the integral ${\cal M}^{(\alpha)}_{1,q}$:
                                    \bea \label{df1a}
                                         \displaystyle \lim_{q\rightarrow 1}{\cal M}^{(\alpha)}_{1,q} &=&\displaystyle
                                        \displaystyle \frac{\nu ^3}{3} \left[\frac{2 e^{2 \pi  \nu } \left(1+\tan\alpha\right)^2 \left\{\nu+\ln \left(1+\tan\alpha\right) \right\}}{e^{2 \pi  \nu } \tan ^2\alpha+2 e^{2 \pi  \nu } \tan \alpha +e^{2 \pi  \nu }-1}-\frac{\ln \left(1-e^{2 \pi  \nu }  \left(1+\tan\alpha\right)^2\right)}{\pi}\right],~~~~~~~~~~~~\\
   \lim_{q\rightarrow 1,|\nu|>>1,f_p\rightarrow 0}&&{\cal M}^{(\alpha)}_{3,q} = 0.                                                                     \eea
                                                    As a result in the large mass limiting situation with $q\rightarrow 1$ the long range correlation can be expressed in terms of R$\acute{e}$nyi entropy as: 
                                          \bea \lim_{q\rightarrow 1,|\nu|>>1, f^{(\alpha)}_p\rightarrow 0}S_{q,\bf intr}(\alpha)&\approx& \frac{\nu ^3}{3} \left[\frac{2 e^{2 \pi  \nu } \left(1+\tan\alpha\right)^2 \left\{\nu+\ln \left(1+\tan\alpha\right) \right\}}{e^{2 \pi  \nu } \tan ^2\alpha+2 e^{2 \pi  \nu } \tan \alpha +e^{2 \pi  \nu }-1}-\frac{\ln \left(1-e^{2 \pi  \nu }  \left(1+\tan\alpha\right)^2\right)}{\pi}\right]\nonumber\\&=& S_{\bf intr}(\alpha)=\lim_{|\nu|>>1, f^{(\alpha)}_p\rightarrow 0}{\bf c_6}(\alpha,\nu).\eea                                      
          In fig.~(\ref{ix5c}), fig.~(\ref{i5xc}), fig.~(\ref{is5xc}), fig.~(\ref{i5xcv}), fig.~(\ref{ci5xcv}), we have demonstrated the behaviour of R$\acute{e}$nyi entropy for $q=0.9$, $q=0.7$, $q=0.5$, $q=0.3$ and $q=0.1$ with respect to the mass parameter $\nu^2$. Here we did the computation in $D=4$ de Sitter space in absence ($f^{(\alpha)}_p=0$) of axionic source.  Similarly in fig.~(\ref{ix5a}), fig.~(\ref{i5xa}), fig.~(\ref{is5xa}), fig.~(\ref{i5xav}), fig.~(\ref{ci5xav}), we have demonstrated the behaviour of R$\acute{e}$nyi entropy for $q=0.9$, $q=0.7$, $q=0.5$, $q=0.3$ and $q=0.1$ with respect to the mass parameter $\nu^2$. Additionally, the largest eigenvalue of the density matrix ($q\rightarrow\infty$) in absence and presence of axionic source are plotted in fig.~(\ref{axz}) and fig.~(\ref{axza}). Here we did the computation in $D=4$ de Sitter space in presence ($f^{(\alpha)}_p=10^{-7}$) of axionic source. In this both the cases we also have normalised the R$\acute{e}$nyi entropy with the result obtained from conformal mass parameter $\nu=1/2$ in presence of $\alpha$ vacuum. For a given value of the parameter $q$
                    we have shown the plots for $\alpha=0$ (\textcolor{red}{\bf red}), $\alpha=0.03$ (\textcolor{blue}{\bf blue}), $\alpha=0.1$ (\textcolor{green}{\bf green}) and $\alpha=0.3$ (\textcolor{violet}{\bf violet}) in both the cases. Here we observe the following features:
                    \begin{itemize}
                    \item For $q=0.9$ case in absence of the axionic source (see fig.~(\ref{ix5c})) in the large mass parameter range ($\nu^2<0$) normalised R$\acute{e}$nyi entropy asymptotically approaches towards zero value. On the other hand in the small mass parameter range ($\nu^2>0$) it show oscillations in a periodic fashion. Here the amplitude of the oscillation is larger for $\alpha=0.3$ compared to the other values of $\alpha$. Also it is important to note that, at $\nu=1/2$, $\nu=3/2$ and $\nu=5/2$ we get extrema for the oscillation. Further in presence of the axionic source (see fig.~(\ref{ix5a})) in the large mass parameter range ($\nu^2<0$) normalised R$\acute{e}$nyi entropy rapidly approaches to zero value for all values of the parameter $\alpha$ considered in this paper. Also in the small mass parameter range ($\nu^2>0$) the amplitude of the oscillation is significantly large for $\alpha=0.3$. Also it is observed that for $\nu^2>0$ the long range correlation is larger in presence of the axionic source. But for $\nu^2<0$ the long range correlation is rapidly decaying with $f^{(\alpha)}_p=10^{-7}$ and asymptotically decaying with $f^{(\alpha)}_p=0$ for all values of $\alpha$.
                    
                    \item For other values of the parameter $q$ i.e. $q=0.7$, $q=0.5$, $q=0.3$ and $q=0.1$ cases in absence of the axionic source (see fig.~(\ref{i5xc}), fig.~(\ref{is5xc}), fig.~(\ref{i5xcv}) and fig.~(\ref{ci5xcv})) in the large mass parameter range ($\nu^2<0$) normalized R$\acute{e}$nyi entropy asymptotically approaches towards zero value. On the other hand in the small mass parameter range ($\nu^2>0$) it show oscillations in a periodic fashion. Here the amplitude of the oscillation is larger for $\alpha=0.3$ compared to the other values of $\alpha$. Also it is important to note that, at $\nu=1/2$, $\nu=3/2$ and $\nu=5/2$ we get extrema for the oscillation. Further in presence of the axionic source (see fig.~(\ref{i5xa}), fig.~(\ref{is5xa}), fig.~(\ref{i5xav}) and fig.~(\ref{ci5xav})) one can observe the exact behaviour as observed without any source contribution. It also implies that for all $q<0.9$ the normalised R$\acute{e}$nyi entropy is insensitive to the source contribution.
                    
                    \item For $q\rightarrow\infty$ case in absence (see fig.~(\ref{axz})) and presence of the axionic source (see fig.~(\ref{axza})) variation of normalised R$\acute{e}$nyi entropy with $\nu^2$ for all values of the parameter $\alpha$ is similar. It is important to note that the amplitudes of the oscillations in $\nu^2>0$ region and the saturation value in $\nu^2<0$ region is larger in presence of axionic source.
                    \end{itemize}

           Next, in fig.~(\ref{zxx5c}), fig.~(\ref{px5a}), fig.~(\ref{pcx5aa}), fig.~(\ref{px5aaa}) and fig.~(\ref{z5c}), fig.~(\ref{p5a}), fig.~(\ref{pc5aa}), fig.~(\ref{p5aaa}), we have depicted the behaviour of R$\acute{e}$nyi entropy with respect to the parameter $\alpha$ in absence ($f^{(\alpha)}_{p}=0$) and presence ($f^{(\alpha)}_{p}=10^{-7}$) of axionic source for the mass parameter $\nu^2>0$. In all figures it is observed that a crossover takes place for $\nu^2=1/4,9/4,25/4$ (\textcolor{green}{\bf green}), $\nu^2=1/16,9/16,25/16$ (\textcolor{blue}{\bf blue}) and $\nu^2=0$ (\textcolor{red}{\bf red}) with small values of the parameter $\alpha$. We also observe that for $\nu^2=1/4,9/4,25/4$ (\textcolor{green}{\bf green}) R$\acute{e}$nyi entropy decreases with increasing value of the parameter $\alpha$. On the other hand, for $\nu^2=1/16,9/16,25/16$ (\textcolor{blue}{\bf blue}) and $\nu^2=0$ (\textcolor{red}{\bf red}) R$\acute{e}$nyi entropy increases with increasing value of the parameter $\alpha$. Additionally, in presence of axionic source the R$\acute{e}$nyi entropy is slightly larger compared to the result obtained in absence of source contribution. Also it is observed that no crossover takes place for $\nu^2=-1/2$ (\textcolor{green}{\bf green}), $\nu^2=-1/4,-9/4,-25/4$ (\textcolor{blue}{\bf blue}) and $\nu^2=-1/16,-9/16,-25/16$ (\textcolor{red}{\bf red}) with all values of the parameter $\alpha$. Also it is important to note that, for all values of $\nu^2<0$ R$\acute{e}$nyi entropy increases with increasing value of the parameter $\alpha$. Further in fig.~(\ref{z5}), fig.~(\ref{z5a}), fig.~(\ref{z5aa}), fig.~(\ref{z5aaa}), we have shown the variation of R$\acute{e}$nyi entropy with respect to the parameter $q$ in absence and presence of axionic source for $\alpha=0$ (\textcolor{purple}{\bf purple}), $\alpha=0.03$ (\textcolor{orange}{\bf orange}), $\alpha=0.1$ (\textcolor{cyan}{\bf cyan}) and $\alpha=0.3$ (\textcolor{brown}{\bf brown}) respectively. It is observed that for small values of the parameter $q$ the value of the R$\acute{e}$nyi entropy for a given value of $\alpha$ always increase. On the other hand for small values of the parameter $q$ R$\acute{e}$nyi entropy saturates to a finite small value.                               
                                                      
          \end{itemize} 
                                                                                                                                                                                                                                                                                                                      

\section{\textcolor{blue}{Summary}}
\label{4}

To summarize, in this paper, we have addressed the following issues:
\begin{itemize}
	\item  First we have presented the computation of entanglement entropy in de Sitter space in presence of axion with a linear source contribution in the effective potential as originating from $\textcolor{red}{\bf Type~IIB}$ string theory. To demonstrate this we have derived the axion wave function in an open chart.
	
	\item Next using the $\alpha$ vacuum state we have expressed the wave function in terms of creation and annihilation operators. Further applying Bogoliubov transformation on $\alpha$ vacuum state we  have constructed the expression for reduced density matrix.
	
	\item  Further, using reduced density matrix we have derived the entanglement entropy, which is consistent with ref.~\cite{Maldacena:2012xp} if we set $\alpha=0$. In the $\nu^2<0$ range we have derived analytical result for the entanglement entropy. Finally, we have used numerical approximations to estimate entanglement entropy with any value of $\nu^2$.
	
	\item  We have also computed the R$\acute{e}$nyi entropy in presence of axion source. In absence of the source this result is consistent with  ref.~\cite{Maldacena:2012xp} in $q\rightarrow 1$ limit. Here in $\nu^2<0$ region we have provided the analytical expression for the R$\acute{e}$nyi entropy. We have also used numerical techniques to study the behaviour of R$\acute{e}$nyi entropy and largest eigenvalue of the density matrix with any value of $\nu^2$.

	\item Our result provides the necessary condition to generate non vanishing entanglement in primordial cosmology due to axion. 
	\end{itemize}

The future directions of this paper are appended below:
\begin{itemize}

\item Using the derived expression for density matrix for generalised $\alpha$ vacua, one can further compute any $n$ point long range quantum correlation to study the implications in the context of primordial cosmology. It is expected that this result will surely help to understand the connection between the Bell's inequality violation, quantum entanglement and primordial non-Gaussianity. 

\item  Till now we have studied the necessary condition for generating non zero entanglement entropy in primordial
cosmology. In this connection one can further compute quantum discord, entanglement negativity etc., which play a significant role to quantify long range quantum correlations without necessarily involving quantum entanglement.
\end{itemize}

	\section*{\textcolor{blue}{Acknowledgments}}
	SC would like to thank Quantum Gravity and Unified Theory and Theoretical Cosmology Group, Max Planck Institute for Gravitational Physics, Albert Einstein Institute and Inter University Center for Astronomy and Astrophysics, Pune for providing the Post-Doctoral
	Research Fellowship. SP acknowledges the J. C. Bose National Fellowship for support of his research. Last but not the least, We would all like to acknowledge our
	debt to the people of India for their generous and steady support for research in natural sciences.
\appendix
\section{\textcolor{blue}{\bf Wave function for Axion using Bunch Davies vacuum}}
        \label{3av1}
        Further using Eqn~(\ref{axi}) the field equation of motion for the axion can be written as \cite{Choudhury:2017bou}:
        \bea \left[(a(t))^{-3}\partial_{t}\left(a^3(t)\partial_{t}\right)-(Ha(t))^{-2}\hat{\bf L}^2_{\bf H^3}+m^2_{axion}\right]\phi&=&\mu^3,\eea
       where the scale factor $a(t)$ in de Sitter open chart is given by, $a(t)=\sinh t/H$.
       Here the Laplacian operator $\hat{\bf L}^2_{\bf H^3}$ in ${\bf H^3}$ satisfies the following eigenvalue equation \cite{Sasaki:1994yt}:
              \bea \hat{\bf L}^2_{\bf H^3}{\rm\cal Y}_{plm}(r,\theta,\phi)&=&\frac{1}{\sinh^2r}\left[\partial_{r}\left(\sinh^2r~\partial_{r}\right)+\frac{1}{\sin\theta}\partial_{\theta}\left(\sin\theta~\partial_{\theta}\right)+\frac{1}{\sin^2\theta}\partial^2_{\phi}\right]{\rm\cal Y}_{plm}(r,\theta,\phi)=-(1+p^2){\rm\cal Y}_{plm}(r,\theta,\phi),~~~~~~~~\eea     
where ${\cal Y}_{plm}(r,\theta,\phi)$ represents orthonormal eigenfunctions which can be written as:
                      \bea  {\cal Y}_{plm}(r,\theta,\phi)&=&\frac{\Gamma\left(ip+l+1\right)}{\Gamma\left(ip+1\right)}~\frac{p}{\sqrt{\sinh r}}~{\cal P}^{-\left(l+\frac{1}{2}\right)}_{\left(ip-\frac{1}{2}\right)}\left(\cosh r\right)Y_{lm}(\theta,\phi).\eea          
The total solution of the equations of motion can be written as:
              \bea \Phi(t,r,\theta,\phi)&=&\int^{\infty}_{0} dp \sum_{\sigma=\pm 1}\sum^{\infty}_{l=0}\sum^{+l}_{m=-l}\left[a_{\sigma p lm}{\cal U}_{\sigma p lm}(t,r,\theta,\phi)+a^{\dagger}_{\sigma p lm}{\cal U}^{*}_{\sigma p lm}(t,r,\theta,\phi)\right],\eea  
              where ${\cal U}_{\sigma p lm}(t,r,\theta,\phi)$ forms complete basis of mode function, $ {\cal U}_{\sigma p lm}(t,r,\theta,\phi)=\frac{H}{\sinh t}\chi_{p,\sigma}(t){\cal Y}_{plm}(r,\theta,\phi).$
              Here $\chi_{p,\sigma}(t)$ forms a complete set of positive frequency function. Also this can be written as a sum of complementary ($\chi^{(c)}_{p,\sigma}(t)$) and particular integral  ($\chi^{(p)}_{p,\sigma}(t)$) part, as given by $\chi_{p,\sigma}(t)=\chi^{(c)}_{p,\sigma}(t)+\chi^{(p)}_{p,\sigma}(t)$.
             Explicitly the
       complementary and particular integral part can be expressed as \cite{Choudhury:2017bou}:
     \bea
 &&\chi^{(c)}_{p,\sigma}(t)=\chi^{(c)}_{-p,\sigma}(t)=
                            \displaystyle \frac{1}{2\sinh\pi p}\left[\frac{\left(e^{\pi p}-i\sigma~e^{-i\pi\nu}\right)}{\Gamma\left(\nu+\frac{1}{2}+ip\right)}{\cal P}^{ip}_{\left(\nu-\frac{1}{2}\right)}(\cosh t_{\bf R/L}) 
                            -\frac{\left(e^{-\pi p}-i\sigma~e^{-i\pi\nu}\right)}{\Gamma\left(\nu+\frac{1}{2}-ip\right)}{\cal P}^{-ip}_{\left(\nu-\frac{1}{2}\right)}(\cosh t_{\bf R/L})\right],~~~~~~~~~
                                                                                               \\
                                                                                                                                                                                       &&\chi^{(p)}_{p,\sigma}(t)=\mu^3\sinh^2 t\sum^{\infty}_{n=0}\frac{1}{\left(p^2-p^2_{n}\right)}\chi^{(c)}_{p_{n},\sigma}(t)\int dt^{'}~\chi^{(c)}_{p_{n},\sigma}(t^{'}), \eea 
    where 
    the parameter $\nu$ is defined as,
   $ \nu=\sqrt{\frac{9}{4}-\frac{m^2_{axion}}{H^2}}=\sqrt{\frac{9}{4}-\frac{\mu^3 b}{f_a H^2}}=\sqrt{\frac{9}{4}-\frac{\Lambda^4_G}{f^2_a H^2}}.$



\end{document}